\documentclass[sigplan,screen]{acmart}
\usepackage{cleveref}
\usepackage{caption}
\usepackage{subcaption}
\usepackage{multirow}
\usepackage{float} 
\usepackage{makecell} 

\AtBeginDocument{%
  \providecommand\BibTeX{{%
    \normalfont B\kern-0.5em{\scshape i\kern-0.25em b}\kern-0.8em\TeX}}}

\newcommand{\tebr}{\textit{Token-EBR}}
\newcommand{\ntebr}{\textit{Naive Token-EBR}}
\newcommand{\pftebr}{\textit{Pass-first Token-EBR}}
\newcommand{\ptebr}{\textit{Periodic Token-EBR}}
\newcommand{\aftebr}{\textit{Amortized-free Token-EBR}}
\newcommand{\myparagraph}[1]{\noindent\textbf{#1}}

\begin{document}

\title{Are Your Epochs Too Epic? Batch Free Can Be Harmful}

% \affiliation{%

% \affiliation{%

% \affiliation{%

% \author{Anonymous authors}

\author{Daewoo Kim}
\affiliation{%
 \institution{University of Waterloo}
 \country{Canada}}
\email{daewoo.kim@uwaterloo.ca}

\author{Trevor Brown}
\affiliation{%
 \institution{University of Waterloo}
 \country{Canada}}
\email{trevor.brown@uwaterloo.ca}

\author{Ajay Singh}
\affiliation{%
 \institution{University of Waterloo}
 \country{Canada}}
\email{ajay.singh1@uwaterloo.ca}

% \affiliation{%

% \affiliation{%

% \affiliation{%

% \affiliation{%

% \renewcommand{\shortauthors}{Brown and Kim}

\begin{abstract}
Epoch based memory reclamation (EBR) is one of the most popular techniques for reclaiming memory in lock-free and optimistic locking data structures, due to its ease of use and good performance in practice.
However, EBR is known to be sensitive to thread delays, which can result in performance degradation.
Moreover, the exact mechanism for this performance degradation is not well understood.

This paper illustrates this performance degradation in a popular data structure benchmark, and does a deep dive to uncover its root cause---a subtle interaction between EBR and state of the art memory allocators.
In essence, modern allocators attempt to reduce the overhead of freeing by maintaining bounded thread caches of objects for local reuse, actually freeing them (a very high latency operation) only when thread caches become too large.
EBR immediately bypasses these mechanisms whenever a particularly large batch of objects is freed, substantially increasing overheads and latencies.
Beyond EBR, many memory reclamation algorithms, and data structures, that reclaim objects in large batches suffer similar deleterious interactions with popular allocators.

We propose a simple algorithmic fix for such algorithms to amortize the freeing of large object batches over time, and apply this technique to ten existing memory reclamation algorithms, observing performance improvements for nine out of ten, and over 50\% improvement for six out of ten in experiments on a high performance lock-free ABtree.
We also present an extremely simple token passing variant of EBR and show that, with our fix, it performs 1.5-2.6$\times$ faster than the fastest known memory reclamation algorithm, and 1.2-1.5$\times$ faster than not reclaiming at all, on a 192 thread four socket Intel system. 
\end{abstract}

\ccsdesc[500]{Computing methodologies~Concurrent algorithms}
\ccsdesc[500]{Computing methodologies~Concurrent programming languages}

\keywords{safe memory reclamation, memory allocator, concurrent data structures, memory management}

\settopmatter{printacmref=false, printfolios=false}



\acmYear{2024}\copyrightyear{2024}
\setcopyright{acmlicensed}
\acmConference[PPoPP '24]{The 29th ACM SIGPLAN Annual Symposium on Principles and Practice of Parallel Programming}{March 2--6, 2024}{Edinburgh, United Kingdom}
\acmBooktitle{The 29th ACM SIGPLAN Annual Symposium on Principles and Practice of Parallel Programming (PPoPP '24), March 2--6, 2024, Edinburgh, United Kingdom}
\acmDOI{10.1145/3627535.3638491}
\acmISBN{979-8-4007-0435-2/24/03}

\maketitle

% reclaiming memory for modern concurrent data structures is hard, needs separate reclamation algorithms

% ebr is one of the most popular (fast and easy to use)

% modern concurrent data structures also need to be paired with fast memory allocators that are designed for highly concurrent systems to achieve their full potential (e.g., jemalloc, tcmalloc)

\section{Introduction}
Concurrent data structures serve as crucial building blocks for high performance multicore applications.
Many concurrent data structures assume that memory is automatically garbage collected, which means that in unmanaged environments such as C/C++, they must be paired with a separate safe memory reclamation (SMR) algorithm.
SMR algorithms ensure that a thread can free an object only if no other thread can possibly access the object (or else a segmentation fault could occur).
Epoch based reclamation (EBR) is one of the most widely used reclamation algorithms for concurrent data structures, largely owing its popularity to its ease of use and extremely low overhead.

To obtain the highest performance from a modern concurrent data structure, one must typically also pair it with a fast memory allocator that is engineered for highly concurrent systems.
JEmalloc~\cite{evans2006scalable} and TCmalloc~\cite{ghemawat2005tcmalloc} are two of the most popular choices.
Both of these allocators have seen widespread use and undergone significant development by industry.
For example, JEmalloc is the standard allocator for the FreeBSD operating system, as well as the allocator for the Firefox web browser.

SMR algorithms and memory allocators have historically been developed and optimized for relatively small scale systems with one or two processor sockets with fairly uniform memory architectures.
However, in recent years, processor and system designs have become increasingly non-uniform.
Recent AMD processors follow a hierarchical chiplet design, wherein a 64 core processor is actually a set of eight interconnected chiplets of eight cores each, with their own internal interconnects and local caches.
In server environments, large multi socket servers are becoming common, with Amazon AWS M7i cloud servers running on four socket, 192 hardware thread Intel configurations similar to the experimental system we use in this paper.
It is important to consider the impact such increasing non-uniformity has on the performance of popular algorithms and system software. %, a range of factors, including memory locality, scheduling and interconnect congestion influence the software performance. This extends to both standalone software and software that interfaces with multiple libraries.
To this end, we conduct a rigorous study of the performance of a fast concurrent data structure paired with a state of the art implementation of EBR called DEBRA~\cite{brown2015reclaiming} and two popular memory allocators\textendash JEmalloc and TCmalloc.
Our experiments show that common workloads that show strong scaling up to moderate thread counts (up to two processor sockets) can experience severe performance degradation when running at very high thread counts (using three or four sockets).
Subsequent experiments delve into the root cause of this performance problem, which turns out to be a subtle interaction between the EBR algorithm and the memory allocator.

With the help of a new visualization technique that we call timeline graphs, we discover that algorithms such as DEBRA that free objects in large batches circumvent a key optimization in JEmalloc.
This optimization is intended to avoid the overhead of returning an object to a remote thread that allocated it (i.e., its \textit{owner}),
by instead placing the object in a local buffer that the local thread can subsequently allocate from. %to satisfy a subsequent (local) allocation.
Every object allocated locally from this buffer is an object that does not need to be freed remotely, back to its owner.
Freeing a large batch of objects can overflow this buffer, triggering an extremely high latency \texttt{free} call (on the order of tens of milliseconds or more) in which many objects are removed from this buffer and freed remotely to their respective owners, incurring extremely high lock contention in the process.
We call this the \textit{remote batch free} (RBF) problem.

Beyond EBR, the RBF problem appears to broadly affect any kind of data structure or algorithm that reclaims memory in batches.
Many algorithms, including nearly all modern SMR algorithms, treat batching objects to free them together as a kind of optimization.
In this work, we argue that this is actually an anti-pattern. %we observe that a typical optimization

We propose a simple algorithmic fix called \textit{amortized free} (AF) to mitigate the RBF problem.
AF essentially reverses the typical batch freeing optimization.
Whenever an algorithm would normally free a batch of nodes, we instead place them in an auxiliary list and free them gradually over time.
Freeing objects gradually creates opportunities for a thread to reallocate these objects locally, rather than returning them to their owners.

To demonstrate the generality of this work, we reproduce the problem in both JEmalloc and TCmalloc, implement our solution in \textit{ten} different SMR algorithms, and leverage new insights to design a new, exceedingly simple, high performance SMR algorithm.
Although our primary goal is to shed light on the subtle interactions between modern algorithms and system software, our experiments show that our solution is strikingly effective, and it results in dramatic improvements over prior art, as we explain below.
%the effectiveness of this approach. results validate its efficacy in significantly improving performance with a synthetic data structure benchmark.
%Encouragingly, \textit{amortized free} applies to to many other memory reclamation algorithms %, and, in fact, , we sjthe fix is general and can applied to many state of the art reclamation algorithms (especially, those which use batching and are sensitive to thread delays) 
%and significantly improve their performance.
%Additionally, we present an extremely simple variant of EBR, \aftebr, that is easier to implement and performs better than the state of the art. %And also design a benchmark to implement and evaluate multiple reclamation algorithms with the fix. 

\medskip

\noindent\textbf{In summary, we make the following contributions.}
\noindent\begin{itemize}
  \item We reveal a subtle and highly impactful negative performance interaction between modern memory allocators and algorithms that free objects in batches, as well as a simple algorithmic fix for this problem called amortized freeing.
  \item We introduce timeline graphs, a new visualization that makes it substantially easier to understand the behaviour of threads in workloads with high latency operations.
  % and modern memory allocators. high impact performance problem in EBR is uncovered through a series of experiments and its root cause is investigated using a new visualization technique\textemdash \textit{timeline graphs}. We propose a simple algorithmic fix of the problem which we call \textit{amortized freeing}. 
  \item We apply amortized freeing to \textit{ten} of the most popular SMR algorithms, resulting in performance improvements in nine of the ten algorithms, \textit{doubling} the performance of six of the algorithms on average with 192 threads.
  \item Leveraging the insights gained in this work, we develop an extremely simple variant of EBR, \aftebr{}, that outperforms the fastest existing SMR algorithm by 2.6$\times$ with 192 threads. %. The latter outperforms DEBRA, which at the time of writing is known to be most efficient EBR algorithm, by 1.8X on average across multiple thread counts and by 2.6X at 192 threads. In general, our best variant, \aftebr{} outperforms the techniques we compared against by 1.2-8.8X when the averaged across multiple thread counts. 
  % \textbf{[[[ trevor: by up to how much? by how much on average? by how much at x threads? ]]]}
  %\item We also test the generality of the \textit{amortized freeing} by applying it to improve the performance of multiple state of the art reclamation algorithms encompassing DEBRA, Neutralization based reclamtion, Interval based reclamation, Wait free eras, Hazard eras, Hazard pointers, RCU, and QSBR. The \textit{amortized free} variants on an average outperform their original versions by $\sim$1.4X across multiple thread counts and in most cases by 2X at 192 threads.
  % \item We apply the same technique used to improve DEBRA to a simpler EBR algorithm called \textit{Token EBR}, improving its performance substantially.
  % \item Finally, we identify several performance problems in a state of art memory allocator, and propose and evaluate solutions to these problems. The solutions result in significant performance improvements in some workloads.
  \item To our knowledge, this is the most rigorous and in depth analysis of the behaviour of threads in EBR algorithms, and of the interaction between SMR and memory allocators, to date.
\end{itemize}

The paper is organized as follows.
We discuss background in \Cref{chap:background}, diagnose the RBF problem, identify its root causes, and describe a simple solution in \Cref{chap:reclamation}.
\Cref{chap:tokenebr} presents our new SMR algorithm as a sequence algorithmic improvements.
Along the way, with the help of our timeline graphs, we describe in detail the lessons learned about the behaviour of threads in each variant that lead to our improvements.
Our evaluation appears in \Cref{chap:evaluation}.
Finally, we survey related work in \Cref{chap:related_work} and conclude in \Cref{chap:conclusion}.

\section{Background}
\label{chap:background}
% \textcolor{red}{detail explanation about EBR and other reclamation algorithms(Hazard pointer, etc...?.}
% \textcolor{red}{Leave only JEmalloc and TCmalloc for allocator part.}
%In this chapter, we briefly present background information on several memory reclamation algorithm and memory allocators.
In this paper, we study two of the most popular allocators, JEmalloc~\cite{evans2006scalable} and TCmalloc~\cite{ghemawat2005tcmalloc}, as well as a new allocator from Microsoft Research called MImalloc~\cite{leijen2019mimalloc}.
Necessary details about the allocators' designs are included inline in the paper, but a more detailed description of the three allocators appears in the supplementary material.

We give a brief primer on EBR.
%
% \subsection{Safe Memory Reclamation} \label{sec:background-reclamation}
%
% single threaded reclamation is easy (why?)
% concurrent reclamation is harder (why?)
%     crashes if you free?
% \textit{Safe Memory Reclamation} solves the challenge of avoiding \textit{use-after-free} errors in lockfree and optimistic data structures 
%
% \subsubsection{Epoch Based Reclamation (EBR)} \label{sec:debra}
%
% In a single threaded system, reclaiming memory is relatively easy, even in languages like C++ that require programmers to explicitly free objects.
% However, reclaiming memory in concurrent data structures is more difficult.
% If a thread calls \texttt{free()} to reclaim an object while another thread can still access that object, this can result in the program crashing.
% 
% 
% ebr is popular (why?)
%   it's easy and fast
% high level description: time is divided into epochs. instead of freeing, threads buffer objects during an epoch to be freed later. when the epoch changes, objects from old epochs can be freed.
%
Epoch based reclamation is one of the most widely used memory reclamation algorithms for concurrent data structures, particularly because it is easy to use and offers high performance.
At a high level, in EBR, instead of freeing objects immediately, threads store these objects in a buffer called a \textit{limbo bag}, to be freed later as a batch.
The program execution is logically divided into \textit{epochs}, and whenever the epoch changes, objects that were placed in a limbo bag in older epochs are freed.
%time is divided into epochs, and whenever a thread removes an object from the data structure, instead of freeing it, the thread places the object in a \textit{limbo bag}.

% \subsubsection{The DEBRA algorithm}

Brown~\cite{brown2015reclaiming} introduced DEBRA, which has been shown to be one of the fastest EBR algorithms.
% The improved algorithms presented in this paper are compared with DEBRA experimentally.
%In this paper, we will compare our improved algorithms with DEBRA experimentally.
A basic understanding of DEBRA will be important for our performance investigation below.
In DEBRA, there is a global epoch number, and a single writer multi reader announcement array with one slot per thread, in which each thread stores the number of the epoch it is currently in.
Threads update their announced epoch number at the start of each data structure operation.
The global epoch can be advanced when all threads have announced it.
To efficiently determine whether all threads have announced the current epoch, each thread periodically (once every $k$ operations) reads \textit{one} other thread's epoch, proceeding in round robin order.
The first thread to notice that all threads have announced the current epoch will update the global epoch.

\section{Diagnosing the RBF Problem} 
\label{chap:reclamation}
\begin{figure}[tb]
    \begin{subfigure}{0.49\linewidth}
        \includegraphics[width=\linewidth]{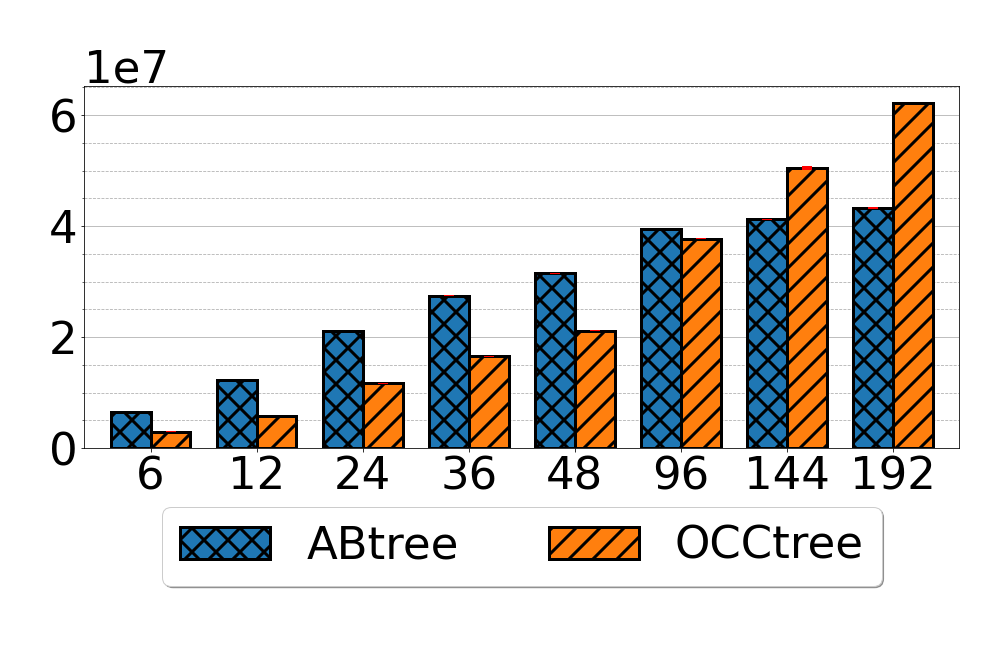}
        \vspace{-6mm}
        \caption{Performance}
        \label{subfig:Brons vs AB w/ recl p}
    \end{subfigure}
    \begin{subfigure}{0.49\linewidth}
        \includegraphics[width=\linewidth]{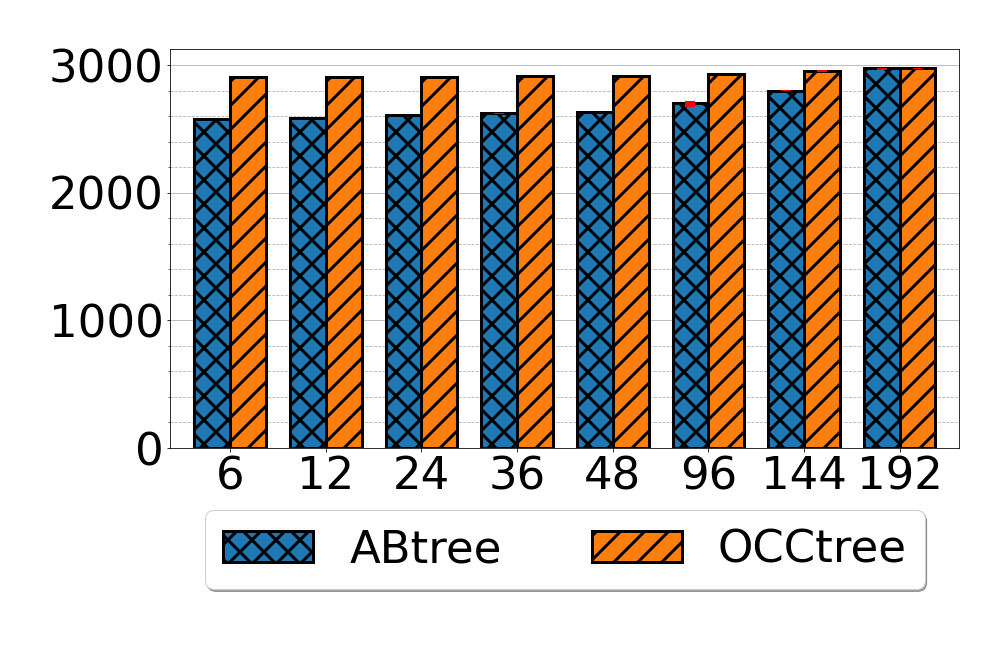}
        \vspace{-6mm}
        \caption{Peak memory usage (MiB)}
        \label{subfig:Brons vs AB w/ recl m}
    \end{subfigure}
    \begin{subfigure}{0.49\linewidth}
        \includegraphics[width=\linewidth]{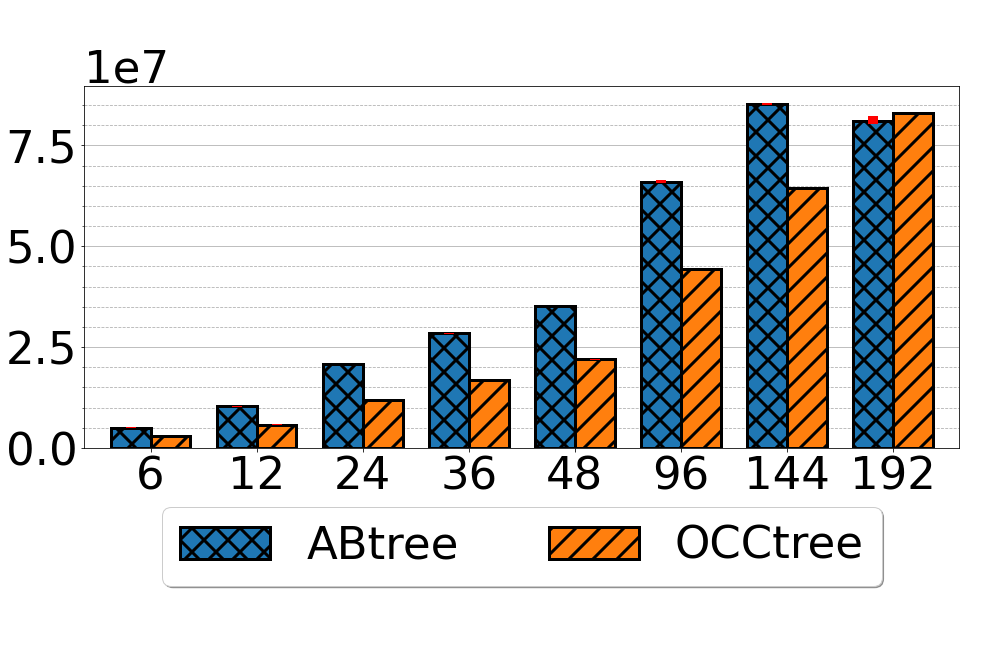}
        \vspace{-6mm}
        \caption{Performance}
        \label{subfig:Brons vs AB w/o recl p}
    \end{subfigure}
    \begin{subfigure}{0.49\linewidth}
        \includegraphics[width=\linewidth]{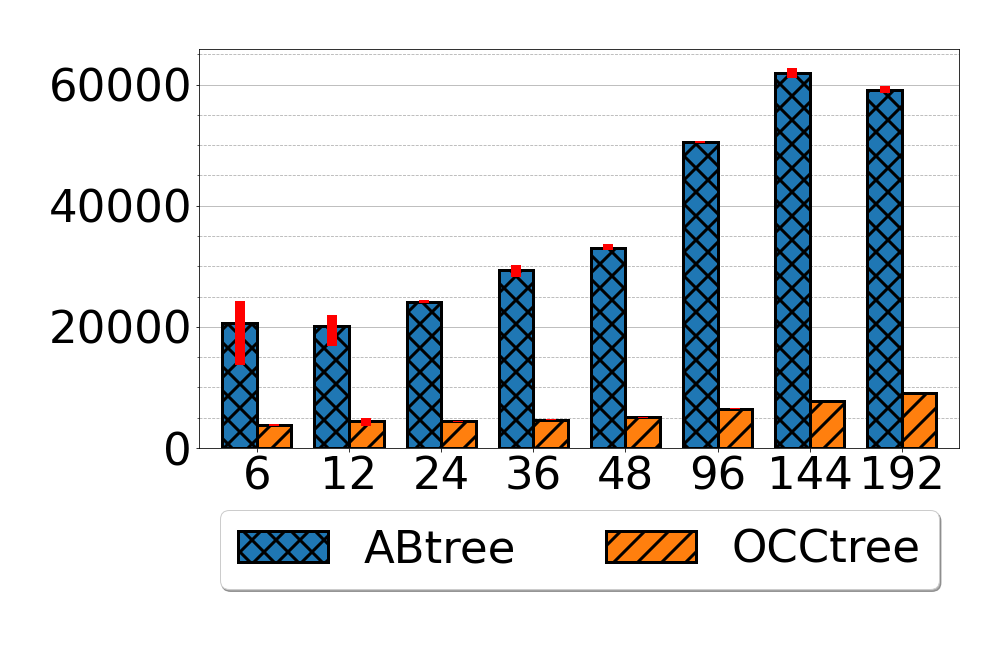}
        \vspace{-6mm}
        \caption{Peak memory usage (MiB)}
        \label{subfig:Brons vs AB w/o recl m}
    \end{subfigure}
    % \vspace{-2mm}
    \caption{Performance (operations per second) and peak memory usage with JEmalloc, for OCCtree and ABtree, using DEBRA (upper) vs leaking memory (lower). 
    }
    \label{fig:Brons vs AB}
\end{figure}

% \subsection{Symptom: Poor Scaling with NUMAs} %Motivation : Identifying the Problem and establishing that it is generic}

In running experiments on concurrent tree data structures, we observed that on large scale NUMA systems with four processor sockets, some data structures that scale similarly when running on a single socket scale drastically differently when running on four sockets.
For example, consider an AVL tree by Bronson et~al.~\cite{bronson2010practical} that uses optimistic concurrency control (hereafter, OCCtree), and a concurrency friendly variant of a B-tree by Brown~\cite{brown2017techniques} that uses lock-free techniques (hereafter, ABtree).
We performed experiments to compare the performance of these data structures in a simple microbenchmark.
Both data structures allocated memory using JEmalloc and reclaimed memory using DEBRA.

\paragraph{Experimental Methodology}
For each thread count $n \in \{6, 12, 24, 36, 48, 96, 144, 192\}$, three trials were performed.
In each trial, $n$ threads access the same data structure, and for five seconds, repeatedly: flip a coin to decide whether to insert or delete a key, and perform the resulting operation on a uniform random key in a fixed key range $[0, 2 \times 10^7)$.
Note that with a fixed key range, and 50 percent insert operations and 50 percent delete operations, in the steady state (after threads have run for a long time), the data structure should contain half of the key range.
To avoid measuring the performance of a data structure as its size is changing at the beginning of the trial, the five second measured portion of the experiment begins once the size of the data structure stabilizes.
In each trial, the total number of insert and delete operations performed per second, across all threads (i.e., \textit{throughput}), is reported.
Each data point shows the average throughput over three trials, and the minimum and maximum throughput over these three trials is shown using error bars.
This experimental methodology is similar to that in other papers that study concurrent memory reclamation (e.g., \cite{brown2015reclaiming, wen2018interval, singh2021nbr, sheffi2021vbr}).

\paragraph{System}
This experiment was run on a four socket Intel Xeon Platinum 8160 with 384 GB of DRAM.
Each socket has 24 cores running at 2.1GHz nominal frequency with turbo boost up to 3.7GHz, and 48 logical processors with hyperthreading enabled, for a total of 96 cores and 192 hyperthreads across all sockets.
The operating system was Ubuntu 20.04 LTS, with kernel version 5.8.0-55.
Code was compiled with \texttt{g++} 9.3.0-17 with \texttt{-O3} optimization and \texttt{std=c++14}.
All experiments were run with \texttt{numactl --interleave=all} and threads were pinned to logical processors such that thread counts 1-24 run on a single socket, without hyperthreading, 25-48 run in a single socket with hyperthreading, and as additional threads are added, the same pinning pattern is applied to additional sockets as needed.
(Sockets are populated with one thread per logical processor before additional sockets are used.) %\footnote{We have included, in the system description, all of the details identified by Grammoli~\cite{gramoli2017information} as crucial for reproducing shared memory multicore experiments. That paper provides, among other things, a concise study of the impact of thread pinning and turbo boost on performance. In our experiments turbo boost is enabled, since it is typically enabled on real systems, and it has a substantial impact not only on low thread counts, but also on high thread counts.

%On the experimental system, when the benchmark was run on one or two cores, the clock speeds at those cores were 3.5-3.6GHz. 4 cores ran at 3.4GHz, 8 at 3.3-3.4GHz, 12 at 3.2-3.3GHz, 24 at 2.8GHz, and the all-core turbo speed was 2.8GHz, despite the base frequency of the processor being 2.1GHz. This can make some of the scaling results in this paper appear somewhat worse than they would be if turbo boost were disabled, however the absolute performance results at every thread count are better than they would be in that case.}

\paragraph{Symptom: poor scaling on NUMA}
%\textcolor{blue}{
\Cref{subfig:Brons vs AB w/ recl p} clearly demonstrates the scaling issue described above.
%shows the results of this experiment. \textbf{<<< we feel that using "this" here is a bit weird and irritating as it is too far from the exp description in Motivation section.>>>}
Both data structures scale well up to 48 threads, but the ABtree stops scaling above 96 threads, whereas the OCCtree continues to scale. % (albeit not as well as it did up to 96 threads).
One significant difference between the ABtree and the OCCtree is that the ABtree allocates one or two large nodes (240 bytes each) per insert or delete operation, whereas the OCCtree only allocates one small node (64 bytes) per insert operation (and does not allocate memory in a delete operation).

\paragraph{Hypothesis: memory reclamation is a bottleneck}
\Cref{subfig:Brons vs AB w/ recl m} shows the peak memory used (on average over three trials) for each of the data points in \Cref{subfig:Brons vs AB w/ recl p}.
Interestingly, peak memory usage for the ABtree is not much higher than for the OCCtree. %, although it is \textit{somewhat} higher at high thread counts, where its scaling is diminished. \textbf{<< feels like saying two things in same sentence, confusing, could just say mem usage is similar as that is the main point we wanna convey to readers. >>}
%
%
% Rebuttal: Sorry this was not very clear. We should have said that each operation on the Bronson et al. tree allocates and frees less memory, on average, than an operation on the ABtree, and thus a smaller fraction of the execution time needs to be spent on memory reclamation. "Peak memory usage" should not have been part of this hypothesis.
%
%
% This is perhaps surprising since the nodes of the ABtree are approximately 3$\times$ larger than the nodes of the OCCtree, and the ABtree allocates at least one node in each update operation, whereas the OCCtree only allocates one node in its insertion operations (and performs no allocation in its deletion operations).
This is perhaps surprising, since the ABtree allocates and reclaims substantially more memory per operation, on average, than the OCCtree.
One might thus expect the ABtree to have substantially higher peak memory usage, especially at low thread counts, where it also performs \textit{more operations} than the OCCtree.
But, that is clearly not the case.
% One possible explanation is that, on average, each operation on the ABtree tree allocates and frees more memory than an operation on the OCCtree.
One possible explanation is that, compared to the OCCtree, the ABtree spends a larger fraction of the execution time allocating and reclaiming memory than performing other useful work in the data structure (traversing and/or modifying it), limiting its performance.

To confirm this hypothesis, we disable memory reclamation for both data structures, and simply leak memory, to see whether this closes the performance gap between them.
The results in \Cref{subfig:Brons vs AB w/o recl p} largely confirm this hypothesis. %$ that performance of Brown ABtree outperformed Bronson AVL tree when there is no memory reclamation.
%Brown ABtree was degraded when it reclaimed memory using DEBRA.
%Memory management policy influenced data structures differently because Brown ABtree used more memory compared to Bronson AVL tree.
It is clear that the ABtree allocates much more memory (\Cref{subfig:Brons vs AB w/o recl m}), which means the allocator must do more work, and the memory reclamation algorithm must do much more work to maintain a small memory footprint.
This suggests the performance degradation comes from memory management, either in the allocator or reclamation algorithm.

\subsection{Investigating the Reclamation Bottleneck} \label{sec:investigating the reclamation overhead}

\begin{figure}[t]
    % \begin{subfigure}{0.99\linewidth}
    %     \includegraphics[width=\linewidth]{figures/freetime_immediate_jemalloc_96_interleave_pinyes.png}
    %     \caption{96 threads}
    %     \label{subfig:freetime_immediate_jemalloc_96_interleave}
    % \end{subfigure}
    % \begin{subfigure}{0.99\linewidth}
    %     \includegraphics[width=\linewidth]{figures/freetime_immediate_jemalloc_192_interleave_pinyes.png}
    %     \caption{192 threads}
    %     \label{subfig:freetime_immediate_jemalloc_192_interleave}
    % \end{subfigure}
    \noindent\begin{subfigure}{0.45\linewidth}
        \includegraphics[width=\linewidth]{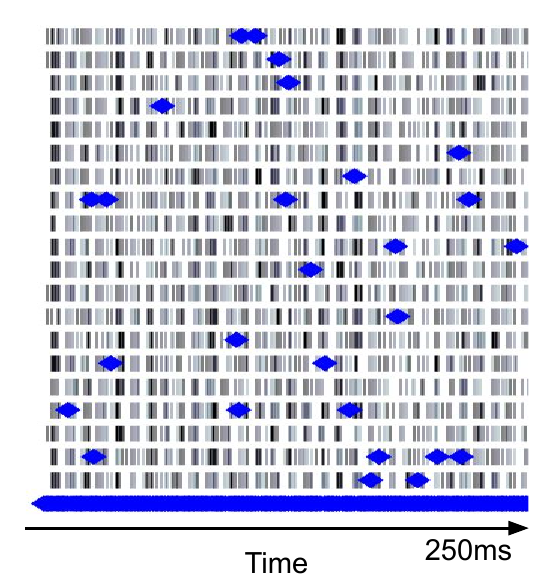}
        \vspace{-6mm}
        \caption{96 threads}
        \label{subfig:freetime_immediate_jemalloc_96_interleave_pinno_zoom}
    \end{subfigure}
    \hspace{0.06\linewidth}
    \begin{subfigure}{0.45\linewidth}
        \includegraphics[width=\linewidth]{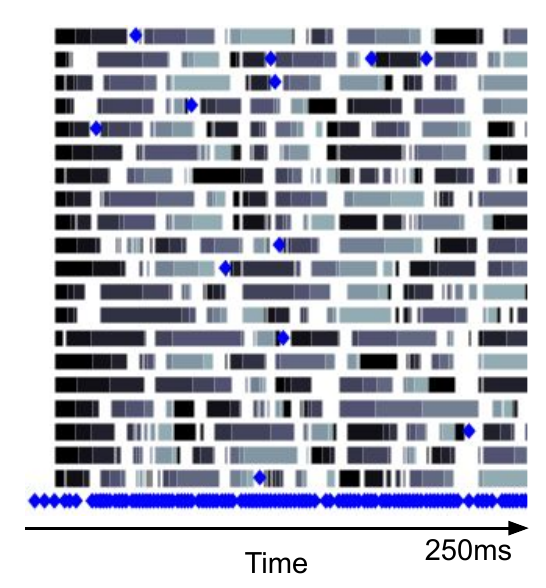}
        \vspace{-6mm}
        \caption{192 threads}
        \label{subfig:freetime_immediate_jemalloc_192_interleave_pinno_zoom}
    \end{subfigure}
    \vspace{-3mm}
    \caption{Timeline graphs showing how much time threads spend freeing \textbf{batches} of nodes as epochs change with JEmalloc. (Y-axis = thread ID, blue dot = epoch change, space between boxes = time spent accessing the data structure.)}
    \label{fig:immediate_jemalloc_240_interleave}
\end{figure}

%\subsubsection{Visualizing thread delays}
% Need Rewrite
Crucially, although DEBRA maintains a similar memory footprint for all of the thread counts in our experiment (\Cref{subfig:Brons vs AB w/ recl m}), the performance of the ABtree substantially flattens at high thread counts in \Cref{subfig:Brons vs AB w/ recl p}.
This suggests that DEBRA is struggling to keep up with the amount of garbage being produced.
DEBRA, like all EBR algorithms, is very sensitive to thread delays: \textit{a single delayed thread can prevent all threads from reclaiming garbage}~\cite{singh2021nbr, singhTPDS}.
To determine if this performance degradation is caused by thread delays, we visualized the behaviour of threads, specifically the time spent freeing objects, in a graph that we call a \textit{timeline}.

\paragraph{Timeline Graphs}
We have not seen timeline graphs used elsewhere in the literature, but our experience with them suggests they are a very useful tool for investigating performance problems caused by thread delays.
We implemented a highly efficient mechanism to allow threads to record data (specifically two time stamps and a user specified value) in memory to be printed to files at the end of an experiment, with very little impact on performance.
We did not measure any significant impact on performance when recording up to 100,000 timeline events per thread.
Our code will be made available publicly when this paper is published.

\Cref{subfig:freetime_immediate_jemalloc_96_interleave_pinno_zoom} and \Cref{subfig:freetime_immediate_jemalloc_192_interleave_pinno_zoom} are timeline graphs showing how much time threads spend freeing nodes as epochs change in the ABtree with JEmalloc for 96 threads (left) and 192 (right). %, and, for clarity, an enlarged region of the graph is shown in \Cref{subfig:freetime_immediate_jemalloc_96_interleave_pinno_zoom}.
Rows represent different threads, and the x-axis represents time.
For clarity, only 20 of the running threads and a representative 250ms of the 5 second trial are shown.
Full graphs showing all threads and the entire 5 seconds of the trial appear in the supplementary material.
%with 96 threads running, but only 20 shown for clarity. %shown in \Cref{subfig:freetime_immediate_jemalloc_96_interleave}.
Each box represents a \textit{reclamation event}, i.e., the time spent freeing a \textit{batch} of objects removed from the data structure in a previous epoch.
Boxes are coloured to make it easier to differentiate neighbouring events. % which epoch a thread is in when it is freeing (i.\textbf{<<<<Q: is it for differentiating the neighbouring freeing events of a thread and not the epoch number in which it is being freed, we could rephrase it better?>>>>}.
Blue dots represent a thread successfully changing the global epoch number. % made by each thread.
All blue dots are also projected at the bottom of the graph to give a visual indication of how often the epoch changes overall. %made by all threads together.
(This makes it easier to identify periods of time during which the epoch is not changed by any thread.)

\paragraph{Comparing Timelines: 96 vs 192 Threads}
% When a \textit{bad} period begins, more and more threads begin having trouble freeing all of their objects before the next epoch begins.
% To understand why, one needs to understand a little bit more about DEBRA.
Comparing \Cref{subfig:freetime_immediate_jemalloc_96_interleave_pinno_zoom} and \Cref{subfig:freetime_immediate_jemalloc_192_interleave_pinno_zoom}, it is clear that more time is spent freeing objects, and individual reclamation events are much longer, with 192 threads than with 96 threads.
(The time scales for these two enlarged graphs are the same.)
%Note that the timeline graph for 192 (\Cref{subfig:freetime_immediate_jemalloc_192_interleave}) is more densely populated, and its reclamation events are longer, which is especially clear in \Cref{subfig:freetime_immediate_jemalloc_192_interleave_pinno_zoom}.
Recall from \Cref{chap:background} that DEBRA effectively amortizes the cost of scanning threads' epoch announcements over many operations.
This keeps per operation overhead low, but it means that doubling the number of threads should, on average, double the length of time needed to advance the epoch.
This should in turn, double the amount of garbage to be reclaimed in each epoch, and double the length of time needed to free a batch, on average.
So, we would \textit{expect} the lengths of the reclamation events to be \textit{twice} as long for 192 threads as for 96 threads.\footnote{And, indeed, the lengths of reclamation events approximately double from 48 threads to 96 threads. See supplementary material for additional JEmalloc, TCmalloc and MImalloc results. Timeline graphs for TCMalloc showed similar behaviour.}
However, we see that these events are many times longer than expected, which suggests there is an additional factor at work.

\subsection{Root Cause of Long Reclamation Events} \label{sec:root cause of the EBR delays}

Further investigation using Linux Perftools led to the realization that poor performance in JEmalloc, such as when running on four sockets, is usually accompanied by a large fraction of the total cycle count being spent in function called \texttt{je\_tcache\_bin\_flush\_small}. % and \texttt{je\_malloc\_mutex\_lock\_slow}.
\Cref{tab:je_free_overhead_vs_threads} summarizes \texttt{perf} results to support the following discussion, and also quantifies how the total number of epochs changes as the thread count increases.
These results confirm that the cost of freeing objects becomes prohibitive at high thread counts, preventing the data structure from scaling.

According to the source code for this version (5.0.1-25) of JEmalloc, when a thread invokes \texttt{free}, it places the freed object in a thread local buffer, and then checks whether the buffer is filled beyond a given threshold.
If so, it takes a large number of objects from that buffer (approximately 3/4 of the buffer), and for each object, does the following.
First, it identifies which \textit{bin} the object belongs to\footnote{In this paper, we are not giving a detailed description of the ``bin an object belongs to,'' but intuitively, one can imagine it is the heap from which the object was originally allocated (and, hence, the heap to which it should be returned).}.
If the object was originally allocated by a different thread, this bin might reside on a remote core, or even a remote socket.
The thread locks the bin, then iterates over all objects in its buffer (while holding the lock), and for each object that belongs to this bin, it performs the necessary bookkeeping to free the object to that bin.

\begin{table}[tb]
\centering
\begin{tabular}{llllll}
\toprule
\textbf{threads} & \textbf{ops/s} & \textbf{epochs} & \textbf{\% free} & \textbf{\% flush} & \textbf{\% lock} \\
\midrule
    48 & 35.9M & 12631 & 11.5 & 9.9 & 4.9 \\
    96 & 45.3M & 5176 & 39.3 & 38.3 & 24.6 \\
    192 & 43.4M & 1980 & 59.5 & 58.8 & 39.8 \\
\bottomrule
\end{tabular}
\vspace{1mm}
\caption{
    JEmalloc \texttt{free} overhead.
    \% free = time spent in \texttt{free}.
    \% flush = time spent in \texttt{je\_tcache\_bin\_flush\_small}.
    \% lock = time spent in \texttt{je\_malloc\_mutex\_lock\_slow}.
}
% \vspace{-5mm}
\label{tab:je_free_overhead_vs_threads}
\end{table}

Freeing a batch, especially the very large batches induced by high thread counts in DEBRA, triggers this mechanism often.
This defeats the purpose of the buffer, which is intended to give a thread an opportunity to reallocate freed objects from the buffer, rather than always performing the bookkeeping required to move objects to remote bins.
%In some timeline graphs with 192 threads, 30 percent or more of the total cycles were spent in \texttt{je\_tcache\_bin\_flush\_small}, and 15 percent or more of the total time in \texttt{je\_malloc\_mutex\_lock} calls that originate from \texttt{je\_tcache\_bin\_flush\_small}.

Most of the overhead of \texttt{je\_tcache\_bin\_flush\_small} comes from lock contention, as we discovered using \texttt{perf}.
With 192 threads, 39.8\% of the total time was spent in the function \texttt{je\_malloc\_mutex\_lock\_slow}, compared to 9.9\% with 48 threads.
After applying the solution we present below in \Cref{sec:amortized-free}, the time spent in this function is reduced to \textbf{5.5\% with 192 threads} and less than 0.1\% with 48 threads.

In summary, it is extremely expensive in JEmalloc for \texttt{free} to return objects to the remote threads that allocated them.
To avoid this overhead, a thread frees to a local buffer, and subsequently allocates from that buffer.
Every object allocated from that buffer is an object that \textit{does not need to be freed to a remote thread in future}.
Freeing a large batch overflows the buffer, forcing all objects to be freed remotely, causing extremely high lock contention.

\subsection{A Simple Solution: Amortized Free} \label{sec:amortized-free}

Given the above, it is clear that freeing large batches should be avoided wherever possible, at least when using JEmalloc. %is likely to reduce performance when using JEmalloc.
%As the results above show, the performance degradation as the number of threads changes from 96 to 192 seems to be correlated with a substantial increase in the length of time needed to reclaim batches.
%This suggests that the degradation could be caused by freeing large batches. %problem in DEBRA seems to be caused by the batched free behaviour \cmnt{we should refer to the observation of longer boxes in timeline graphs}.
Although batching is inherent in EBR (and is also common in many other memory reclamation algorithms), once a batch of nodes has been \textit{identified as safe to free}, one does not necessarily need to \textit{free them immediately} as a batch.
One could instead, for example, %To address this issue, we %had the idea to
place the batch in a thread local \textit{freeable} list, and gradually \texttt{free} objects one by one, each time a data structure operation is performed.
We call this approach \textit{amortized free} (AF).\footnote{As we will see, the amortized free approach is quite effective in improving multiple allocators and memory reclamation algorithms.
However, another option is to modify the allocator itself to be sensitive to the possibility of batch frees coming from the reclamation algorithm. This would be an interesting direction for future work.}

In addition to gradually freeing objects from this list, one \textit{could} optimize further by \textit{allocating} objects from this list directly.
This would essentially turn this approach into object pooling, avoiding most interaction with the allocator altogether.
We want to show that we can make interaction with the allocator fast---not avoid it---so we do \textbf{not} perform this optimization.\footnote{The results in this paper can also help to explain why memory reclamation algorithms that free large batches, but use object pooling, such as Version Based Reclamation (VBR)~\cite{sheffi2021vbr}, have been shown to outperform some prior EBR algorithms that interact directly with the allocator~\cite{wen2018interval, harris2001pragmatic}.} %: They avoid the high cost of actually freeing their batches.}
%\cmnt{We need to justify why we think immed batch freeing is problematic and why it  is worth trying freeing them slowly. Think we just need to refer back to main conclusions of the prev (investigation) section.}

% [CODE FOR DEBRA RETIRE AND ROTATE BAGS WITHOUT AMORTIZED FREE]

% [CODE WITH AMORTIZED FREE]

% [GRAPH FOR DEBRA SHOWING PERFORMANCE DIFFERENCE]
% \begin{figure}[th]
%     % \begin{subfigure}{0.99\linewidth}
%     %     \includegraphics[width=\linewidth]{figures/freetime_immediate_jemalloc_240_interleave_pinno.png}
%     %     \caption{X-axis = time since first epoch, Y-axis = thread, colour = epoch, blue dots = epoch change, between boxes (white space) = time spent accessing ABtree}
%     %     \label{subfig:freetime_immediate_jemalloc_240_interleave_pinno}
%     % \end{subfigure}
%     % \begin{subfigure}{0.99\linewidth}
%         \includegraphics[width=\linewidth]{figures/unreclaimed_immediate_jemalloc_240_interleave_pinno.png}
%         % \caption{}
%         % \label{subfig:unreclaimed_immediate_jemalloc_240_interleave_pinno}
%     % \end{subfigure}
%     \caption{Average number of unreclaimed objects per thread of the original DEBRA in the ABtree with JEmalloc. X-axis = epoch number, Y-axis = average number of unreclaimed objects per thread}
%     \label{fig:immediate_jemalloc_240_interleave_pinno}
% \end{figure}

\begin{figure}[t]
    \centering
    \noindent\begin{subfigure}{0.45\linewidth}
        \includegraphics[width=\linewidth]{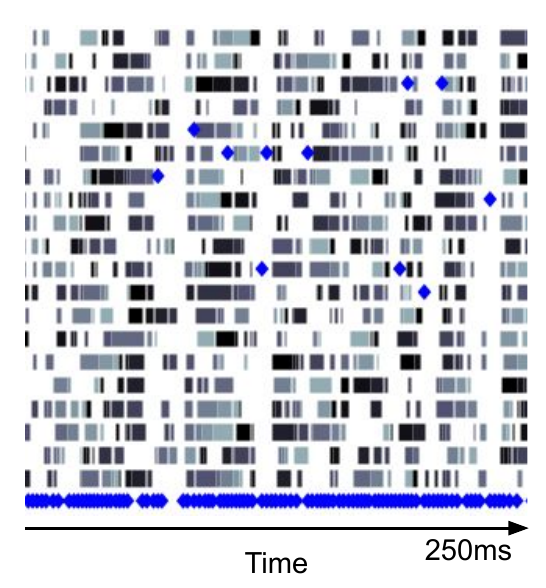}
        \vspace{-6mm}
        \caption{Batch free} \label{subfig:freeone-je192-batch}
    \end{subfigure}
    \hspace{0.06\linewidth}
    \begin{subfigure}{0.45\linewidth}
        \includegraphics[width=\linewidth]{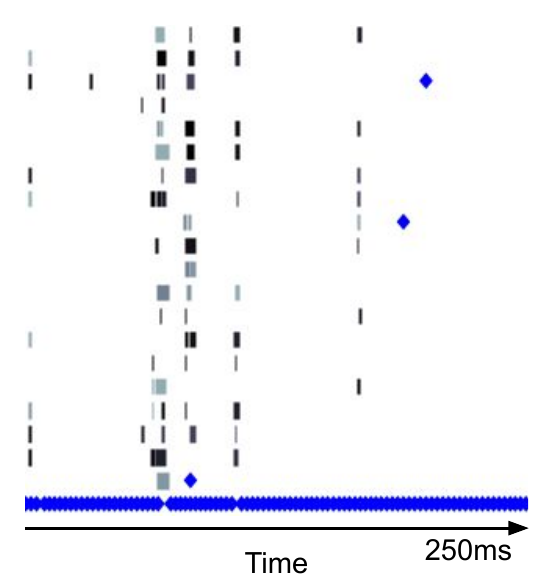}
        \vspace{-6mm}
        \caption{Amortized free} \label{subfig:freeone-je192-amortized}
    \end{subfigure}
\vspace{-3mm}
\caption{
Timeline graphs showing how long \textbf{individual} \texttt{free} calls take for batch free vs amortized free. 192 threads.
%Individual free calls are visualized.
%99.99\% are too small to see.
} %with ABtree, JEmalloc and DEBRA\_AF}
\label{fig:timeline_amortized_jemalloc}
\end{figure}

\begin{table}[tb]
\centering
\noindent\begin{tabular}{llllll}
\toprule
\textbf{approach} & \textbf{ops/s} & \textbf{freed} & \textbf{\% free} & \textbf{\% flush} & \textbf{\% lock} \\
\midrule
JE batch & 43.4M & 114M & 59.5 & 58.8 & 39.8 \\
JE amort. & 111.3M & 292M & 19.2 & 17.6 & 5.5 \\
\bottomrule
\end{tabular}
\vspace{1mm}
\caption{Amortized free vs. batch free. 192 threads. freed = number of objects freed. \% free = time spent freeing. \% flush = time spent in \texttt{je\_tcache\_bin\_flush\_small}. \% lock = time spent in \texttt{je\_malloc\_mutex\_lock\_slow}.}
\label{tab:je_192_amortized_vs_batch}
\end{table}

In the rest of the paper, unless otherwise stated, \textbf{every graph shows results for the ABtree and JEmalloc with 192 threads}, and follows the same methodology as described at the beginning of \Cref{chap:reclamation}.

\paragraph{Effect on the Overhead of Reclamation}

\Cref{subfig:freeone-je192-batch} and \Cref{subfig:freeone-je192-amortized} illustrate the impact of amortized freeing on a timeline graph.
%Compared to \Cref{subfig:freetime_immediate_jemalloc_192_interleave}, the graph is much less dense.
Whereas \Cref{fig:immediate_jemalloc_240_interleave} visualized the time to free \textit{batches} of objects, these new timelines show \textbf{individual free calls}.
In both \Cref{subfig:freeone-je192-batch} and \Cref{subfig:freeone-je192-amortized} the vast majority of \texttt{free} calls are too short to be visible on the graph.
However, there is a clear difference: the batch free approach performs many more high latency \texttt{free} calls.
The small number of high latency \texttt{free} calls that remain in the amortized free graph are further analyzed in the supplementary material.

The supporting \texttt{perf} results in \Cref{tab:je_192_amortized_vs_batch} show that the amortized free algorithm spends $\approx$ 3$\times$ fewer cycles freeing memory and $\approx$ 7$\times$ fewer spinning on locks.
However, the amortized free algorithm is 2.6$\times$ faster.
%the total time spent freeing is reduced by approximately $3\times$.
%
%
%\textcolor{blue}{
% \Cref{subfig:freetime_immediate_jemalloc_240_interleave} also shows the total time spent freeing.
%We ran the same experiments using Linux perftools \texttt{perf record} to profile the benchmark and allocator, to determine what percentage of the total execution time in our experiments were spent freeing objects, and to analyze the impact of amortized freeing.
%The results appear in \Cref{tab:alloc_immed_amort}. % in the rows containing ``JE batch'' and ``TC batch''
%With 240 threads, 60.8\% of the total cycles were spent freeing.
%\textbf{<<as a reader I cannot convince myself that time spent freeing is  60\%. From fig I think it is ~90\%. all black columns constitute 90\% of timeline.>>}
%just for freeing.
%}
%
% The results, which appear in \Cref{tab:je_192_amortized_vs_batch}, show that 59.5\% of the total execution time is spent freeing objects with the batch free approach, resulting in a total throughput of 43.4M ops/second.
% In contrast, the version with amortized free spends only 19.2\% of its time freeing, resulting in a total throughput of 111.3M ops/second.
These results suggest that amortized freeing can yield substantial performance improvements by reducing the overhead of freeing nodes.

% \begin{table}[tb]
% \centering
% \begin{tabular}{llll}
% \toprule
% \textbf{Parameters} & \textbf{Throughput}  & \textbf{\makecell{\% free time}} & \textbf{\makecell{\# freed}} \\
% \midrule
% % \rowcolor[rgb]{0.753,0.753,0.753}\nbrp\cite{singh2021nbr}        & Yes                           & cond. lock-free   & OS + \hp \\ 
% % \nbr\cite{singh2021nbr}                                             & Yes                           & cond. lock-free   & OS + \hp \\
% JE batch&43.4M&59.5&114M\\
% JE amort.&111.3M&19.2&291M\\
% TC batch&25.7M&52.5&68.6M\\
% TC amort.&83.5M&11.8&219M\\
% \bottomrule
% \end{tabular}
% \vspace{1mm}
% \caption{Improvement in throughput of allocators when amortized freeing is used over when batch freeing is used. 192 threads. 5 second runs.}
% \label{tab:alloc_immed_amort}
% \end{table}

Crucially, note that the algorithms that perform amortized freeing spend a third as much time freeing nodes (or less), even though they allocate and free more than twice as many nodes (because of the increased throughput).
Effectively, JE batch spends $5 \times 0.595 \approx 2.98$ seconds freeing 114M objects, averaging 38.3M objects freed/second, whereas JE amortized averages 304M objects freed/second.
This suggests the improvement in the overhead of managing memory is on the order of 8$\times$.
%\cmnt{I guess 400\% may not be correct (?). We wanna say that reducing the time spent in freeing (memory managing overhead) by half leads to doubling of throughput\Cref{tab:alloc_immed_amort}}

\begin{figure}[t]
    \centering
    % \begin{subfigure}{0.99\linewidth}
    %     \centering
        \includegraphics[width=\linewidth]{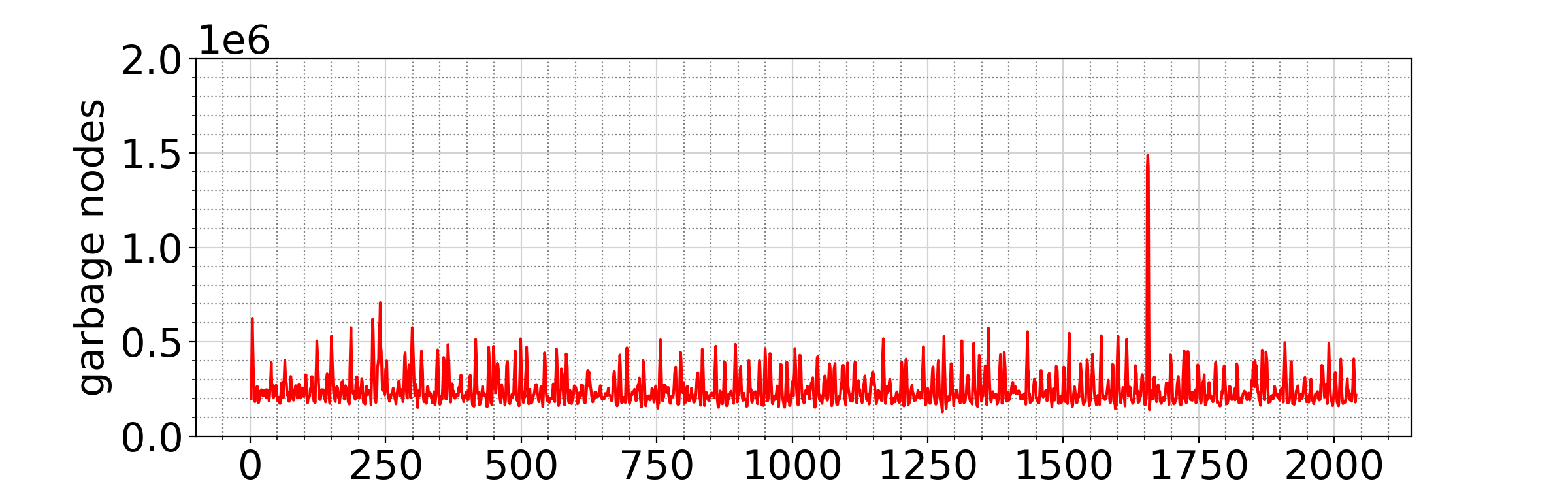}
        % \includegraphics[width=\linewidth]{figures/ext_fig/debras_bar.png}
        % \subcaption{immediate batch freeing}
        % \label{subfig:unreclaimed_immediate_jemalloc_192_interleave_pinyes}
    % \end{subfigure}
    % \vfill
    % \begin{subfigure}{0.99\linewidth}
        % \centering
        \vspace{-2mm}
        \includegraphics[width=\linewidth]{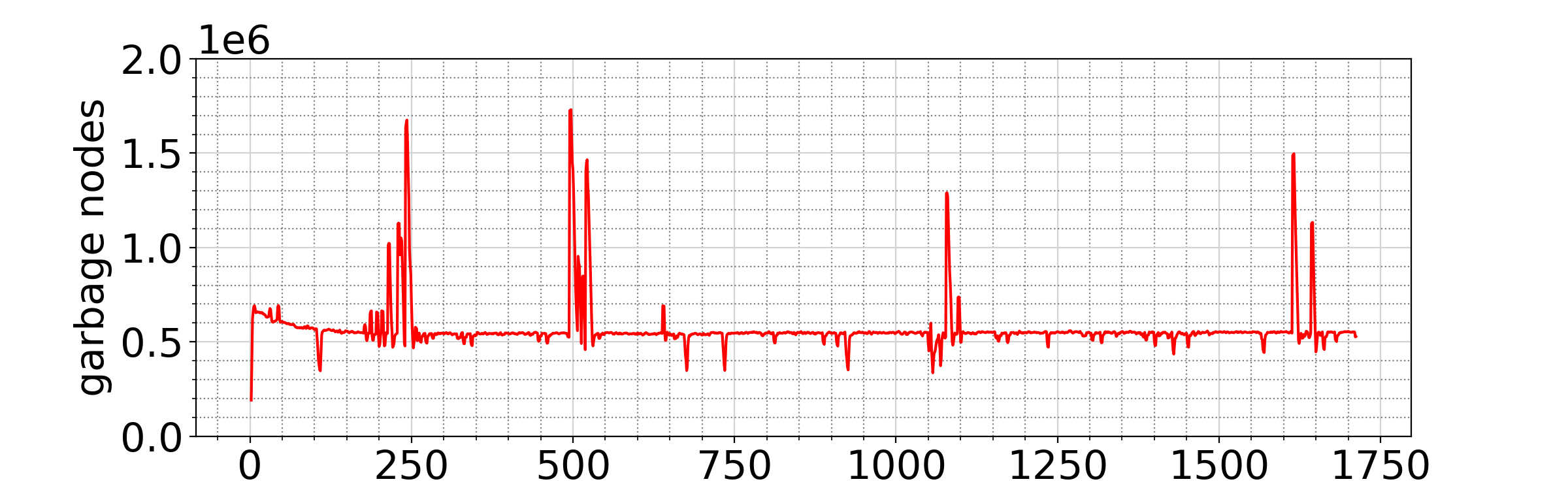}
        % \subcaption{amortized freeing}
        % \label{subfig:unreclaimed_amortized_jemalloc_192_interleave_pinyes}
    % \end{subfigure}
    \vspace{-6mm}
    \caption{Comparing the number of garbage nodes in each epoch for batch free (upper) and amortized free (lower) reveals the latter has a smoothing effect on memory usage.} %  on the number of garbage nodes in each epoch.} %Number of unreclaimed objects in each epoch in the ABtree with DEBRA and JEmalloc, for batch freeing vs amortized freeing.} % X-axis = epoch number, Y-axis = garbage nodes) objects per thread.}
    \label{fig:immediate_amortized_jemalloc}
\end{figure}

\paragraph{Effect on Unreclaimed Garbage}%
\Cref{fig:immediate_amortized_jemalloc} shows how the amount of unreclaimed garbage changes as the epoch advances for the original batch free approach (upper) and the amortized free approach (lower).
To be precise, these graphs show: for each epoch, the sum, over all threads $t$, of the number of unreclaimed nodes $t$ had in its limbo bag when $t$ began that epoch.
%Since threads advance to a new epoch only when they receive the token, a point on the x-axis does not directly correspond to a single point in time.
Amortized freeing substantially reduces the number of peaks in the graph while maintaining only a slightly larger amount of garbage on average.
% . the changes in the number of unreclaimed nodes over the increasing epoch numbers as the experiment progresses for the original immediate batch freeing approach (\Cref{subfig:unreclaimed_immediate_jemalloc_192_interleave_pinyes}) and for our amortized freeing approach (\Cref{subfig:unreclaimed_amortized_jemalloc_192_interleave_pinyes}). 
% % The impact of this change on the average number of unreclaimed objects per thread (i.e., the amount of garbage waiting to be freed) is shown in \Cref{subfig:immediate_jemalloc_240_interleave_pinno} and \Cref{subfig:amortized_jemalloc_240_interleave_pinno}.
% %These figures show average number of unreclaimed objects per thread.
% % original text
% The amounts of unreclaimed objects in \Cref{subfig:unreclaimed_immediate_jemalloc_192_interleave_pinyes} and \Cref{subfig:unreclaimed_amortized_jemalloc_192_interleave_pinyes} are similar, but in the original DEBRA (\Cref{subfig:unreclaimed_immediate_jemalloc_192_interleave_pinyes}), the number of unreclaimed objects drops much more sharply as a batch is immediately freed. % has sharper drops from after spikes.

\paragraph{These Results Generalize to TCmalloc}

To determine whether these results are specific to JEmalloc, or are more general, we repeated all of the above experimentation and analysis with TCmalloc, another popular allocator.
We found that the same problem arose and the same solution yielded large improvements.
A small preview of the results appears in \Cref{tab:other-allocators}, where amortized freeing can be seen to improve performance by 3.25$\times$.
Additional results appear in the supplementary material.
%To demonstrate that amortized batch freeing is not specific to JEmalloc, we repeated the same experiment with TCmalloc.
% As one can see in \Cref{tab:alloc_immed_amort}, amortized batch freeing also improves total throughput drastically (by 3.25$\times$) when TCmalloc is used.

% [mention the caveat that freeing only one object each time might not be enough for some data structures. mention catchup freeing.]
% need timeline graph

% Through further experimentation, we realized that similar effects also occur in TCmalloc.
% The fact that such anti-synergies between the allocator and reclamation algorithm are possible (and possibly are even common) leads naturally to a deeper study of the structure of modern allocators.

\paragraph{MImalloc Sidesteps the Problem Altogether}

MImalloc, on the other hand, is essentially immune to the problem we describe above by design. % avoids the root cause of delays due to EBR batch freeing.
In MImalloc, a remote free synchronizes on a particular \textbf{page's} free list.
In contrast, in JEmalloc, a remote batch free synchronizes on one of 4T \textit{arenas}, where $T$ is the number of hardware threads, meaning if $T$ threads all perform remote free operations, we would expect approximately 1/4 of them to contend with one another.
In TCmalloc, a remote batch free synchronizes on a global cache, which is even worse.
%\footnote{In JEmalloc, technically the number of free lists (arenas in JEmalloc terminology) is $4T$ where $T$ is the number of hardware threads, and thread IDs are hashed to determine which free list a thread will access. The simplified discussion here still applies in this more complex setting. If two or more threads hash to the same free list, the performance degradation is even worse.}
%That is, free operations in JEmalloc and TCmalloc synchronize on a per-thread object, and MImalloc synchronizes on a per-page object.
This makes it relatively inexpensive to immediately free an individual object to a remote thread in MImalloc, as doing so will cause contention only if another thread is simultaneously freeing another object that was allocated \textit{from the same page}.
On the other hand, if a non-trivial number of threads are performing remote batch frees in JEmalloc or TCmalloc, they are much more likely to experience contention. % if $T$ is concurrently allocating or freeing memory, or if another thread is also remotely freeing an object to $T$.)
% object to the remote free list of the \textit{thread} that allocated the object, MImalloc frees an object to the original free list that the object is allocated from.
% MImalloc has three object lists per MImalloc page: allocation free list, local free list, and cross thread free list.
% A thread frees object to the cross thread free list of the remote page, if the object belongs to the remote page.
% Due to page-level freelist, MImalloc has low probability of contention on cross free lists.
% This unique design allows threads to immediately free objects to the page the objects belongs to, without huge delays caused by contention.
%
As expected, \Cref{tab:other-allocators} shows that amortized free does not improve performance in MImalloc, and in fact worsens it slightly. % the original DEBRA has high throughput, and the amortized batch freeing DEBRA does not improve performance.

MImalloc is quite unique in its approach.
To our knowledge, no other allocator maintains per-page free lists.
As a result, we expect that many other allocators would suffer similar RBF problems to JEmalloc and TCmalloc. % negative interactions with RBF algorithms.
It is worth noting that the JEmalloc results above with amortized freeing are faster than MImalloc, and that programmers are not always free to change the allocator in their environment, so finding workarounds for deficiencies in the most popular allocators is worthwhile.

% immediate       tcmalloc        192     interleave      pinyes  25.7    52.56   672     68625955
% amortized       tcmalloc        192     interleave      pinyes  83.5    11.76   1287    218896820

% immediate       mimalloc        192     interleave      pinyes  104.4   15.55   1483    273050257
% amortized       mimalloc        192     interleave      pinyes  95.0    17.23   1401    249047378

\begin{table}[tb]
\centering
\begin{tabular}{llll}
\toprule
\textbf{approach} & \textbf{ops/s} & \textbf{freed} & \textbf{\% free} \\
\midrule
TC batch & 25.7M & 69M & 52.6 \\
TC amort. & 83.5M & 219M & 11.8 \\
MI batch & 104M & 273M & 15.6 \\
MI amort. & 95.0M & 249M & 17.2 \\
\bottomrule
\end{tabular}
% \vspace{1mm}
\caption{Analysis for additional allocators. 192 threads.}
\label{tab:other-allocators}
\end{table}

% \cmnt{Point worth noting when talking about supremacy of MImalloc: We observe in experiments that best debra (orig or amortized) with MI-MAlloc is slower than best debra-amortized with JEmalloc. For aj debra-AF is 10\% and nbrplus is 13\% faster for jemalloc than best debra with MImalloc.}

% we chose a token based implementation of EBR which to best of our knowledge hasn't been well studied and it's characteristics are not well understood. 

\section{Token-EBR: A Simpler EBR Algorithm}
\label{chap:tokenebr}
% Armed with a new understanding of just how much the performance of an EBR algorithm depends on its interaction with the allocator, we conjecture that some existing EBR algorithms may be overcomplicated and overtuned due to an incomplete understanding of which factors are most important. %their interaction the underlying system.
%Amongst the EBR class of algorithms, DEBRA has been one of the most efficient implementations on modern multicore shared memory machines\cite{brown2015reclaiming, singh2021nbr}, which is fairly complex in its implementation.

In this section, we investigate the question of whether an extremely simple EBR algorithm, paired with amortized freeing, can match or exceed the performance of the state of the art.
To this end, we revisit an old idea: token rings.
To our knowledge, the idea of passing a token around a ring to establish an epoch (which we call \tebr) has not been implemented or evaluated in the peer reviewed literature on safe memory reclamation, appearing only in a thesis by Tam~\cite{tam2006qdo} where the idea is simply sketched at a high level, then dismissed as inefficient.

We consider a sequence of possible implementations of the abstract algorithm, study their characteristics, show that all of the implementations that do not use amortized freeing are unusable in practice, and derive a final implementation that outperforms the state of the art in concurrent memory reclamation.
It was surprising to us that such a simple EBR implementation outperformed the state of the art when paired with \textit{Amortized Freeing}.

We start by explaining the abstract algorithm and why it is correct.

\paragraph{Abstract Algorithm}
Conceptually, the algorithm is straightforward.
%it is an alternative simplest design of EBR where
All threads, $T_1$, $T_2$, $\cdots$ $T_n$, are arranged in a ring.
The token is passed around the ring to define epochs.
More specifically, all threads begin in epoch zero, and enter a new epoch whenever they receive the token.
Each thread has two limbo bags: one for the current epoch (the \textit{current bag}), and one for the previous epoch (the \textit{previous bag}), both initially empty.
%Whenever thread $T_i$ receives a token from its predecessor $T_{i-1}$, it forwards it to its successor $T_{i+1}$. %, as shown in \Cref{Fig:Token_diagram}.
%A thread currently having the token is said to be the \textit{owner}.
Objects that a thread unlinks from the data structure are always placed in the thread's current bag. %Whenever a thread $T_i$ unlinks an object from the data structure, it places it in its current bag.

Whenever $T_i$ begins a new data structure operation, it checks whether it has received the token.
If so, it conceptually enters a new epoch, and can (1) pass the token to the next thread, and (2) free all objects in its previous bag. % and (2) pass the token to the next thread.
(We will discuss the safety of freeing objects below.)
Since its previous bag is now empty, it can simply swap its current and previous bags.
The result is an empty current bag, and a previous bag that contains objects unlinked in the previous epoch (as it should).

\paragraph{Correctness Sketch}
To understand why this algorithm ensures safe memory reclamation (i.e., ensures that a thread frees an object only if no other thread has a pointer to it), consider a time interval $I$ during which the token makes its way around the entire ring.
During this interval, each thread has passed the token, so each thread has begun a new data structure operation.
Suppose an object $O$ is unlinked from the data structure (but not yet freed) before the interval $I$ \textit{begins}.
We argue that $O$ is safe to free at the end of $I$.

Each thread that begins a data structure operation before $O$ is unlinked could potentially still access $O$ until it finishes its operation.
However, at the \textit{end} of interval $I$, each thread has finished its current operation and started a new operation.
And, each thread's new operation began after $O$ was already unlinked from the data structure, and so cannot have obtained a pointer to $O$ by accessing the data structure.
So, no thread can access $O$, and thus $O$ is safe to free.
More broadly, every object that was removed from the data structure before $I$ is safe to free after $I$.

\subsection{Naive \tebr}
In our first implementation of \tebr, which we call \ntebr, at the start of a data structure operation, a thread first (1) checks whether it has received the token, (2) if so frees the contents of its previous bag, swaps its bags and passes the token, in that order. %, and (3) finally passes the token.  the following steps in order. %each thread maintains two bags of objects which contain objects removed (retired) from the data structure, a \textit{current} bag and an \textit{old} bag. Initially, both bags are empty.
%All threads that remove an object from the data structure only place the object in its \textit{current} bag.
%And at the start of a new data structure operation, each thread executes following three steps:
% \begin{enumerate}
%     \item[S1] Checks whether it has received the token.
%     \item[S2] If so, it first frees all objects in its \textit{old} bag, and swaps its \textit{previous} and \textit{current} bags.
%     \item[S3] Forwards the token to the next thread.
% \end{enumerate}
%Swapping, efficiently empties the \textit{current} bag by moving all the retired objects to the empty \textit{old} bag, which would be eventually freed the next time token is received.

% And, whenever a thread begins a new data structure operation, it checks whether it has received the token.
% If so, it first frees all objects in its \textit{old} bag, and swaps its \textit{old} and \textit{current} bags.
% Swapping, efficiently empties the \textit{current} bag by moving all the retired objects to the empty \textit{old} bag, which would be eventually freed the next time token is received.
% % After the swap, the \textit{current} bag is empty, and the \textit{old} bag contains objects waiting to be freed the next time the token is received.
% Finally, the thread passes the token to the next thread, and then proceeds to perform the data structure operation.

\begin{figure}[tb]
    \begin{subfigure}{0.49\linewidth}
        \includegraphics[width=\linewidth]{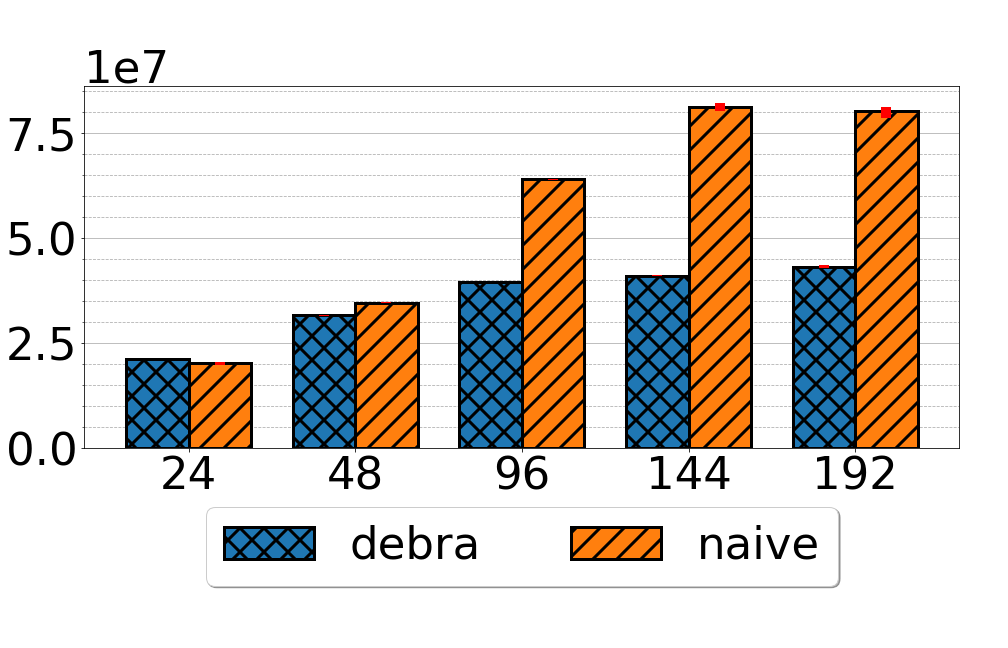}
        \vspace{-7mm}
        \caption{Performance}
        \label{subfig:naive_token_p}
    \end{subfigure}
    \begin{subfigure}{0.49\linewidth}
        \includegraphics[width=\linewidth]{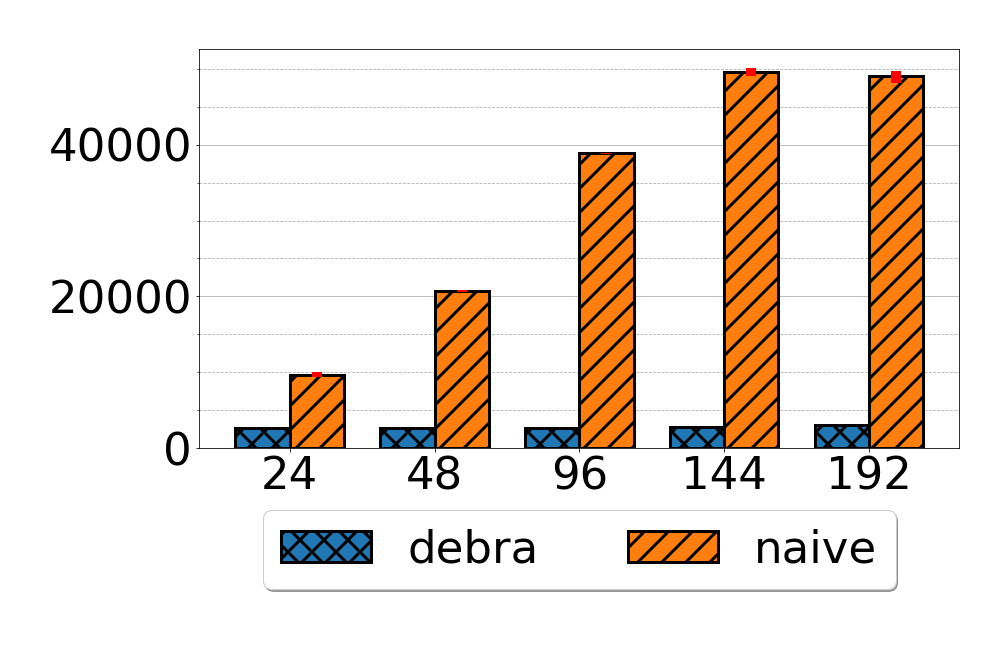}
        \vspace{-7mm}
        \caption{Peak memory usage (MiB)}
        \label{subfig:naive_token_m}
    \end{subfigure}
    % \vspace{-2mm}
    \caption{Performance and peak memory usage with JEmalloc, for ABtree, using \ntebr{}.}
    \label{fig:naive_token}
\end{figure}

% [MULTWe STAGE IMPROVEMENTS TO TOKEN EBR WITH GRAPHS... MAYBE EACH IMPROVEMENT STAGE IN A SUBSUBSECTION OR PARAGRAPH]

\begin{figure}[tb]
    \begin{subfigure}{0.99\linewidth}
        \includegraphics[width=\linewidth]{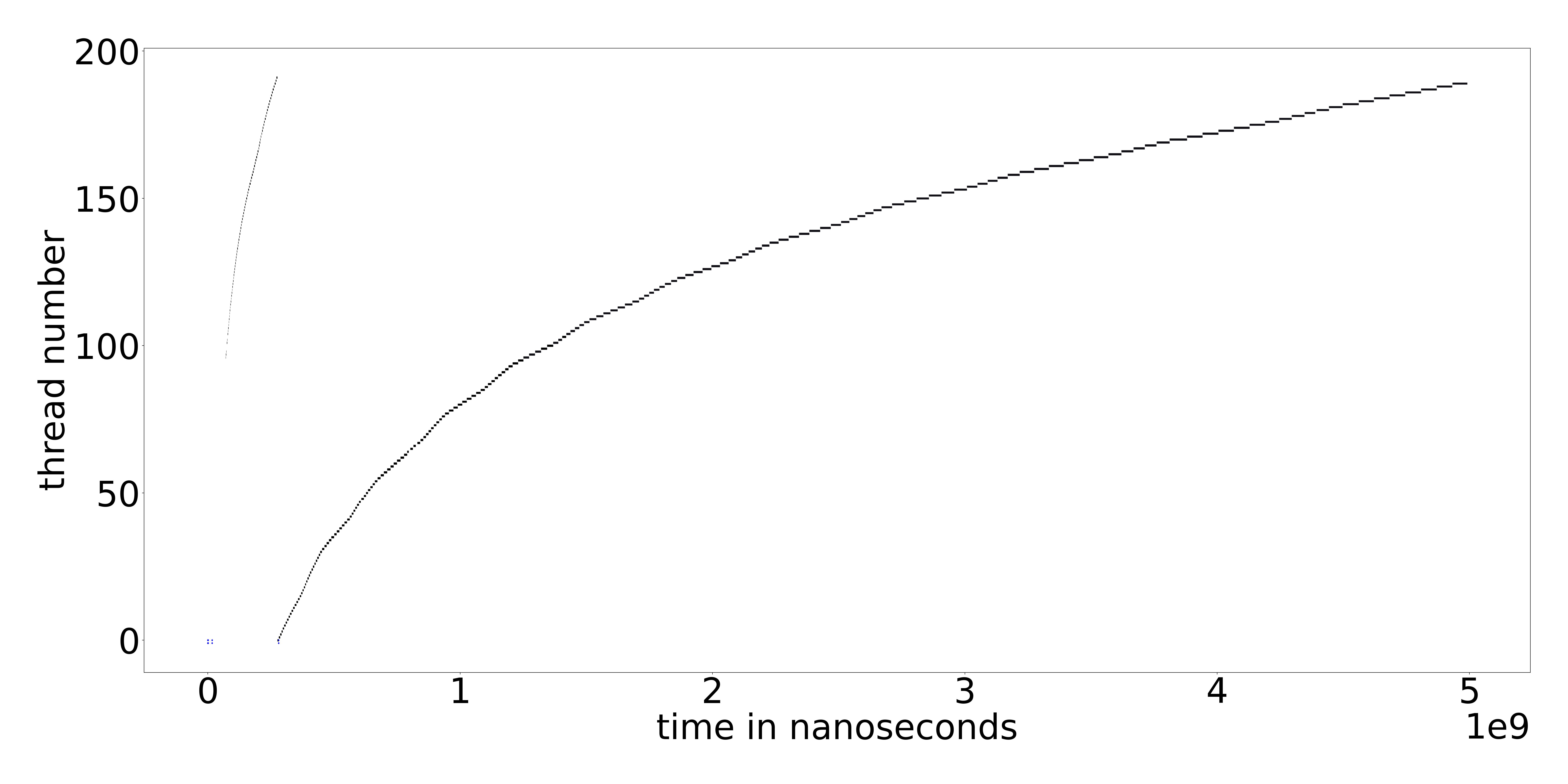}
        %\caption{Timeline graph}
        %\label{subfig:freetime_token1_jemalloc_192_interleave_pinyes}
    \end{subfigure}
    \begin{subfigure}{0.99\linewidth}
        \vspace{-2mm}
        \includegraphics[width=\linewidth]{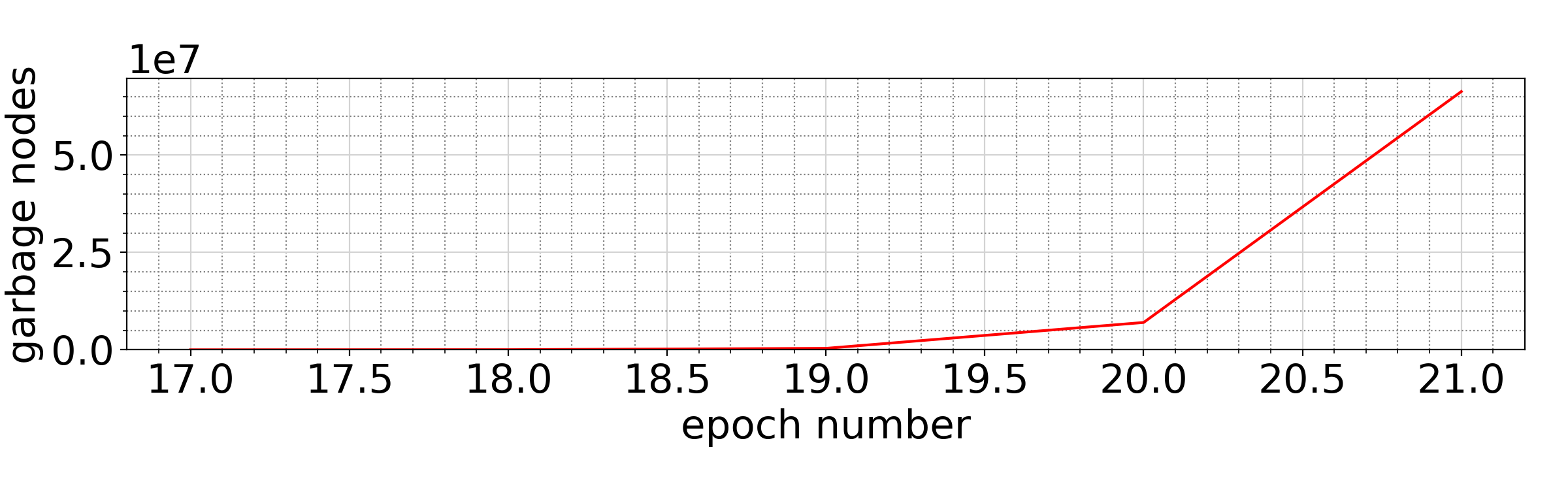}
        %\caption{Average number of unreclaimed objects per thread}
        %\label{subfig:unreclaimed_token1_jemalloc_192_interleave_pinyes}
    \end{subfigure}
    % \vspace{-5mm}
    \caption{Timeline graph showing \textbf{batch} frees (upper) and number of garbage nodes (lower) for \ntebr{}.}
    \label{fig:token1_jemalloc_192_interleave_pinyes}
\end{figure}

At first glance, \Cref{subfig:naive_token_p} suggests that \ntebr{} is a better algorithm than DEBRA.
However, \Cref{subfig:naive_token_m} reveals that it does a terrible job of actually reclaiming memory.
And, all of that time spent not reclaiming memory can be directed towards performing data structure operations, artificially inflating its throughput---at least until the system runs out of memory.

The problem is illustrated in \Cref{fig:token1_jemalloc_192_interleave_pinyes}, which looks quite different from the timeline graphs shown  in \Cref{chap:reclamation}.
Visually, the graph looks like a continuous curve, but the ``curve'' is in fact a sequence of batch frees performed by individual threads, one after another, with no two threads freeing objects concurrently.

This serialization is a direct consequence of the decision to free the contents of the previous bag \textit{before} passing the token.
The next thread cannot free until it receives the token---after the previous thread finishes freeing.
Worse still, while one thread is freeing, $n-1$ other threads are continually performing data structure operations without freeing, accumulating more and more garbage.
Consequently, each thread finds itself with more garbage to free than the previous thread, and the problem compounds in each epoch.
We call this the \textit{garbage pile up} problem.

This problem is so pronounced, that the last epoch dominates the graph, lasting from $\approx$ $0.25$ seconds until 5 seconds.
The previous epoch lasts approximately a fifth of a second, and is visible as a very faint curve near the beginning of the time axis.
There are 21 epochs in total, and the first 19 are too short to see.

The bottom plot in \Cref{fig:token1_jemalloc_192_interleave_pinyes} shows the drastic increase in the amount of garbage accumulated in each epoch.
To understand why the garbage pile up is so large in the final epoch, note that more than 90\% of the execution is spent in that epoch, and after the first thread reclaims memory at the \textit{start} of that epoch, it never reclaims again (and similarly for the second thread, and so on).

\paragraph{\pftebr}

In our second algorithm, \pftebr, the token is passed \textit{before} the contents of the previous bag are freed.
Note that, in terms of correctness, it does not matter whether the thread frees and then passes the token, or passes then frees---it is the \textit{receipt} of the token that tells the thread it can safely free its previous bag.

\begin{figure}[t]
    \begin{subfigure}{0.99\linewidth}
        \includegraphics[width=\linewidth]{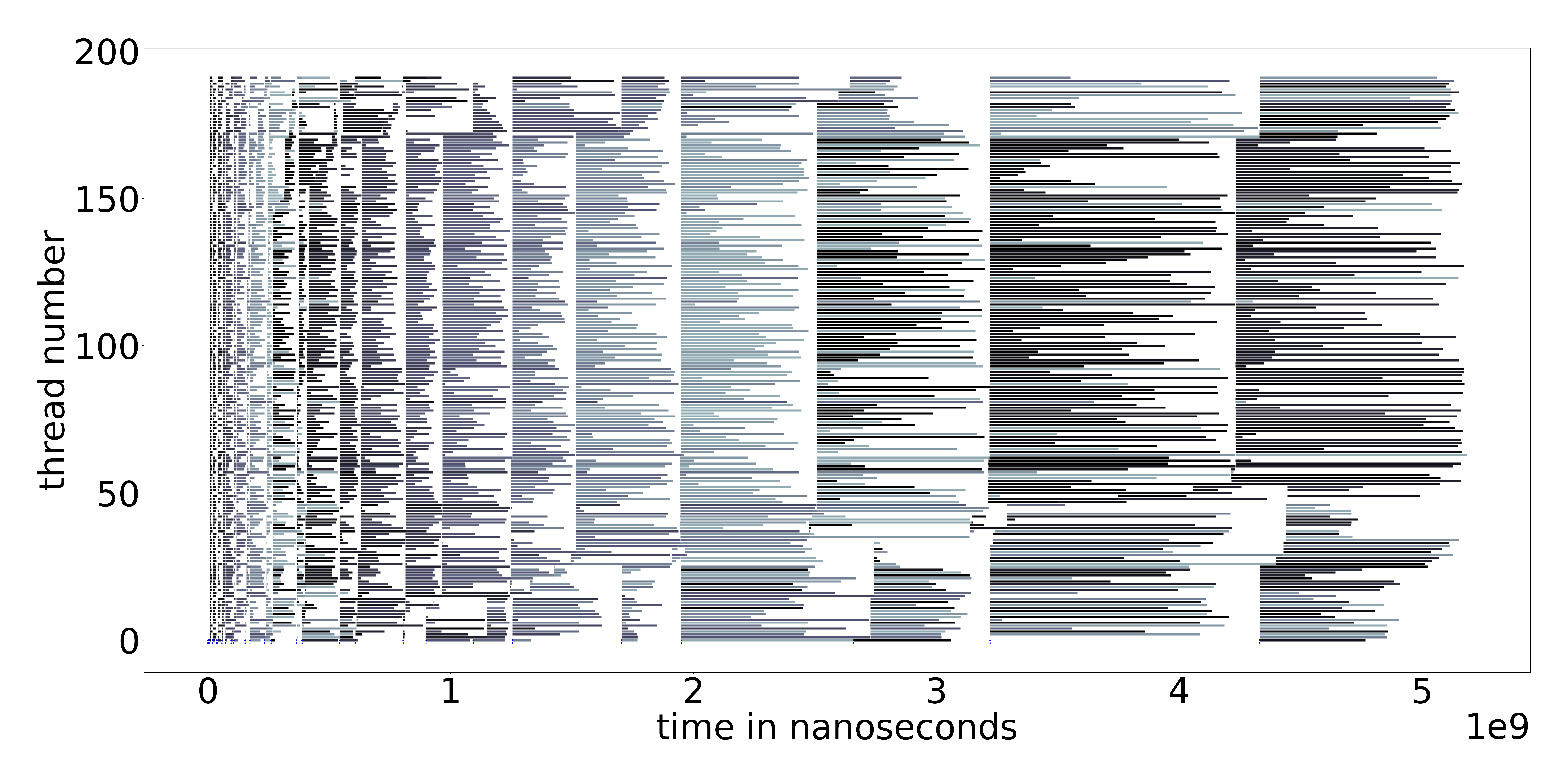}
        % \caption{Timeline graph}
        % \label{subfig:freetime_token2_jemalloc_192_interleave_pinyes}
    \end{subfigure}
    \begin{subfigure}{0.99\linewidth}
        \vspace{-2mm}
        \includegraphics[width=\linewidth]{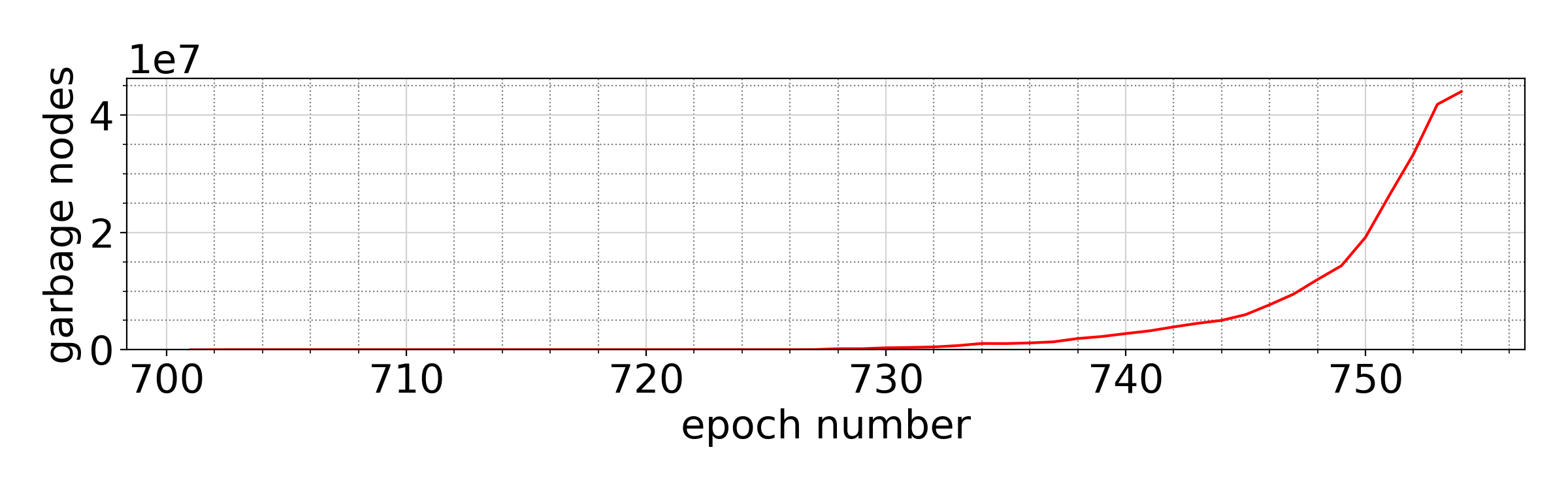}
        % \caption{Average number of unreclaimed objects per thread}
        % \label{subfig:unreclaimed_token2_jemalloc_192_interleave_pinyes}
    \end{subfigure}
    % \vspace{-5mm}
    \caption{Timeline graph showing \textbf{batch} frees (upper) and number of garbage nodes (lower) for \pftebr{}.}
    \label{fig:token2_jemalloc_192_interleave_pinyes}
\end{figure}

\begin{figure}[H]
    \begin{subfigure}{0.49\linewidth}
        \includegraphics[width=\linewidth]{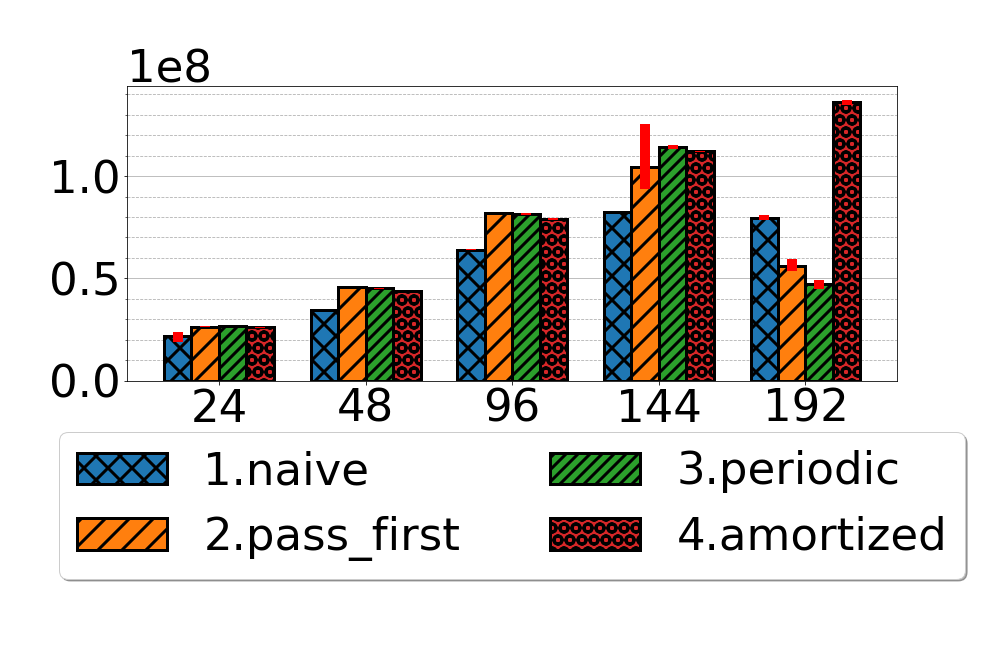}
        \vspace{-7mm}
        \caption{Performance}
        \label{subfig:amortized_token_p}
    \end{subfigure}
    \begin{subfigure}{0.49\linewidth}
        %\hspace{-2mm}%
        \includegraphics[width=\linewidth]{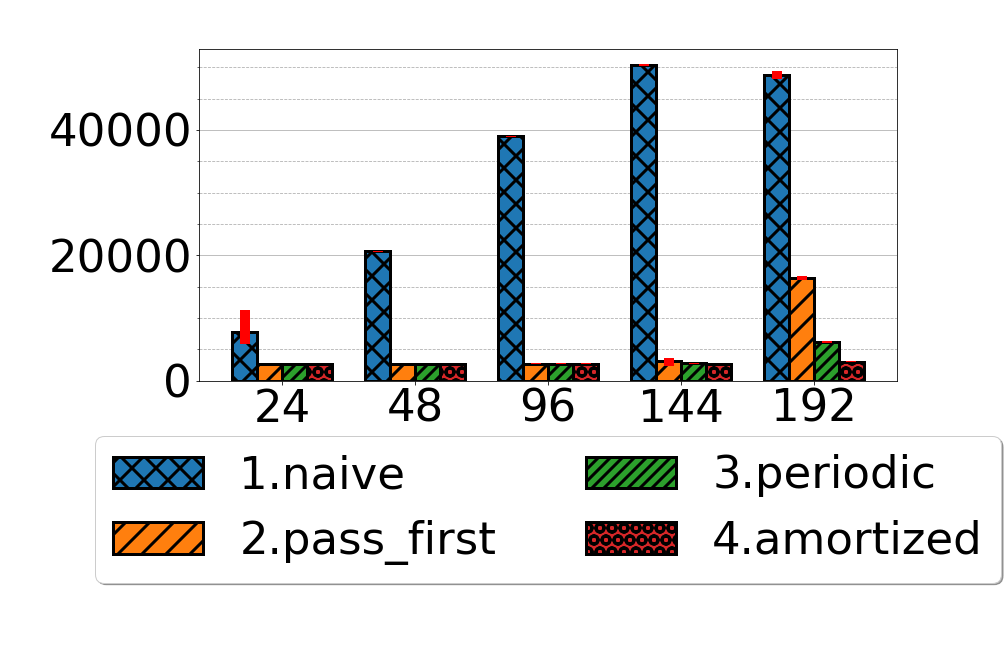}
        \vspace{-7mm}
        \caption{Peak memory usage (MiB)}
        \label{subfig:amortized_token_m}
    \end{subfigure}
    % \vspace{-2mm}
    \caption{Performance and peak memory usage with JEmalloc, for ABtree, using \aftebr{}.}
    \label{fig:amortized_token}
\end{figure}

% \textbf{[[[performance and timeline graphs to justify this?]]]}
\Cref{fig:amortized_token} shows that \pftebr{} improves performance and reduces peak memory usage compared to \ntebr{}.
%is less susceptible to the garbage pile up problem.
Now that threads can free their bags concurrently, \Cref{fig:token2_jemalloc_192_interleave_pinyes} looks a little bit more like the timeline graphs in \Cref{chap:reclamation}.
% clearly shows threads now can now reclaim concurrently and the rounds (epochs) have become relatively shorter which shows in the completion of higher number of epochs (750) in bottom plot of \Cref{fig:token2_jemalloc_192_interleave_pinyes}) in \pftebr{} versus the 21 epochs in bottom plot of \Cref{fig:token1_jemalloc_192_interleave_pinyes} in \ntebr.
However, the algorithm is clearly still susceptible to garbage pile up, as the lengths of batch free operations are increasing over time.
This is further confirmed in \Cref{fig:token2_jemalloc_192_interleave_pinyes}. % at high thread counts.
%It is perhaps surprising at first that garbage pile up is not prevented altogether.

One reason for this is that a thread that receives the token, passes it to the next thread, and begins freeing a large bag of objects, may actually receive the token again before it is finished freeing. %We explain this through an example. 
It will then hold onto the token until it has finished freeing, which can potentially be a long time.
%Suppose a thread receives the token, passes it to the next thread, and then begins freeing a large bag of objects.
%If, while the thread is still freeing its objects, it receives the token again, it will hold onto the token (potentially for a long time) until it has finished freeing all of its objects, and only then pass the token to the next thread.
%So, if a thread ever has a particularly large bag of objects to free, it can still needlessly prevent other threads from freeing their objects.
%In \Cref{fig:token2_jemalloc_192_interleave_pinyes}, the amount of unreclaimed garbage can be seen to increase rapidly towards the end of the execution. %there is still  garbage growth.
% Due to this increase in unreclaimed garbage, the reclamation event boxes in the timeline graph (\Cref{fig:token2_jemalloc_192_interleave_pinyes}) grow longer over time which translates to sudden increase in the number of garbage nodes. 

\paragraph{\ptebr}

% \begin{figure}[th]
%     \begin{subfigure}{0.49\linewidth}
%         \includegraphics[width=\linewidth]{figures/token_p.png}
%         \caption{Performance}
%         \label{subfig:token_p}
%     \end{subfigure}
%     \begin{subfigure}{0.49\linewidth}
%         \hspace{-2mm}%
%         \includegraphics[width=\linewidth]{figures/token_m.png}
%         \caption{Peak memory usage (MiB)}
%         \label{subfig:token_m}
%     \end{subfigure}
%     \caption{Performance and peak memory usage w/JEmalloc, for ABtree, using Token EBR}
%     \label{fig:token}
% \end{figure}

In our third algorithm, \ptebr{}, a thread passes the token, then begins freeing the contents of its previous bag.
However, as it is freeing objects, it periodically checks, every $k$ \texttt{free} calls (100 in our experiments), whether it has received the token, and if so passes it along.
% 
% The observation that threads could still hold a token while freeing a large bag from the previous round leads to our next improved implementation called \ptebr.
% In \ptebr, while a thread is freeing a bag of objects, it periodically checks (in our experiments, after every 100 \texttt{free()} calls) to see if it has received the token.
% If so, it immediately passes the token to the next thread, and then resumes freeing its objects. Thus, even though threads could be freeing large bags from previous rounds they keep on passing the token around giving chance for other threads to keep freeing their bags.  
% 

\begin{figure}[t]
    \begin{subfigure}{0.99\linewidth}
        \includegraphics[width=\linewidth]{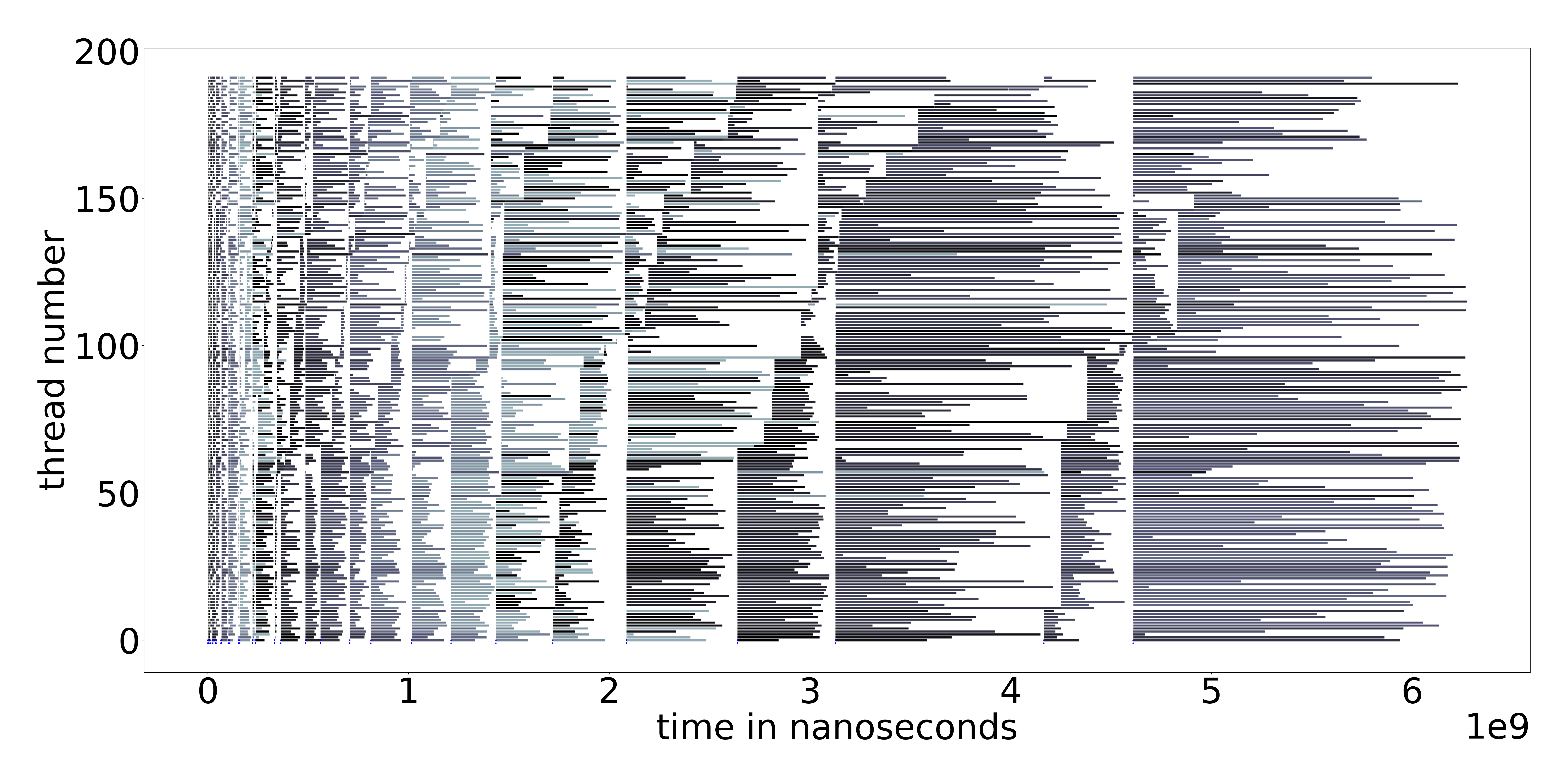}
        % \caption{Timeline graph}
        % \label{subfig:freetime_token3_jemalloc_192_interleave_pinyes}
    \end{subfigure}
    \begin{subfigure}{0.99\linewidth}
        \vspace{-2mm}
        \includegraphics[width=\linewidth]{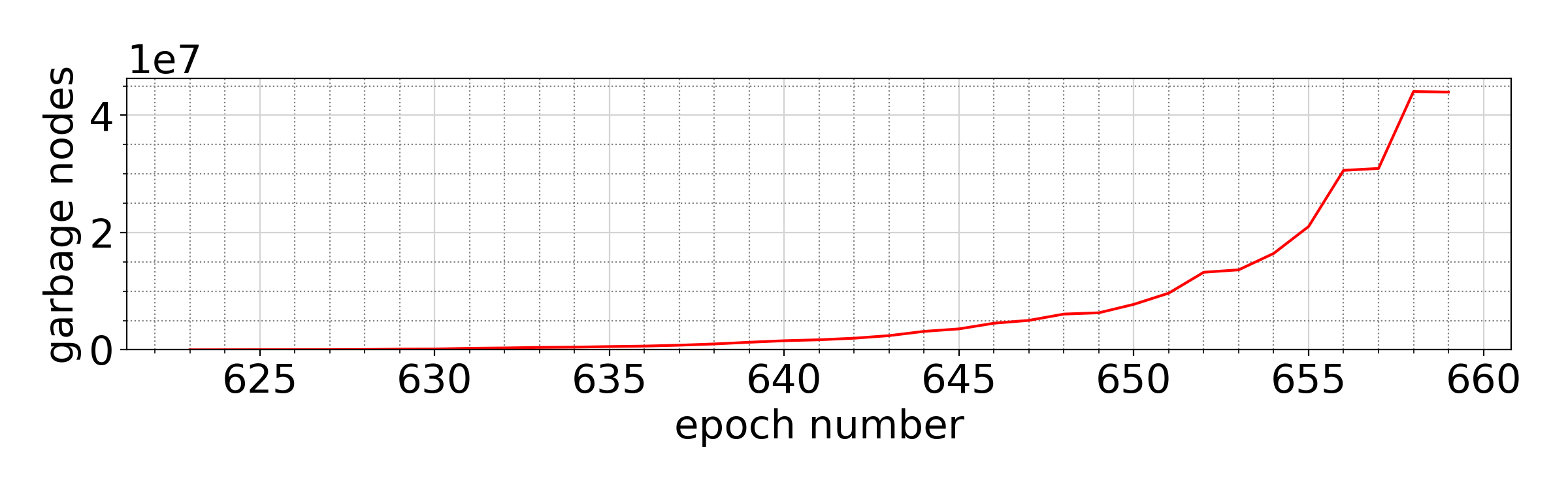}
        % \caption{Average number of unreclaimed objects per thread}
        % \label{subfig:unreclaimed_token3_jemalloc_192_interleave_pinyes}
    \end{subfigure}
    % \vspace{-5mm}
    \caption{Timeline graph showing \textbf{batch} frees (upper) and number of garbage nodes (lower) for \ptebr{}.}
    \label{fig:token3_jemalloc_192_interleave_pinyes}
\end{figure}
% \textbf{[[[performance and timeline graphs to justify this]]]}

As \Cref{fig:amortized_token} shows, this approach performs somewhat similarly to \pftebr{}, but has significantly lower peak memory usage with 192 threads.
Unfortunately, as \Cref{fig:token3_jemalloc_192_interleave_pinyes} shown, there is \textit{still} a garbage pile up problem.
This result may seem surprising at first.
If we \texttt{free} no more than 100 objects while holding the token before passing it, how can threads have an opportunity to accumulate more than 40 million garbage nodes without passing the token all the way around the ring and advancing the epoch?

It turns out we have already learned the answer to this question above.
In JEmalloc, whenever a thread overflows its internal buffer of freed nodes, and triggers a high latency remote free operation, a \textbf{single free call} can run for a very long time---on the order of milliseconds.
As a result, it does not matter whether a thread checks if it holds the token after every 100 \texttt{free} calls, or after \textit{every} \texttt{free} call.
Unless it can somehow perform this check \textit{during} a single long free call (which presumably would require modifying the allocator itself), the epoch cannot advance until that long free call has finished.
And, as we saw in \Cref{subfig:freeone-je192-batch}, large batch free operations can and do frequently cause such long free calls.

One might also wonder why peak memory usage is lower for this approach, even through the epoch counts in \Cref{fig:token2_jemalloc_192_interleave_pinyes} and \Cref{fig:token3_jemalloc_192_interleave_pinyes} are similar, and the problem of high latency free operations still exists.
Since epochs are increasing in length as time passes, peak memory usage is essentially dictated solely by the final epoch.
Moreover, a thread that begins freeing in the final epoch has typically accumulated so much garbage that it will not finish freeing until the experiment is over (as evidenced by the fact that batch frees do not finish until after the 5 second time limit).
In the final epoch in \pftebr, such a thread will not pass the token, so all threads after it in the token ring will accumulate garbage in the final epoch, increasing peak memory usage.
On the other hand, in \ptebr, although such a thread will be unable to pass the token \textit{during} a high latency free operation, if the thread performs at least \textit{two} high latency free operations in the final epoch, then it can still pass the token in between high latency free operations.
This enables additional threads to free their limbo bags concurrently in the final epoch, reducing peak memory usage.
%That said, performance remains largely unchanged, since the fundamental problem of increasing epoch lengths still exists.

\begin{figure}[t]
    % \begin{subfigure}{0.99\linewidth}
         \hspace{-2mm}
         \includegraphics[width=\linewidth]{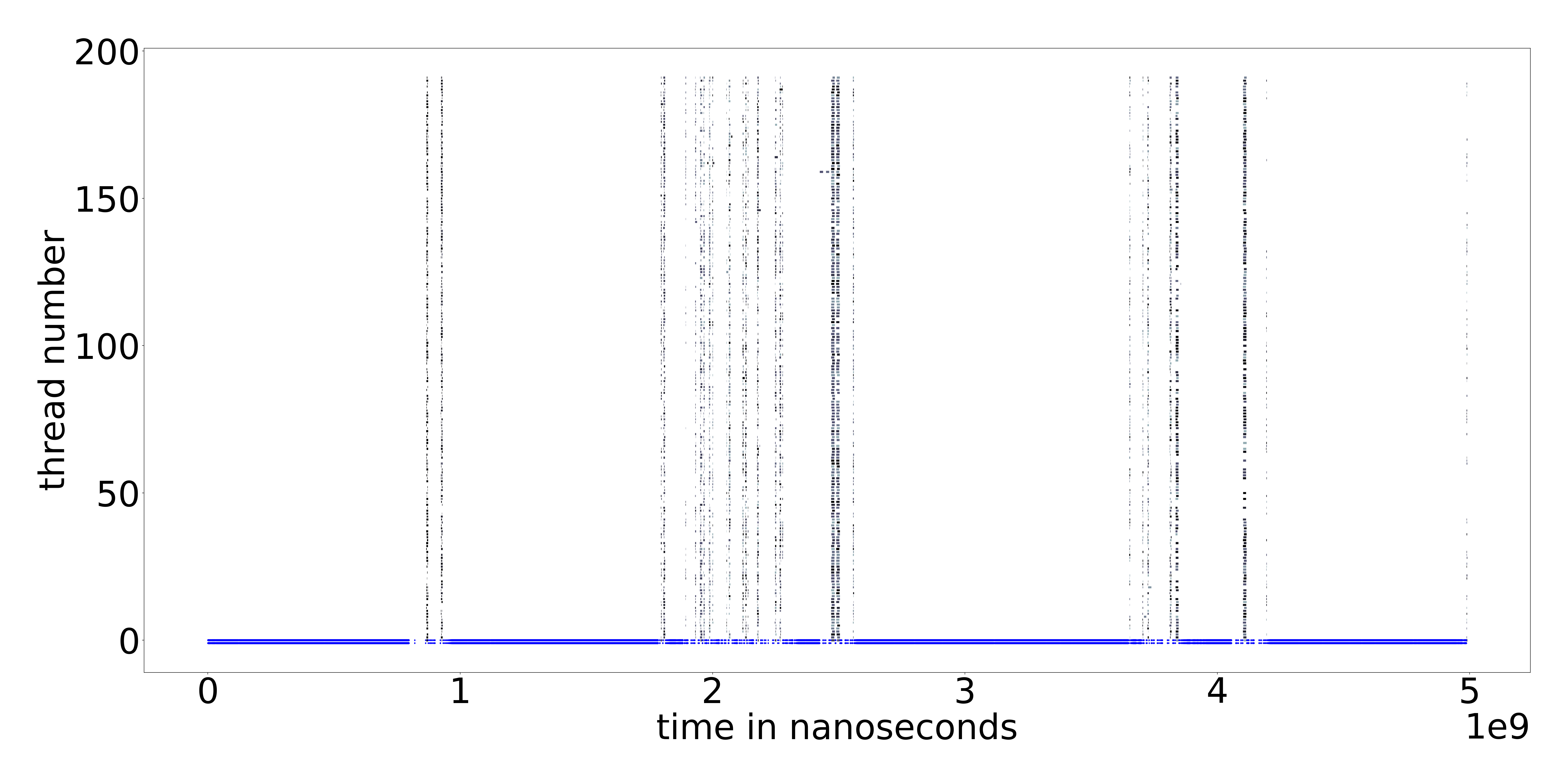}
    %     \caption{Timeline graph}
    %     \label{subfig:freetime_token4_jemalloc_192_interleave_pinyes}
    % \end{subfigure}
    % \begin{subfigure}{0.99\linewidth}
        \vspace{-2mm}
        \includegraphics[width=\linewidth]{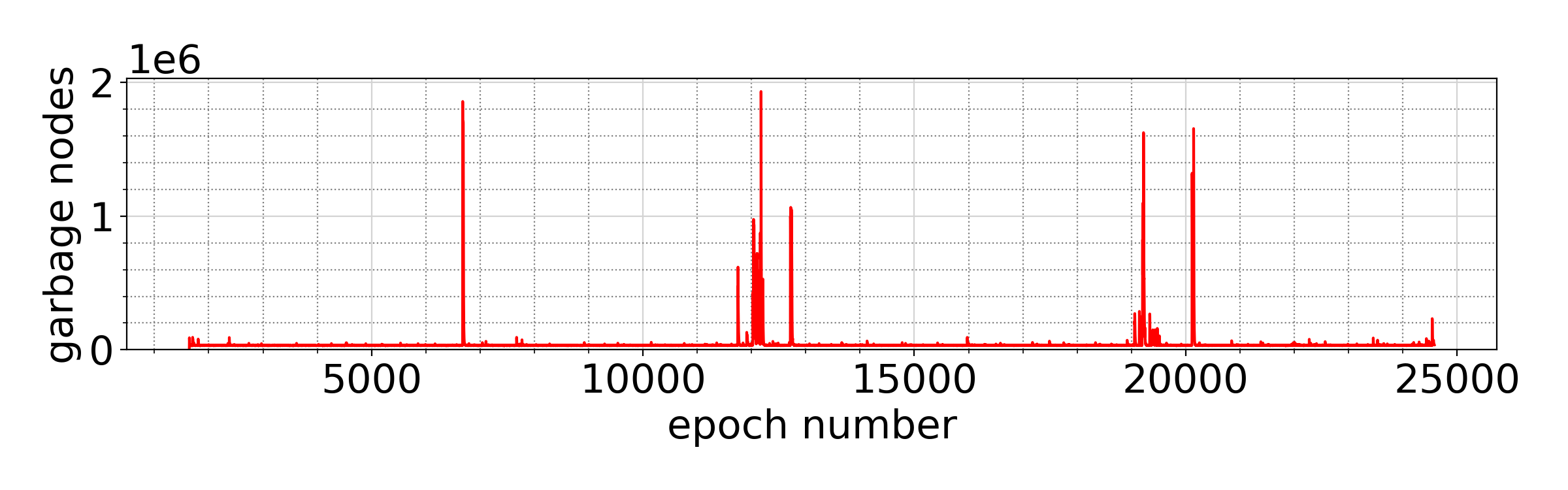}
        % \caption{Average number of unreclaimed objects per thread}
        % \label{subfig:unreclaimed_token4_jemalloc_192_interleave_pinyes}
    % \end{subfigure}
    % \vspace{-7mm}
    % \vspace{-5mm}
    \caption{Timeline graph (upper) and number of garbage nodes (lower) for \aftebr{}.
    This timeline graph shows \textbf{individual} \texttt{free} calls longer than 0.1ms.}
    \label{fig:token4_jemalloc_192_interleave_pinyes}
\end{figure}

\paragraph{Amortized-Free Token-EBR}
Our final algorithm applies the amortized freeing approach to \ptebr{}. %Hoping to keep the bag sizes small enough to not cause buffer overflow in JEmalloc that leads to \textit{garbage pile up}, we apply \textit{amortized freeing} to \ptebr{}.
\Cref{fig:amortized_token} shows this offers drastic improvements in both performance and peak memory usage with 192 threads. %pecifically, \aftebr{} is $\sim$1.75X faster and has $\sim$5X lower peak memory usage. 
As \Cref{fig:token4_jemalloc_192_interleave_pinyes} shows, amortized free mitigates the garbage pile up problem %, and effectively controls the amount of unreclaimed garbage 
(as well as greatly increasing the epoch count).
\Cref{tab:token_throughput} summarizes the impact of each algorithmic change on performance, time spent freeing, and the total number of objects freed.
%The amortized free approach reduces the number eliminates the garbage pile up problem
%The average lower bag size reduces chances of frequent buffer overflow in JEmalloc which in turn more likely avoids \textit{high latency free} calls which avoid stopping of token from being forwarded for a long time that induces the \textit{garbage pile up} cycle for rest of the experiment.
%The \Cref{tab:token_throughput} shows in \aftebr{} a strikingly higher number of object are freed in very low amount of time which translates in its higher throughput. 

\begin{table}[h]
\centering
\begin{tabular}{llll}
\toprule
\textbf{algorithm} & \textbf{ops/s}  & \textbf{\% free} & \textbf{freed}\\
\midrule
% \rowcolor[rgb]{0.753,0.753,0.753}\nbrp\cite{singh2021nbr}        & Yes                           & cond. lock-free   & OS + \hp \\ 
% \nbr\cite{singh2021nbr}                                             & Yes                           & cond. lock-free   & OS + \hp \\
Naive&73.7M&3.3&7M\\
Pass-first&52.4M&45.4&98M\\
Periodic&54.4M&47.1&118M\\
Amortized&123.7M&14.7&323M\\
\bottomrule
\end{tabular}
% \vspace{1mm}
\caption{Analysis of Token EBR variants. 192 threads.
% Throughputs are lower than \Cref{fig:amortized_token} due to \texttt{perf} overhead.
}
\label{tab:token_throughput}
% \vspace{-6mm}
\end{table}

%%%%%%%%%%%%%%%%%%%%%%%%%%%%%%%%%%%%%%%%%%%%%%%%%%%%%%%%%%%%%%%%%%%%%%%%
%   Experiment Section Begin
%%%%%%%%%%%%%%%%%%%%%%%%%%%%%%%%%%%%%%%%%%%%%%%%%%%%%%%%%%%%%%%%%%%%%%%%

\section{Evaluation} \label{chap:evaluation}

We evaluate the impact of the amortized freeing technique using a benchmark that implements the ABtree with 10 memory reclamation algorithms, encompassing \aftebr{} (\texttt{token\_af}), \texttt{debra}~\cite{brown2015reclaiming} and its amortized free version (debra\_af), hazard eras (\texttt{he})~\cite{ramalhete2017brief}, hazard pointers (\texttt{hp})~\cite{michael2004hazard}, interval based reclamation (\texttt{ibr})~\cite{wen2018interval}, two neutralization based reclamation techniques~\cite{singh2021nbr}, \texttt{nbr} and \texttt{nbr+}, quiescent state based reclamation (\texttt{qsbr})~\cite{hart2007performance}, read-copy-update (\texttt{rcu})~\cite{hart2007performance} and wait free eras (\texttt{wfe})~\cite{nikolaev2020universal} in two different experiments.
The methodology and system are the same as in ~\Cref{chap:reclamation}, except for the selected thread counts.

Note that our code will be submitted for artifact evaluation, and will be made public when the paper is published.

\begin{figure*}[ht]
     % \begin{minipage}{\textwidth}
        \begin{subfigure}{0.49\textwidth}
            \centering
            \includegraphics[width=1\linewidth, height=5cm, keepaspectratio]{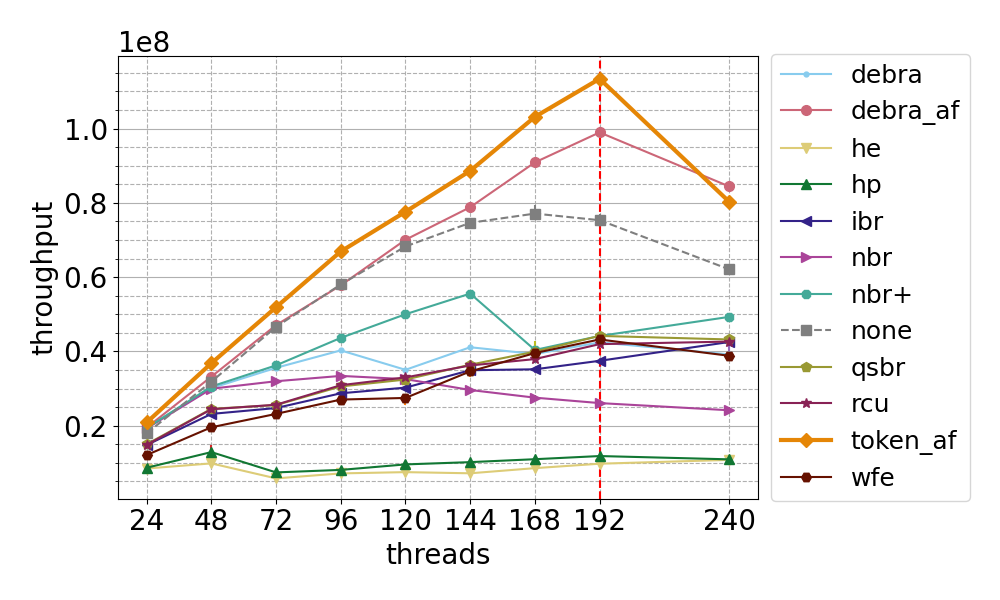}
            % \vspace{-3mm}
            \subcaption{Experiment 1: Comparison of proposed \aftebr{} (token\_af) with other reclamation techniques across threads.}
            \label{fig:exp1}
        \end{subfigure}
        \hfill
        \begin{subfigure}{0.49\textwidth}
            \includegraphics[width=1\linewidth, height=5cm, keepaspectratio]{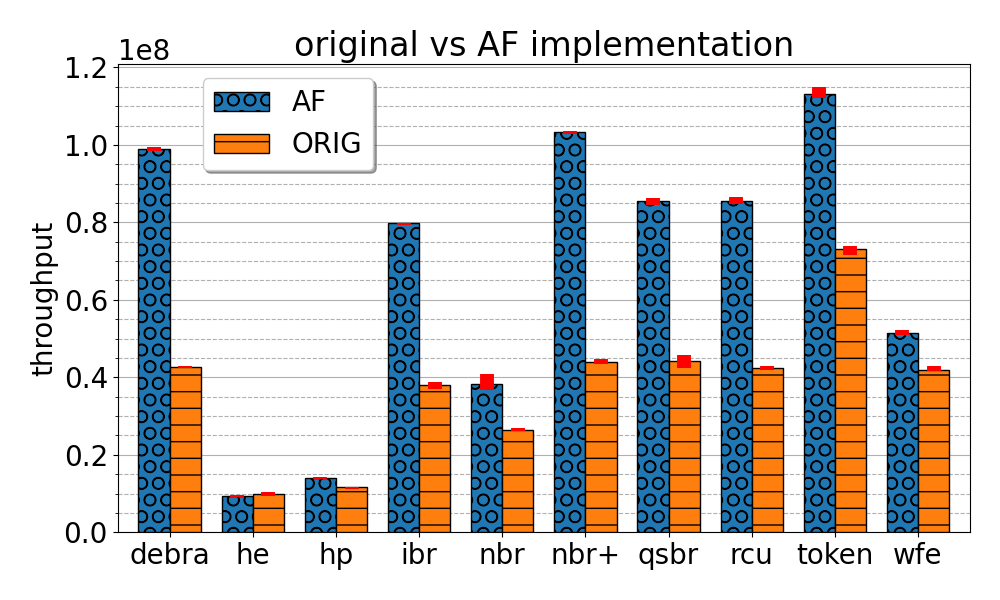}
            % \vspace{-3mm}
            \subcaption{Experiment 2: Comparison with amortized free versions of various reclaimers at 192 threads.}
            \label{fig:exp2}
        \end{subfigure}
     
     % \end{minipage}
    % \vspace{-2mm}
    \caption{Data structure: Brown's ABtree. Workload: 50\% inserts and 50\% deletes. Size: 20M. Allocator:JEmalloc.
    }
    \label{fig:exp12}
\end{figure*}

\paragraph{Experiment 1:} We compare the performance of \texttt{token\_af} and \texttt{debra\_af} against a broad cross section of the state of the art in reclamation techniques (in~\Cref{fig:exp1}).
The vertical red line denotes the number of hardware threads in the system.
A leaky implementation (none) is also included.
%With more than above which , becoming oversubscribed.
%Additionally, we compare with the enhanced \textit{amortized free} version of DEBRA (\texttt{debra\_AF}) and a leaky implementation (none). 

Our \texttt{token\_af} algorithm %, designed to address the RBF issue, 
outperforms all other techniques.
Averaging the results across all thread counts, it is $\approx$1.7$\times$ fast than the next fastest algorithm, nbr+, and $\approx$7-9$\times$ better than the slowest algorithms, \texttt{hp} and \texttt{he}. %, with averaged speedups across all threads.
Surprisingly, both of our amortized free algorithms \texttt{token\_af} and \texttt{debra\_af} significantly outperform \texttt{none}, which is often (incorrectly) described as an upper bound on the performance of a reclamation algorithm in the literature. % benchmark for evaluating reclaimer performance in literature.
Of course, prior reclamation algorithms have been shown to outperform leaky implementations previously~\cite{brown2015reclaiming}, citing improved locality. % that with improved memory and cache locality, reclaimers could scale and outperform better than not reclaiming at all. 
%In this context, \texttt{token\_af} sets an even higher standard for reclamation throughput.

\paragraph{Experiment 2:}
We assess the potential for the amortized free technique to improve other reclamation algorithms by implementing amortized free (AF) versions of \textit{all 10} of the reclamation algorithms studied in Experiment 1.
We then compare their throughput with their original (ORIG) implementations with 192 threads (\Cref{fig:exp2}).
All of these algorithms accumulate a batch of garbage before freeing, and we use a uniform batch size of 32K nodes for all algorithms.
This particular size was chosen because (1) each algorithm performed best with this size, and (2) the nbr algorithm already uses this bag size by default~\cite{singh2021nbr}.

\Cref{fig:exp2} shows that the AF algorithms outperform their ORIG counterparts for 9 algorithms: debra, ibr, nbr/nbr+, qsbr, rcu, token, hp and wfe, demonstrating up to 2.3$\times$ improvements in throughput.
The \texttt{he} algorithm does not improve, whereas \texttt{hp} and \texttt{wfe} show modest improvements of $\approx$1.2$\times$.
The lack of improvement (or limited improvement) in these algorithms is likely due to the fact that they have much higher synchronization overhead than the other algorithms, which prevents any improvement from manifesting. %and this overhead likely bottlenecks.
%incurred by these reclaimers, potentially overshadowing the gains from \textit{amortized freeing}.
Additional experiments showing AF vs ORIG improvements at various thread counts appear in the %these reclaimer's original and \textit{amortized free} versions across different thread counts are presented in the 
supplementary material.

\section{Related Work} \label{chap:related_work}

% Several popular memory allocators, and some memory reclamation algorithms, have already been described and cited above.
% So, this chapter focuses on describing the most closely related work that has not yet been addressed.
% Broadly, this related work falls into two categories: token based memory reclamation algorithms, and work that studies the impact of peak memory usage on performance.

% \subsection{Token Based Memory Reclamation}
The idea of passing a token around to establish a time when it is safe to free is not new.
Practitioners have discussed implementing EBR in this way on online discussion forums~\cite{ycombinator}, and similar algorithms have been used to reclaim memory in operating system kernels~\cite{krieger2006k42}.
Token based algorithms have also been used as part of complex garbage collection algorithms in the distributed setting~\cite{hudson1997training}, and token based EBR has been described in a thesis due to Tam in the shared memory multicore setting~\cite{tam2006qdo}, where it was dismissed as inefficient.
%The distributed algorithm is complicated than is needed for a multicore system.
%
% The multicore token based EBR algorithm in~\cite{tam2006qdo} essentially matches the description above.
% However, the experiments in that thesis show that the token based algorithm is inefficient, and the authors dismiss the design, partly on the ground that it is inefficient. %, proceeding with other designs instead.
% This paper shows it does not have to be inefficient.
Our new token based algorithm surpasses the state of the art, which is especially impressive considering there has been more than 15 years of progress in EBR algorithms since the publication of~\cite{tam2006qdo}.

The study of concurrent memory reclamation saw multiple fundamental results appear in the early 2000s, including EBR algorithms~\cite{fraser2004practical, mckenney1998read}, pointer based reclamation techniques~\cite{michael2004hazard, herlihy2005nonblocking} and reference counting techniques~\cite{detlefs2001lock}.
% Since then, many new algorithms have appeared~\cite{wen2018interval, brown2015reclaiming, hart2007performance,gidenstam2008efficient, nikolaev2019hyaline, ramalhete2017brief, nikolaev2020universal, ramalhete2017brief, nikolaev2020universal, nikolaev2021crystalline,dice2016fast, BlellochDRC21, correia2021orcgc,moreno2023releasing, jungHP2023, anderson2022turning} some of which are hybrid of these two and others use novel hw/sw mechanisms~\cite{alistarh2018threadscan, alistarh2017forkscan, alistarh2014stacktrack}. 
% The two important category of reclamation algorithms which appeared more recently are optimistic~\cite{sheffi2021vbr, cohen2015automatic, cohen2015efficient} and neutralization based techniques~\cite{singh2021nbr, kang2020marriage}.
Since then, many new algorithms have appeared~\cite{wen2018interval, alistarh2018threadscan, alistarh2017forkscan, alistarh2014stacktrack, brown2015reclaiming, hart2007performance,gidenstam2008efficient, nikolaev2019hyaline, sheffi2021vbr, cohen2015automatic, singh2021nbr, ramalhete2017brief, nikolaev2020universal, ramalhete2017brief, nikolaev2020universal, nikolaev2021crystalline,dice2016fast, BlellochDRC21, correia2021orcgc,moreno2023releasing, jungHP2023, cohen2015efficient, kang2020marriage, anderson2022turning}, nearly all of which free batches of objects. % some of which are hybrid of these two and others use novel hw/sw mechanisms~\cite{}. 
%The two important category of reclamation algorithms which appeared more recently are optimistic~\cite{ } and neutralization based techniques~\cite{, }.

%Each of these memory reclamation algorithms promotes the utilization of batch freeing to diminish the overhead arising from frequent reclamation. 
Most recently, %And, we are aware of one recently introduced technique, known as
a hardware-software co-design called \textit{Conditional Access}~\cite{singhCA2023} has appeared that facilitates the immediate reclamation of individual object, rather than batches.
Experiments using JEmalloc showed significant performance improvement vs traditional memory reclamation algorithms that free batches of objects, but the authors did not establish a concrete reason for the improvement.
Our work sheds new light on why techniques like \textit{Conditional Access} that immediately free individual objects perform well in practice.

Mitake et~al~\cite{mitake2019looking} studied the impact of peak memory usage on in-memory database transaction latencies that used epoch based reclamation.
They suggested freeing batches using a separate background thread to increase the frequency with which the main worker threads could participate in advancing the epoch.
In light of our work, moving batch freeing to a background thread appears to be insufficient to avoid the RBF problem.
Batch freeing is, itself, the problem.

\section{Conclusion} \label{chap:conclusion}

In this paper, we identify a performance problem that limits the performance and scaling of concurrent memory reclamation algorithms on modern multi socket systems.
We perform a rigorous study of the root causes of this problem, and identify subtle interactions between popular allocators and memory reclamation algorithms that free batches of objects.

We present a simple workaround, amortized freeing, and demonstrate its effectiveness by applying it to ten existing memory reclamation algorithms, and a new reclamation algorithm \texttt{token\_af}. %representing a broad cross section of the state of the art.
%We also %additionally leverage the insights gained in this work to 
%develop a new memory reclamation algorithm that is substantially faster than prior art.
As our experiments show, the potential impact of the presented ideas on the memory reclamation literature is substantial.
%the impact of amortized freeing is substantial.
With 192 concurrent threads, amortized freeing on average \textit{doubles the performance of the six fastest existing memory reclamation algorithms}, and \texttt{token\_af} \textit{outperforms the state of the art by 2.6$\times$}.

Amortized freeing was a natural fit for the ABtree data structure that we used in our experiments, since each operation frees at most one object on average.
In data structures that free more than one object per operation on average, amortized freeing should be tuned to free more than one object per operation.
Amortized freeing will be most effective if the number of objects freed and allocated per operation is similar, so that objects gradually freed to internal thread-local buffers in the allocator can be reallocated locally. % without communicating with remote threads.

Finally, we have timeline graphs to thank for many of our insights.
We hope this new visualization tool will prove useful in designing and profiling new concurrent algorithms.

\begin{acks}
This work was supported by the \grantsponsor{NSERC}{Natural Sciences and Engineering Research Council of Canada (NSERC) Collaborative Research and Development grant}{https://www.nserc-crsng.gc.ca/professors-professeurs/rpp-pp/crd-rdc_eng.asp}: 
\grantnum{CRDPJ}{539431-19}, the \grantsponsor{JELF}{Canada Foundation for Innovation John R. Evans Leaders Fund}{https://www.innovation.ca/apply-manage-awards/funding-opportunities/john-r-evans-leaders-fund} (\grantnum{JELF}{38512}) with equal support from the \grantsponsor{Ontario Research Fund}{Ontario Research Fund CFI Leaders Opportunity Fund}{https://www.ontario.ca/page/ontario-research-fund-research-infrastructure}, \grantsponsor{NSERC}{NSERC Discovery Program Grant}{https://www.nserc-crsng.gc.ca/professors-professeurs/grants-subs/dgigp-psigp_eng.asp}: \grantnum{RGPIN}{2019-04227}, \grantsponsor{NSERC}{NSERC Discovery Launch Grant}{https://www.nserc-crsng.gc.ca/professors-professeurs/grants-subs/DLS-STVLD_eng.asp}: \grantnum{DGECR}{2019-00048}, and the \grantsponsor{University of Waterloo}{University of Waterloo}{}.
The findings and opinions expressed in this paper are those of the authors and do not necessarily reflect the views of the funding agencies.
We also thank the anonymous reviewers for their thoughtful comments and insights. %helpful suggestions. % helping us to improve the manuscript.
% }
\end{acks}
\bibliographystyle{ACM-Reference-Format}
\bibliography{sample-base}


\begin{thebibliography}{40}

%%% the Web address.

\ifx \showCODEN    \undefined \def \showCODEN     #1{\unskip}     \fi
\ifx \showDOI      \undefined \def \showDOI       #1{#1}\fi
\ifx \showISBNx    \undefined \def \showISBNx     #1{\unskip}     \fi
\ifx \showISBNxiii \undefined \def \showISBNxiii  #1{\unskip}     \fi
\ifx \showISSN     \undefined \def \showISSN      #1{\unskip}     \fi
\ifx \showLCCN     \undefined \def \showLCCN      #1{\unskip}     \fi
\ifx \shownote     \undefined \def \shownote      #1{#1}          \fi
\ifx \showarticletitle \undefined \def \showarticletitle #1{#1}   \fi
\ifx \showURL      \undefined \def \showURL       {\relax}        \fi
% invisible to TeX
\providecommand\bibfield[2]{#2}
\providecommand\bibinfo[2]{#2}
\providecommand\natexlab[1]{#1}
\providecommand\showeprint[2][]{arXiv:#2}

\bibitem[nik(2021)]%
        {nikolaev2021crystalline}
 \bibinfo{year}{2021}\natexlab{}.
\newblock \showarticletitle{Crystalline: Fast and Memory Efficient Wait-Free Reclamation}, \bibfield{editor}{\bibinfo{person}{Ruslan Nikolaev} {and} \bibinfo{person}{Binoy Ravindran}} (Eds.).
\newblock \bibinfo{journal}{\emph{CoRR}}  \bibinfo{volume}{abs/2108.02763}.
\newblock
\showeprint[arXiv]{2108.02763}
\urldef\tempurl%
\url{https://arxiv.org/abs/2108.02763}
\showURL{%
\tempurl}


\bibitem[Alistarh et~al\mbox{.}(2014)]%
        {alistarh2014stacktrack}
\bibfield{author}{\bibinfo{person}{Dan Alistarh}, \bibinfo{person}{Patrick Eugster}, \bibinfo{person}{Maurice Herlihy}, \bibinfo{person}{Alexander Matveev}, {and} \bibinfo{person}{Nir Shavit}.} \bibinfo{year}{2014}\natexlab{}.
\newblock \showarticletitle{Stacktrack: An automated transactional approach to concurrent memory reclamation}. In \bibinfo{booktitle}{\emph{Proceedings of the Ninth European Conference on Computer Systems}}. \bibinfo{pages}{1--14}.
\newblock


\bibitem[Alistarh et~al\mbox{.}(2017)]%
        {alistarh2017forkscan}
\bibfield{author}{\bibinfo{person}{Dan Alistarh}, \bibinfo{person}{William Leiserson}, \bibinfo{person}{Alexander Matveev}, {and} \bibinfo{person}{Nir Shavit}.} \bibinfo{year}{2017}\natexlab{}.
\newblock \showarticletitle{Forkscan: Conservative memory reclamation for modern operating systems}. In \bibinfo{booktitle}{\emph{Proceedings of the Twelfth European Conference on Computer Systems}}. \bibinfo{pages}{483--498}.
\newblock


\bibitem[Alistarh et~al\mbox{.}(2018)]%
        {alistarh2018threadscan}
\bibfield{author}{\bibinfo{person}{Dan Alistarh}, \bibinfo{person}{William Leiserson}, \bibinfo{person}{Alexander Matveev}, {and} \bibinfo{person}{Nir Shavit}.} \bibinfo{year}{2018}\natexlab{}.
\newblock \showarticletitle{Threadscan: Automatic and scalable memory reclamation}.
\newblock \bibinfo{journal}{\emph{ACM Transactions on Parallel Computing (TOPC)}} \bibinfo{volume}{4}, \bibinfo{number}{4} (\bibinfo{year}{2018}), \bibinfo{pages}{1--18}.
\newblock


\bibitem[Anderson et~al\mbox{.}(2021)]%
        {BlellochDRC21}
\bibfield{author}{\bibinfo{person}{Daniel Anderson}, \bibinfo{person}{Guy~E. Blelloch}, {and} \bibinfo{person}{Yuanhao Wei}.} \bibinfo{year}{2021}\natexlab{}.
\newblock \showarticletitle{Concurrent Deferred Reference Counting with Constant-Time Overhead}. In \bibinfo{booktitle}{\emph{Proceedings of the 42nd ACM SIGPLAN International Conference on Programming Language Design and Implementation}} (Virtual, Canada) \emph{(\bibinfo{series}{PLDI 2021})}. \bibinfo{publisher}{Association for Computing Machinery}, \bibinfo{address}{New York, NY, USA}, \bibinfo{pages}{526–541}.
\newblock
\showISBNx{9781450383912}
\urldef\tempurl%
\url{https://doi.org/10.1145/3453483.3454060}
\showDOI{\tempurl}


\bibitem[Anderson et~al\mbox{.}(2022)]%
        {anderson2022turning}
\bibfield{author}{\bibinfo{person}{Daniel Anderson}, \bibinfo{person}{Guy~E Blelloch}, {and} \bibinfo{person}{Yuanhao Wei}.} \bibinfo{year}{2022}\natexlab{}.
\newblock \showarticletitle{Turning manual concurrent memory reclamation into automatic reference counting}. In \bibinfo{booktitle}{\emph{Proceedings of the 43rd ACM SIGPLAN International Conference on Programming Language Design and Implementation}}. \bibinfo{pages}{61--75}.
\newblock


\bibitem[Bronson et~al\mbox{.}(2010)]%
        {bronson2010practical}
\bibfield{author}{\bibinfo{person}{Nathan~G Bronson}, \bibinfo{person}{Jared Casper}, \bibinfo{person}{Hassan Chafi}, {and} \bibinfo{person}{Kunle Olukotun}.} \bibinfo{year}{2010}\natexlab{}.
\newblock \showarticletitle{A practical concurrent binary search tree}.
\newblock \bibinfo{journal}{\emph{ACM Sigplan Notices}} \bibinfo{volume}{45}, \bibinfo{number}{5} (\bibinfo{year}{2010}), \bibinfo{pages}{257--268}.
\newblock


\bibitem[Brown(2017)]%
        {brown2017techniques}
\bibfield{author}{\bibinfo{person}{Trevor Brown}.} \bibinfo{year}{2017}\natexlab{}.
\newblock \emph{\bibinfo{title}{Techniques for Constructing Efficient Lock-free Data Structures}}.
\newblock \bibinfo{thesistype}{Ph.\,D. Dissertation}. \bibinfo{school}{University of Toronto}.
\newblock


\bibitem[Brown(2015)]%
        {brown2015reclaiming}
\bibfield{author}{\bibinfo{person}{Trevor~Alexander Brown}.} \bibinfo{year}{2015}\natexlab{}.
\newblock \showarticletitle{Reclaiming memory for lock-free data structures: There has to be a better way}. In \bibinfo{booktitle}{\emph{Proceedings of the 2015 ACM Symposium on Principles of Distributed Computing}}. \bibinfo{pages}{261--270}.
\newblock


\bibitem[Cohen and Petrank(2015a)]%
        {cohen2015automatic}
\bibfield{author}{\bibinfo{person}{Nachshon Cohen} {and} \bibinfo{person}{Erez Petrank}.} \bibinfo{year}{2015}\natexlab{a}.
\newblock \showarticletitle{Automatic memory reclamation for lock-free data structures}.
\newblock \bibinfo{journal}{\emph{ACM SIGPLAN Notices}} \bibinfo{volume}{50}, \bibinfo{number}{10} (\bibinfo{year}{2015}), \bibinfo{pages}{260--279}.
\newblock


\bibitem[Cohen and Petrank(2015b)]%
        {cohen2015efficient}
\bibfield{author}{\bibinfo{person}{Nachshon Cohen} {and} \bibinfo{person}{Erez Petrank}.} \bibinfo{year}{2015}\natexlab{b}.
\newblock \showarticletitle{Efficient memory management for lock-free data structures with optimistic access}. In \bibinfo{booktitle}{\emph{Proceedings of the 27th ACM symposium on Parallelism in Algorithms and Architectures}}. \bibinfo{pages}{254--263}.
\newblock


\bibitem[Correia et~al\mbox{.}(2021)]%
        {correia2021orcgc}
\bibfield{author}{\bibinfo{person}{Andreia Correia}, \bibinfo{person}{Pedro Ramalhete}, {and} \bibinfo{person}{Pascal Felber}.} \bibinfo{year}{2021}\natexlab{}.
\newblock \showarticletitle{Orcgc: automatic lock-free memory reclamation}. In \bibinfo{booktitle}{\emph{Proceedings of the 26th ACM SIGPLAN Symposium on Principles and Practice of Parallel Programming}}. \bibinfo{pages}{205--218}.
\newblock


\bibitem[Detlefs et~al\mbox{.}(2001)]%
        {detlefs2001lock}
\bibfield{author}{\bibinfo{person}{David~L Detlefs}, \bibinfo{person}{Paul~A Martin}, \bibinfo{person}{Mark Moir}, {and} \bibinfo{person}{Guy~L Steele~Jr}.} \bibinfo{year}{2001}\natexlab{}.
\newblock \showarticletitle{Lock-free reference counting}. In \bibinfo{booktitle}{\emph{Proceedings of the twentieth annual ACM symposium on Principles of distributed computing}}. \bibinfo{pages}{190--199}.
\newblock


\bibitem[Dice et~al\mbox{.}(2016)]%
        {dice2016fast}
\bibfield{author}{\bibinfo{person}{Dave Dice}, \bibinfo{person}{Maurice Herlihy}, {and} \bibinfo{person}{Alex Kogan}.} \bibinfo{year}{2016}\natexlab{}.
\newblock \showarticletitle{Fast non-intrusive memory reclamation for highly-concurrent data structures}. In \bibinfo{booktitle}{\emph{Proceedings of the 2016 ACM SIGPLAN International Symposium on Memory Management}}. \bibinfo{pages}{36--45}.
\newblock


\bibitem[Evans(2006)]%
        {evans2006scalable}
\bibfield{author}{\bibinfo{person}{Jason Evans}.} \bibinfo{year}{2006}\natexlab{}.
\newblock \showarticletitle{A scalable concurrent malloc (3) implementation for FreeBSD}. In \bibinfo{booktitle}{\emph{Proc. of the bsdcan conference, ottawa, canada}}.
\newblock


\bibitem[Fraser(2004)]%
        {fraser2004practical}
\bibfield{author}{\bibinfo{person}{Keir Fraser}.} \bibinfo{year}{2004}\natexlab{}.
\newblock \bibinfo{booktitle}{\emph{Practical lock-freedom}}.
\newblock \bibinfo{type}{{T}echnical {R}eport}. \bibinfo{institution}{University of Cambridge, Computer Laboratory}.
\newblock


\bibitem[Ghemawat and Menage(2005)]%
        {ghemawat2005tcmalloc}
\bibfield{author}{\bibinfo{person}{Sanjay Ghemawat} {and} \bibinfo{person}{Paul Menage}.} \bibinfo{year}{2005}\natexlab{}.
\newblock \showarticletitle{TCMalloc: Thread-caching malloc}.
\newblock \bibinfo{journal}{\emph{Retrieved from \url{http://goog-perftools.sourceforge.net/doc/tcmalloc.html} on January 27, 2023}} (\bibinfo{year}{2005}).
\newblock


\bibitem[Gidenstam et~al\mbox{.}(2008)]%
        {gidenstam2008efficient}
\bibfield{author}{\bibinfo{person}{Anders Gidenstam}, \bibinfo{person}{Marina Papatriantafilou}, \bibinfo{person}{H{\aa}kan Sundell}, {and} \bibinfo{person}{Philippas Tsigas}.} \bibinfo{year}{2008}\natexlab{}.
\newblock \showarticletitle{Efficient and reliable lock-free memory reclamation based on reference counting}.
\newblock \bibinfo{journal}{\emph{IEEE Transactions on Parallel and Distributed Systems}} \bibinfo{volume}{20}, \bibinfo{number}{8} (\bibinfo{year}{2008}), \bibinfo{pages}{1173--1187}.
\newblock


\bibitem[Harris(2001)]%
        {harris2001pragmatic}
\bibfield{author}{\bibinfo{person}{Timothy~L Harris}.} \bibinfo{year}{2001}\natexlab{}.
\newblock \showarticletitle{A pragmatic implementation of non-blocking linked-lists}. In \bibinfo{booktitle}{\emph{International Symposium on Distributed Computing}}. Springer, \bibinfo{pages}{300--314}.
\newblock


\bibitem[Hart et~al\mbox{.}(2007)]%
        {hart2007performance}
\bibfield{author}{\bibinfo{person}{Thomas~E Hart}, \bibinfo{person}{Paul~E McKenney}, \bibinfo{person}{Angela~Demke Brown}, {and} \bibinfo{person}{Jonathan Walpole}.} \bibinfo{year}{2007}\natexlab{}.
\newblock \showarticletitle{Performance of memory reclamation for lockless synchronization}.
\newblock \bibinfo{journal}{\emph{J. Parallel and Distrib. Comput.}} \bibinfo{volume}{67}, \bibinfo{number}{12} (\bibinfo{year}{2007}), \bibinfo{pages}{1270--1285}.
\newblock


\bibitem[Herlihy et~al\mbox{.}(2005)]%
        {herlihy2005nonblocking}
\bibfield{author}{\bibinfo{person}{Maurice Herlihy}, \bibinfo{person}{Victor Luchangco}, \bibinfo{person}{Paul Martin}, {and} \bibinfo{person}{Mark Moir}.} \bibinfo{year}{2005}\natexlab{}.
\newblock \showarticletitle{Nonblocking memory management support for dynamic-sized data structures}.
\newblock \bibinfo{journal}{\emph{ACM Transactions on Computer Systems (TOCS)}} \bibinfo{volume}{23}, \bibinfo{number}{2} (\bibinfo{year}{2005}), \bibinfo{pages}{146--196}.
\newblock


\bibitem[Hudson et~al\mbox{.}(1997)]%
        {hudson1997training}
\bibfield{author}{\bibinfo{person}{Richard~L Hudson}, \bibinfo{person}{Ron Morrison}, \bibinfo{person}{J~Eliot~B Moss}, {and} \bibinfo{person}{David~S Munro}.} \bibinfo{year}{1997}\natexlab{}.
\newblock \showarticletitle{Training distributed garbage: The DMOS collector}.
\newblock \bibinfo{journal}{\emph{Object-Oriented Programming Systems, Language and Applications}} (\bibinfo{year}{1997}).
\newblock


\bibitem[Jung et~al\mbox{.}(2023)]%
        {jungHP2023}
\bibfield{author}{\bibinfo{person}{Jaehwang Jung}, \bibinfo{person}{Janggun Lee}, \bibinfo{person}{Jeonghyeon Kim}, {and} \bibinfo{person}{Jeehoon Kang}.} \bibinfo{year}{2023}\natexlab{}.
\newblock \showarticletitle{Applying Hazard Pointers to More Concurrent Data Structures}. In \bibinfo{booktitle}{\emph{Proceedings of the 35th {ACM} Symposium on Parallelism in Algorithms and Architectures, {SPAA} 2023, Orlando, FL, USA, June 17-19, 2023}}, \bibfield{editor}{\bibinfo{person}{Kunal Agrawal} {and} \bibinfo{person}{Julian Shun}} (Eds.). \bibinfo{publisher}{{ACM}}, \bibinfo{pages}{213--226}.
\newblock
\urldef\tempurl%
\url{https://doi.org/10.1145/3558481.3591102}
\showDOI{\tempurl}


\bibitem[Kang and Jung(2020)]%
        {kang2020marriage}
\bibfield{author}{\bibinfo{person}{Jeehoon Kang} {and} \bibinfo{person}{Jaehwang Jung}.} \bibinfo{year}{2020}\natexlab{}.
\newblock \showarticletitle{A marriage of pointer-and epoch-based reclamation}. In \bibinfo{booktitle}{\emph{Proceedings of the 41st ACM SIGPLAN Conference on Programming Language Design and Implementation}}. \bibinfo{pages}{314--328}.
\newblock


\bibitem[Krieger et~al\mbox{.}(2006)]%
        {krieger2006k42}
\bibfield{author}{\bibinfo{person}{Orran Krieger}, \bibinfo{person}{Marc Auslander}, \bibinfo{person}{Bryan Rosenburg}, \bibinfo{person}{Robert~W Wisniewski}, \bibinfo{person}{Jimi Xenidis}, \bibinfo{person}{Dilma Da~Silva}, \bibinfo{person}{Michal Ostrowski}, \bibinfo{person}{Jonathan Appavoo}, \bibinfo{person}{Maria Butrico}, \bibinfo{person}{Mark Mergen}, {et~al\mbox{.}}} \bibinfo{year}{2006}\natexlab{}.
\newblock \showarticletitle{K42: building a complete operating system}.
\newblock \bibinfo{journal}{\emph{ACM SIGOPS Operating Systems Review}} \bibinfo{volume}{40}, \bibinfo{number}{4} (\bibinfo{year}{2006}), \bibinfo{pages}{133--145}.
\newblock


\bibitem[Leijen et~al\mbox{.}(2019)]%
        {leijen2019mimalloc}
\bibfield{author}{\bibinfo{person}{Daan Leijen}, \bibinfo{person}{Benjamin Zorn}, {and} \bibinfo{person}{Leonardo de Moura}.} \bibinfo{year}{2019}\natexlab{}.
\newblock \showarticletitle{Mimalloc: Free list sharding in action}. In \bibinfo{booktitle}{\emph{Programming Languages and Systems: 17th Asian Symposium, APLAS 2019, Nusa Dua, Bali, Indonesia, December 1--4, 2019, Proceedings 17}}. Springer, \bibinfo{pages}{244--265}.
\newblock


\bibitem[McKenney and Slingwine(1998)]%
        {mckenney1998read}
\bibfield{author}{\bibinfo{person}{Paul~E McKenney} {and} \bibinfo{person}{John~D Slingwine}.} \bibinfo{year}{1998}\natexlab{}.
\newblock \showarticletitle{Read-copy update: Using execution history to solve concurrency problems}. In \bibinfo{booktitle}{\emph{Parallel and Distributed Computing and Systems}}, Vol.~\bibinfo{volume}{509518}. Citeseer, \bibinfo{pages}{509--518}.
\newblock


\bibitem[Michael(2004)]%
        {michael2004hazard}
\bibfield{author}{\bibinfo{person}{Maged~M Michael}.} \bibinfo{year}{2004}\natexlab{}.
\newblock \showarticletitle{Hazard pointers: Safe memory reclamation for lock-free objects}.
\newblock \bibinfo{journal}{\emph{IEEE Transactions on Parallel and Distributed Systems}} \bibinfo{volume}{15}, \bibinfo{number}{6} (\bibinfo{year}{2004}), \bibinfo{pages}{491--504}.
\newblock


\bibitem[Mitake et~al\mbox{.}(2019)]%
        {mitake2019looking}
\bibfield{author}{\bibinfo{person}{Hitoshi Mitake}, \bibinfo{person}{Hiroshi Yamada}, {and} \bibinfo{person}{Tatsuo Nakajima}.} \bibinfo{year}{2019}\natexlab{}.
\newblock \showarticletitle{Looking into the Peak memory consumption of epoch-based reclamation in scalable in-memory database systems}. In \bibinfo{booktitle}{\emph{Database and Expert Systems Applications: 30th International Conference, DEXA 2019, Linz, Austria, August 26--29, 2019, Proceedings, Part II 30}}. Springer, \bibinfo{pages}{3--18}.
\newblock


\bibitem[Moreno and Rocha(2023)]%
        {moreno2023releasing}
\bibfield{author}{\bibinfo{person}{Pedro Moreno} {and} \bibinfo{person}{Ricardo Rocha}.} \bibinfo{year}{2023}\natexlab{}.
\newblock \showarticletitle{Releasing Memory with Optimistic Access: A Hybrid Approach to Memory Reclamation and Allocation in Lock-Free Programs}. In \bibinfo{booktitle}{\emph{Proceedings of the 35th ACM Symposium on Parallelism in Algorithms and Architectures}}. \bibinfo{pages}{177--186}.
\newblock


\bibitem[Nikolaev and Ravindran(2019)]%
        {nikolaev2019hyaline}
\bibfield{author}{\bibinfo{person}{Ruslan Nikolaev} {and} \bibinfo{person}{Binoy Ravindran}.} \bibinfo{year}{2019}\natexlab{}.
\newblock \showarticletitle{Hyaline: fast and transparent lock-free memory reclamation}. In \bibinfo{booktitle}{\emph{Proceedings of the 2019 ACM Symposium on Principles of Distributed Computing}}. \bibinfo{pages}{419--421}.
\newblock


\bibitem[Nikolaev and Ravindran(2020)]%
        {nikolaev2020universal}
\bibfield{author}{\bibinfo{person}{Ruslan Nikolaev} {and} \bibinfo{person}{Binoy Ravindran}.} \bibinfo{year}{2020}\natexlab{}.
\newblock \showarticletitle{Universal wait-free memory reclamation}. In \bibinfo{booktitle}{\emph{Proceedings of the 25th ACM SIGPLAN Symposium on Principles and Practice of Parallel Programming}}. \bibinfo{pages}{130--143}.
\newblock


\bibitem[Ramalhete and Correia({[n.\,d.]})]%
        {ramalhete2017brief}
\bibfield{author}{\bibinfo{person}{Pedro Ramalhete} {and} \bibinfo{person}{Andreia Correia}.} \bibinfo{year}{[n.\,d.]}\natexlab{}.
\newblock \showarticletitle{Brief announcement: Hazard eras-non-blocking memory reclamation}. In \bibinfo{booktitle}{\emph{Proceedings of the 29th ACM Symposium on Parallelism in Algorithms and Architectures}}. \bibinfo{pages}{367--369}.
\newblock


\bibitem[Sheffi et~al\mbox{.}(2021)]%
        {sheffi2021vbr}
\bibfield{author}{\bibinfo{person}{Gali Sheffi}, \bibinfo{person}{Maurice Herlihy}, {and} \bibinfo{person}{Erez Petrank}.} \bibinfo{year}{2021}\natexlab{}.
\newblock \showarticletitle{Vbr: Version based reclamation}. In \bibinfo{booktitle}{\emph{Proceedings of the 33rd ACM Symposium on Parallelism in Algorithms and Architectures}}. \bibinfo{pages}{443--445}.
\newblock


\bibitem[Singh et~al\mbox{.}(2021)]%
        {singh2021nbr}
\bibfield{author}{\bibinfo{person}{Ajay Singh}, \bibinfo{person}{Trevor Brown}, {and} \bibinfo{person}{Ali Mashtizadeh}.} \bibinfo{year}{2021}\natexlab{}.
\newblock \showarticletitle{Nbr: neutralization based reclamation}. In \bibinfo{booktitle}{\emph{Proceedings of the 26th ACM SIGPLAN Symposium on Principles and Practice of Parallel Programming}}. \bibinfo{pages}{175--190}.
\newblock


\bibitem[Singh et~al\mbox{.}(2023)]%
        {singhCA2023}
\bibfield{author}{\bibinfo{person}{Ajay Singh}, \bibinfo{person}{Trevor Brown}, {and} \bibinfo{person}{Michael Spear}.} \bibinfo{year}{2023}\natexlab{}.
\newblock \showarticletitle{Efficient Hardware Primitives for Immediate Memory Reclamation in Optimistic Data Structures}. In \bibinfo{booktitle}{\emph{2023 IEEE International Parallel and Distributed Processing Symposium (IPDPS)}}. \bibinfo{pages}{112--122}.
\newblock
\urldef\tempurl%
\url{https://doi.org/10.1109/IPDPS54959.2023.00021}
\showDOI{\tempurl}


\bibitem[Singh et~al\mbox{.}(2024)]%
        {singhTPDS}
\bibfield{author}{\bibinfo{person}{Ajay Singh}, \bibinfo{person}{Trevor~Alexander Brown}, {and} \bibinfo{person}{Ali~José Mashtizadeh}.} \bibinfo{year}{2024}\natexlab{}.
\newblock \showarticletitle{Simple, Fast and Widely Applicable Concurrent Memory Reclamation via Neutralization}.
\newblock \bibinfo{journal}{\emph{IEEE Transactions on Parallel and Distributed Systems}} \bibinfo{volume}{35}, \bibinfo{number}{2} (\bibinfo{year}{2024}), \bibinfo{pages}{203--220}.
\newblock
\urldef\tempurl%
\url{https://doi.org/10.1109/TPDS.2023.3335671}
\showDOI{\tempurl}


\bibitem[Tam(2006)]%
        {tam2006qdo}
\bibfield{author}{\bibinfo{person}{Adrian Tam}.} \bibinfo{year}{2006}\natexlab{}.
\newblock \emph{\bibinfo{title}{QDo: A Quiescent State Callback Facility}}.
\newblock \bibinfo{thesistype}{Ph.\,D. Dissertation}. \bibinfo{school}{University of Toronto}.
\newblock


\bibitem[Wen et~al\mbox{.}(2018)]%
        {wen2018interval}
\bibfield{author}{\bibinfo{person}{Haosen Wen}, \bibinfo{person}{Joseph Izraelevitz}, \bibinfo{person}{Wentao Cai}, \bibinfo{person}{H~Alan Beadle}, {and} \bibinfo{person}{Michael~L Scott}.} \bibinfo{year}{2018}\natexlab{}.
\newblock \showarticletitle{Interval-based memory reclamation}.
\newblock \bibinfo{journal}{\emph{ACM SIGPLAN Notices}} \bibinfo{volume}{53}, \bibinfo{number}{1} (\bibinfo{year}{2018}), \bibinfo{pages}{1--13}.
\newblock


\bibitem[ycombinator(2017)]%
        {ycombinator}
\bibfield{author}{\bibinfo{person}{ycombinator}.} \bibinfo{year}{2017}\natexlab{}.
\newblock \bibinfo{title}{Why is memory reclamation so important for lock-free algorithms?}
\newblock \bibinfo{howpublished}{Retrieved from \url{https://web.archive.org/web/20200223075152/https://news.ycombinator.com/item?id=15269628} on January 27, 2023}.
\newblock


\end{thebibliography}

% \newpage
\onecolumn
\appendix
\section{Artifact Description}
The artifact containing the source code, scripts to run all experiments, and a detailed readme file is present at the URL: \url{https://doi.org/10.5281/zenodo.10226261}.
If you prefer to run the artifact locally on your machine (without using the docker container) please directly refer to the accompanying readme file in the source code.

The following instructions will help you load and run the provided Docker image within the artifact.
Once the docker container starts you can use the accompanying readme file to compile and run the experiments on the benchmark.
\\
\\
\myparagraph{Steps to load and run the provided Docker image in the artifact:}

Note: Sudo permission may be required to execute the following instructions. The following instructions will help you install and directly run the docker container for amortizedfree-setbench.

\begin{enumerate}
    \item Install the latest version of Docker on your system. We tested the artifact with the Docker version 20.10.2, build 20.10.2-0ubuntu1~20.04.2. Instructions to install Docker may be found at\\ \url{https://docs.docker.com/engine/install/ubuntu}.
\begin{verbatim}
$ docker -v
\end{verbatim}

    \item  Download the artifact named amortizedfree-setbench.zip from the artifact submission link:\\
    \url{https://doi.org/10.5281/zenodo.10226261}.
    % Alternatively, most recent version could be downloaded at the gitlab repo, URL: \url{https://gitlab.com/aajayssingh/nbr_setbench}

    \item Extract the downloaded folder and move to \\
    \textit{amortizedfree-setbench/} directory using $cd$ command.
    \item Find docker image named \textit{amortizedfree\_docker.tar.gz}\\
    in amortizedfree-setbench/ directory. And load the downloaded docker image with the following command:
\begin{verbatim}
 $ sudo docker load -i amortizedfree_docker.tar.gz
\end{verbatim}
\item 
Verify that image was loaded:
\begin{verbatim}
 $ sudo docker images
\end{verbatim}
\item Start a docker container from the loaded image:
\begin{verbatim}
 $  sudo docker run --name amortizedfree -it \
 --privileged amortizedfree-setbench /bin/bash
\end{verbatim}
\item Invoke $ls$ to see several files/folders of the artifact: Dockerfile README.md, common, ds, install.sh, lib, microbench, af\_experiments,
tools.
\end{enumerate}

Now, to compile and run the experiments you could follow the instructions in the readme file.
You may need to change the thread counts in scripts to suit the configuration of your machine.

\newpage
\section{Allocator Background}
\paragraph{TCmalloc}

        TCmalloc (which is short for \textit{thread caching malloc}) was developed by Google researchers, and released in 2005.
        Huge objects are allocated an appropriate number of whole pages from a central page heap.
        Small objects are allocated in 170 different size classes.
        For each size class, there is a \textit{central free list} and a \textit{thread local cache} for each thread.
        The central free list is protected by a lock.

        When a thread allocates memory, it first tries to serve the allocation from its thread local cache (for the appropriate size class).
        If there is no available object in the thread local cache, it repopulates the thread local cache using a batch of objects from the central free list (for that size class).
        %Memory is mapped and partitioned into objects at the central cache level.
%
        When a thread frees an object, if the thread local cache is full, a batch of objects is moved from the thread local cache to the central free list.
        Accesses to the central free list can result in substantial contention in systems with many cores.
    
\paragraph{JEmalloc}

        JEmalloc is an extremely popular allocator.
        It is the default allocator for FreeBSD, Firefox, and many other large software packages, and it has seen active development by large companies such as Facebook.

        Whereas TCmalloc uses a central free list that is shared by all threads, JEmalloc uses \textit{arenas}, which \textit{can be} shared among threads, but the number of arenas is made fairly large in an effort to reduce contention among threads.
        Specifically, by default there are $4n$ arenas, where $n$ is the number of processors.
        A given thread is essentially hashed to an arena, and each arena is protected by a lock.

        % \textbf{[some conjectures to confirm about JEmalloc]}
        Many size classes reside in an arena.
        An arena has one chunk that it allocates 4KB pages from.
        If the chunk becomes empty, a new chunk is \texttt{mmap}ed.
        When an allocation happens in size class $x$, it first checks the corresponding thread cache for that size class.
        If the thread cache is empty, it checks if there is an active page in the arena for that size class (containing some free space).
        If there is a free object, JEmalloc bump allocates from the page.
        If the page is empty (no free space), JEmalloc bump allocates one or more new pages from the arena's chunk (and makes those the active page(s) for this size class).
        %
        % Details of how JEmalloc implements \texttt{free} appear in \Cref{sec:root cause of the EBR delays}.

\paragraph{MImalloc}

        MImalloc is a recent allocator that was developed by Microsoft Research.
        Whereas other allocators use per thread cache (free list), MImalloc has page-level free lists. This helps MiMalloc to reduce contention as thread share at smaller granularity. 
        % Whereas other allocators use per thread cache (free list) as the lowest level of cache, MImalloc has page-level free lists.
        Furthermore, in a MImalloc page, there are three different freelists: allocation free list, local free list, cross-thread free list.
        A thread allocates an object from its allocation free lists, and frees an object to its local free list if the object belongs to it.
        Otherwise it frees to the cross-thread free list of the remote page that it originally allocates the object.
        Accessing cross-free list should be atomic.
        When the allocation free list is empty, the thread moves objects from cross-thread free list to local thread free list and swap allocation free list and local free list.
        
        A thread can use multiple pages for allocation, and the thread manages pages with a list.
        If there is no page that has free object, the thread gets a new page from a segment.
        A segment is 4MB chunk of memory that is created after \texttt{mmap}ed.

\clearpage
\section{ORIG vs AF for ABtree}

\begin{figure}[h]
    \centering
    \begin{subfigure}{0.3\linewidth}
        \centering
        \includegraphics[width=\linewidth]{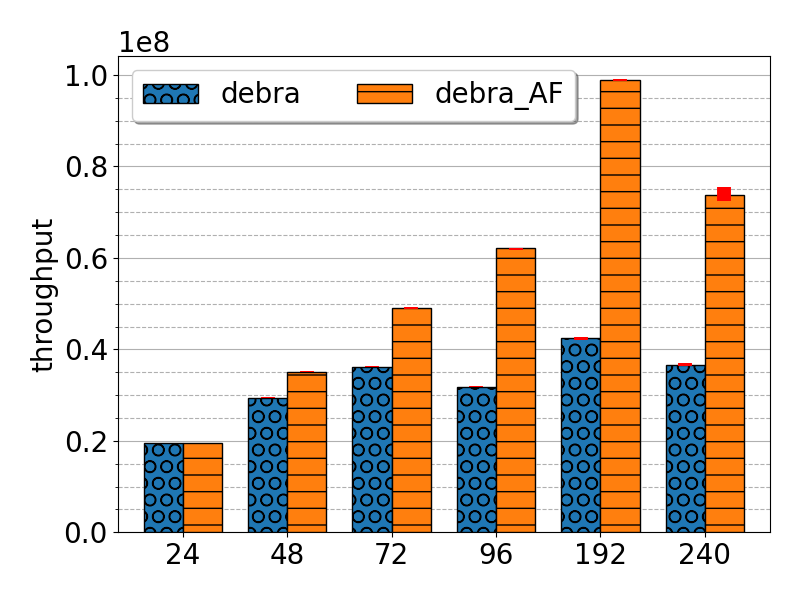}
        \subcaption{debra}
        \label{fig:indiv1}
    \end{subfigure}
    \hfill
    \begin{subfigure}{0.3\linewidth}
        \centering
        \includegraphics[width=\linewidth]{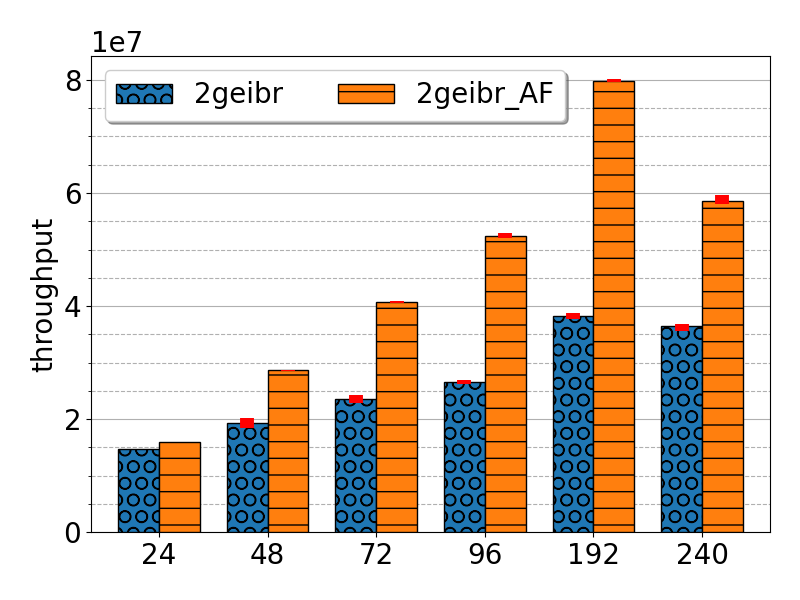}
        \subcaption{ibr}
        \label{fig:indiv2}
    \end{subfigure}
    \hfill
    \begin{subfigure}{0.3\linewidth}
        \centering
        \includegraphics[width=\linewidth]{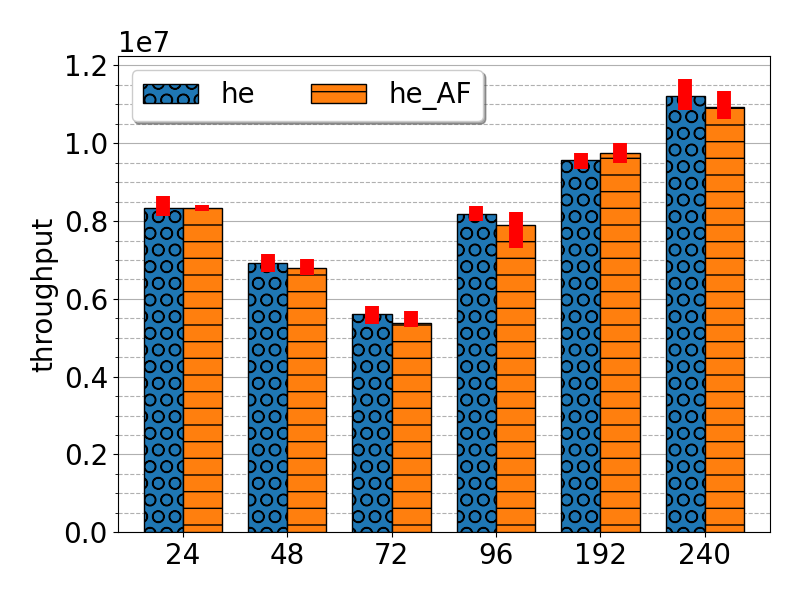}
        \subcaption{he}
        \label{fig:indiv3}
    \end{subfigure}
    \vfill
%     \caption{}
%     \label{fig:indiv-exp1}
% \end{figure*}
% \begin{figure*}[h]
% \centering
    \begin{subfigure}{0.3\linewidth}
        \centering
        \includegraphics[width=\linewidth]{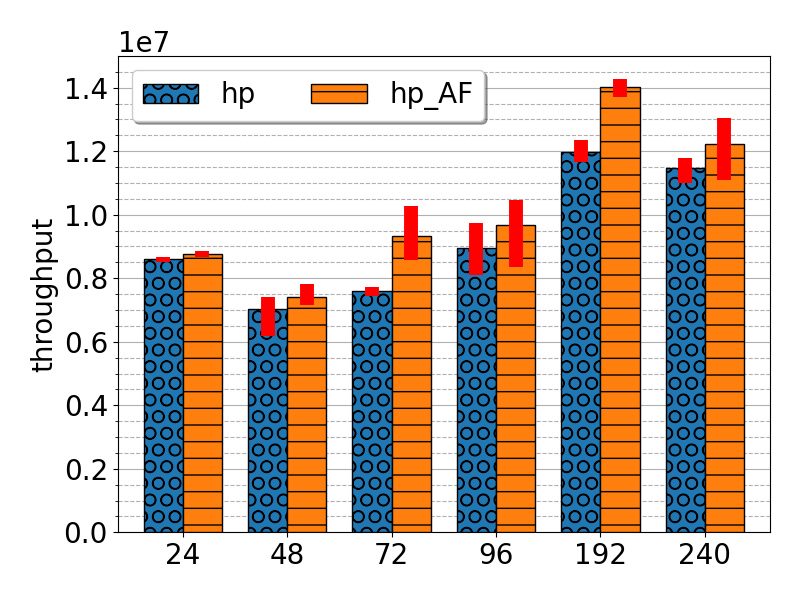}
        \subcaption{hp}
        \label{fig:indiv4}
    \end{subfigure}
    \hfill
    \begin{subfigure}{0.3\linewidth}
        \centering
        \includegraphics[width=\linewidth]{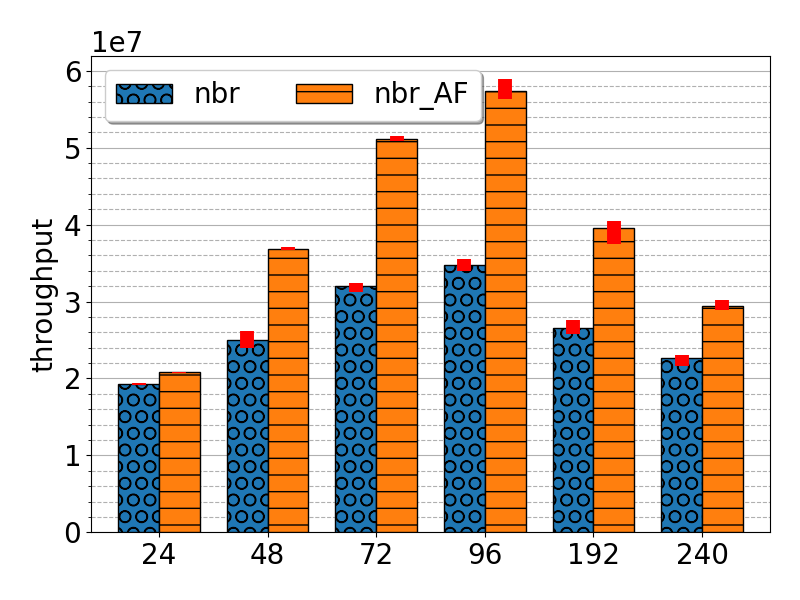}
        \subcaption{nbr}
        \label{fig:indiv5}
    \end{subfigure}
    \hfill
    \begin{subfigure}{0.3\linewidth}
        \centering
        \includegraphics[width=\linewidth]{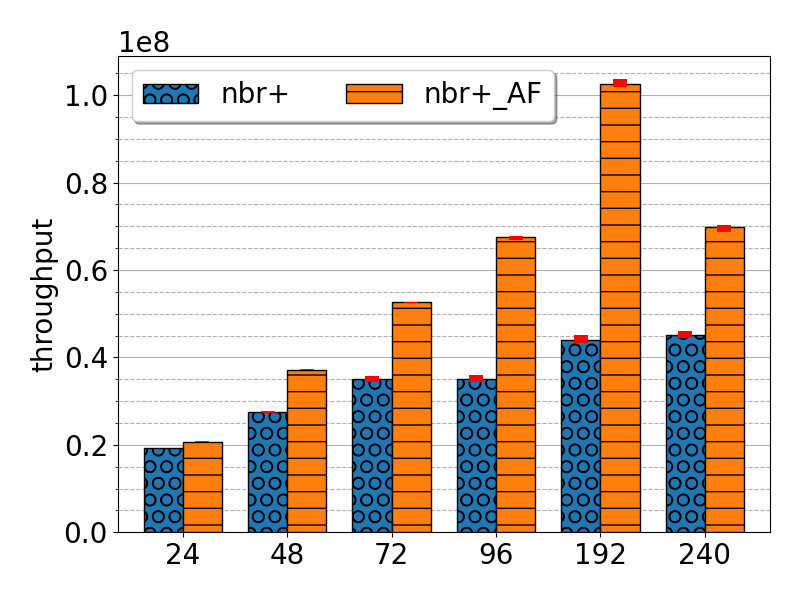}
        \subcaption{nbr+}
        \label{fig:indiv6}
    \end{subfigure}
    \vfill
    \begin{subfigure}{0.3\linewidth}
        \centering
        \includegraphics[width=\linewidth]{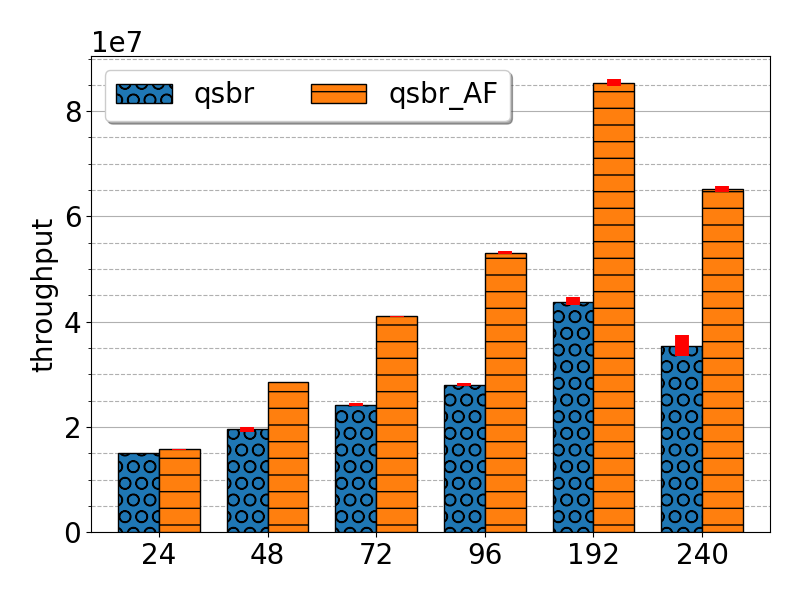}
        \subcaption{qsbr}
        \label{fig:indiv7}
    \end{subfigure}    
    \hfill
    \begin{subfigure}{0.3\linewidth}
        \centering
        \includegraphics[width=\linewidth]{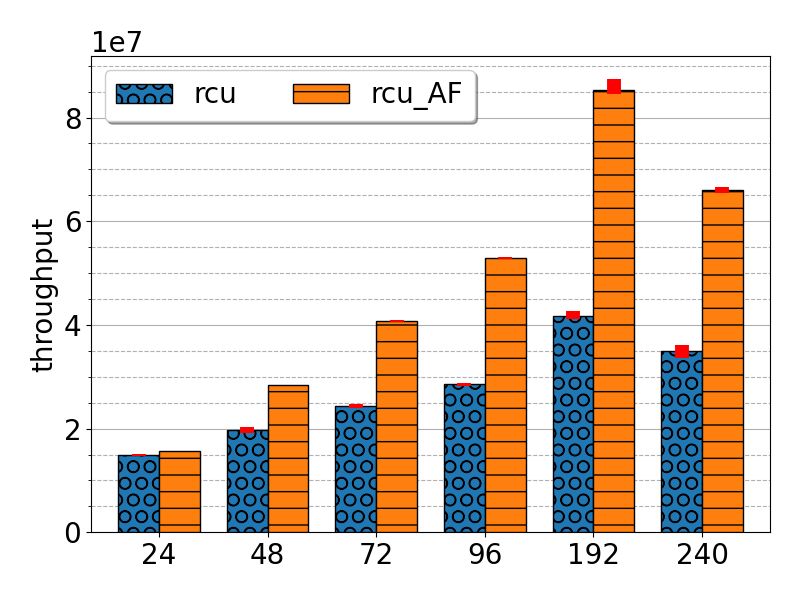}
        \subcaption{rcu}
        \label{fig:indiv8}
    \end{subfigure}        
    \hfill
    \begin{subfigure}{0.3\linewidth}
        \centering
        \includegraphics[width=\linewidth]{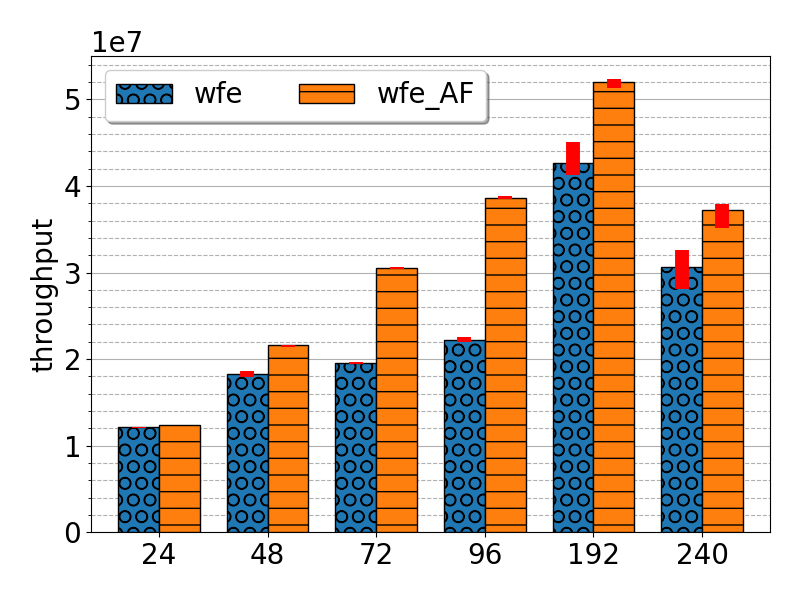}
        \subcaption{wfe}
        \label{fig:indiv9}
    \end{subfigure}        
    \caption{Throughput (operations/second) across varying threads of amortized free versions of reclaimers with their original implementations. Data structure: ABtree. Workload: 50\% inserts and 50\% deletes. Size: 20M. Allocator:JEmalloc.}
    \label{fig:indiv-exp2}
\end{figure}

\clearpage
\section{Comparison of Amortized Freeing using Ticket locking External Binary Search Tree}
In this section we use a ticket locking external binary search tree due to David, Guerraoui and Trigonakis to evaluate amortized freeing.
\begin{figure}[h]
    \centering
    \begin{subfigure}{0.3\linewidth}
        \centering
        \includegraphics[width=\linewidth]{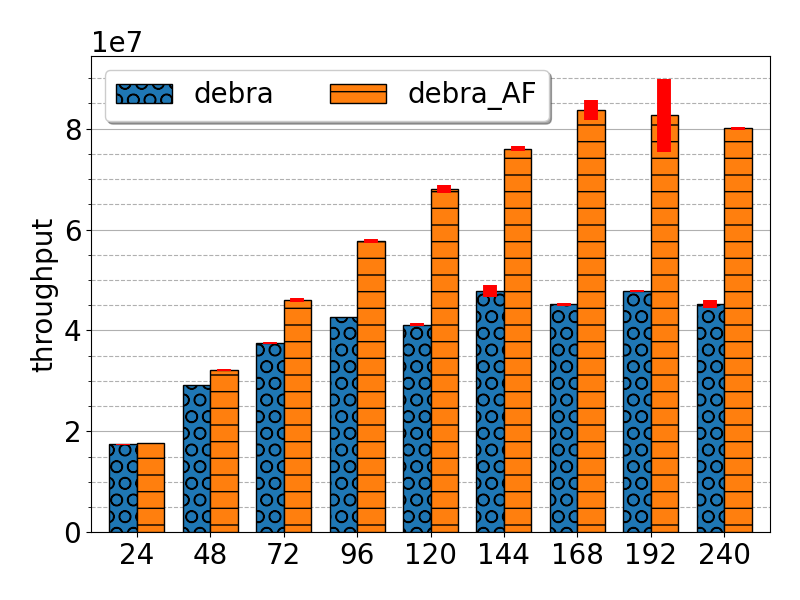}
        \subcaption{debra}
        \label{fig:indiv-dgt1}
    \end{subfigure}
    \hfill
    \begin{subfigure}{0.3\linewidth}
        \centering
        \includegraphics[width=\linewidth]{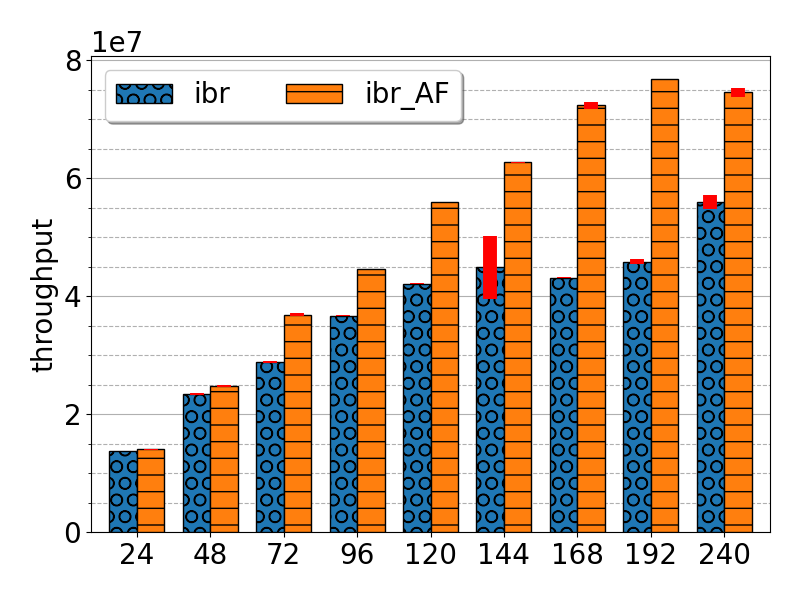}
        \subcaption{ibr}
        \label{fig:indiv-dgt2}
    \end{subfigure}
    \hfill
    \begin{subfigure}{0.3\linewidth}
        \centering
        \includegraphics[width=\linewidth]{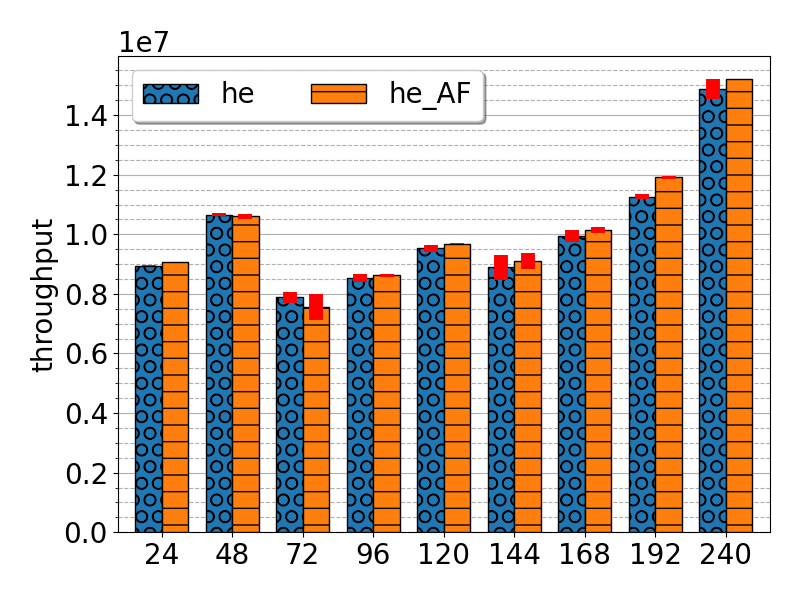}
        \subcaption{he}
        \label{fig:indiv-dgt3}
    \end{subfigure}
    \vfill
%     \caption{}
%     \label{fig:indiv-exp1}
% \end{figure*}
% \begin{figure*}[h]
% \centering
    \begin{subfigure}{0.3\linewidth}
        \centering
        \includegraphics[width=\linewidth]{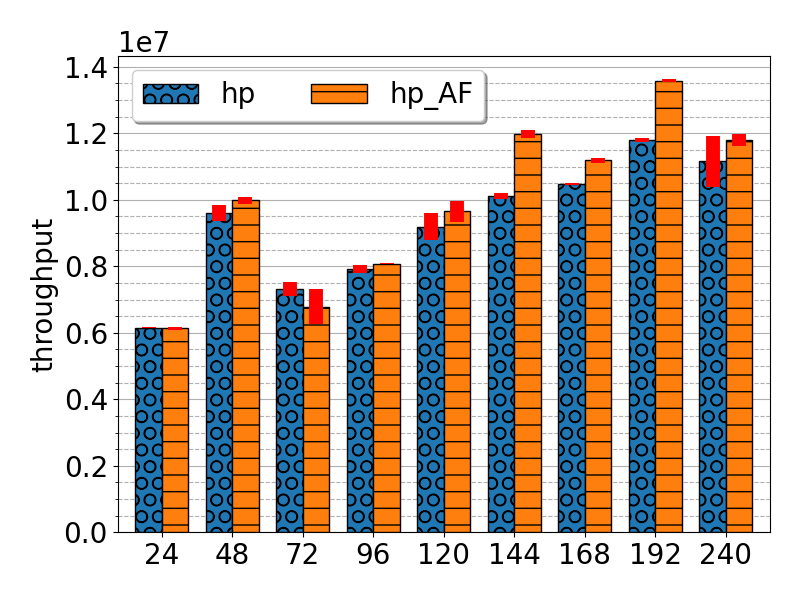}
        \subcaption{hp}
        \label{fig:indiv-dgt4}
    \end{subfigure}
    \hfill
    \begin{subfigure}{0.3\linewidth}
        \centering
        \includegraphics[width=\linewidth]{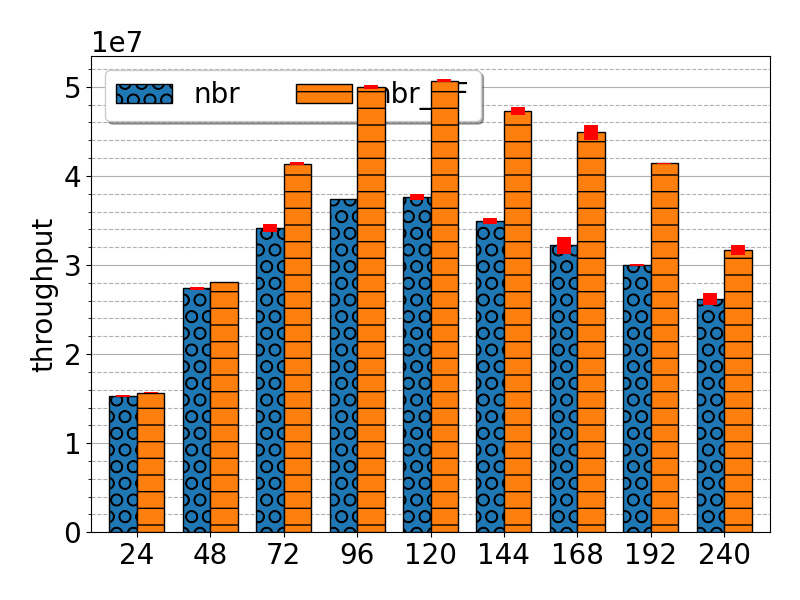}
        \subcaption{nbr}
        \label{fig:indiv-dgt5}
    \end{subfigure}
    \hfill
    \begin{subfigure}{0.3\linewidth}
        \centering
        \includegraphics[width=\linewidth]{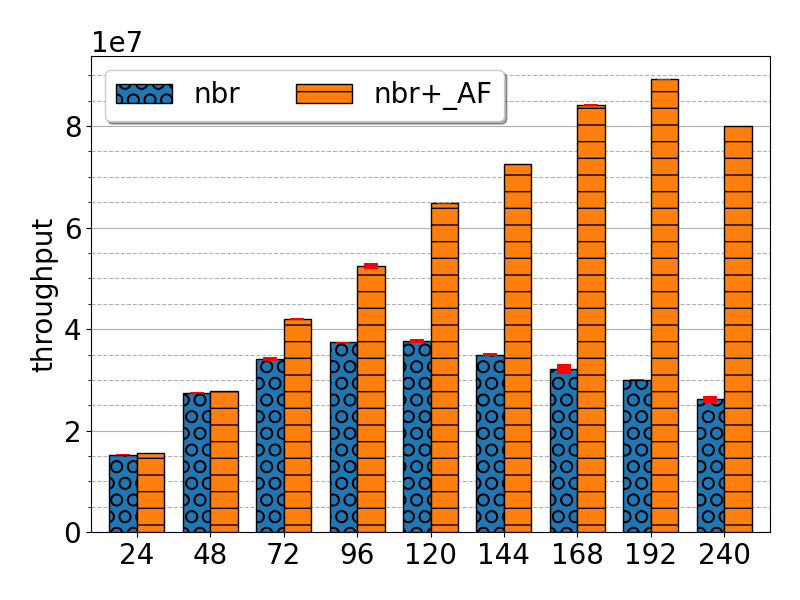}
        \subcaption{nbr+}
        \label{fig:indiv-dgt6}
    \end{subfigure}
    \vfill
    \begin{subfigure}{0.3\linewidth}
        \centering
        \includegraphics[width=\linewidth]{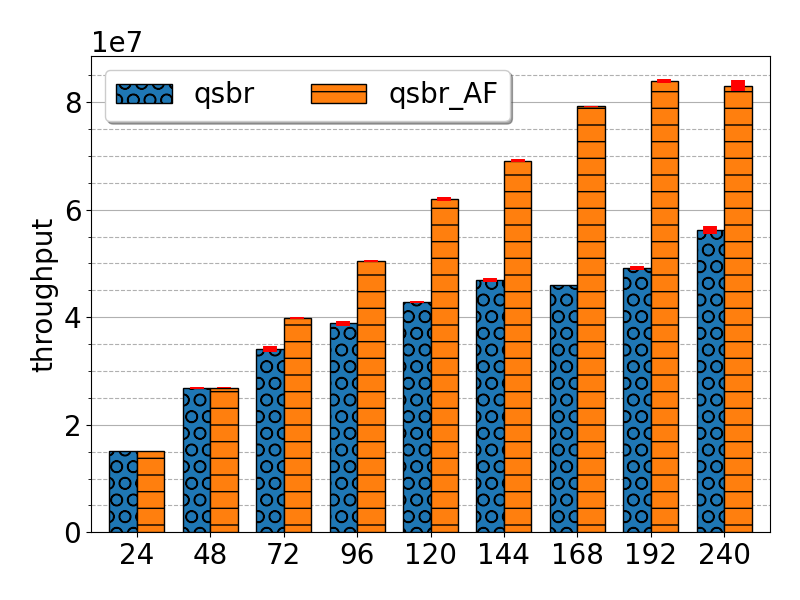}
        \subcaption{qsbr}
        \label{fig:indiv-dgt7}
    \end{subfigure}    
    \hfill
    \begin{subfigure}{0.3\linewidth}
        \centering
        \includegraphics[width=\linewidth]{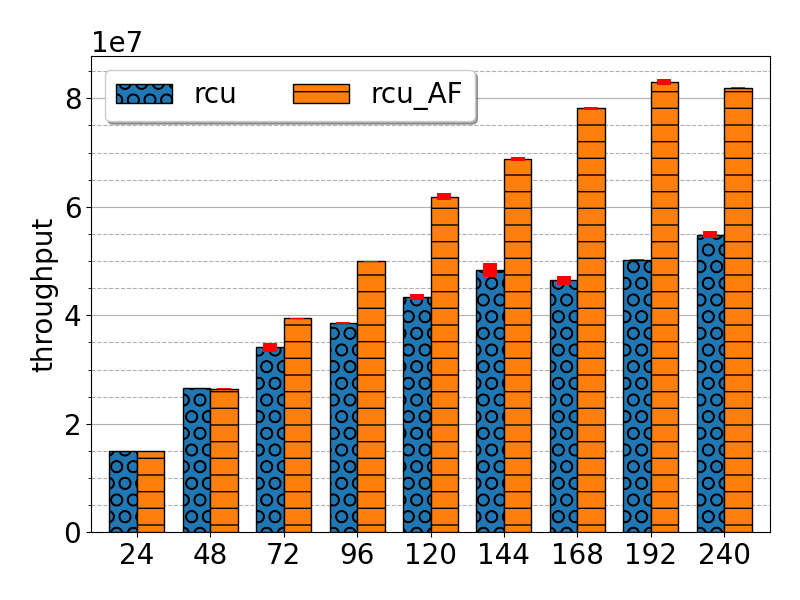}
        \subcaption{rcu}
        \label{fig:indiv-dgt8}
    \end{subfigure}        
    \hfill
    \begin{subfigure}{0.3\linewidth}
        \centering
        \includegraphics[width=\linewidth]{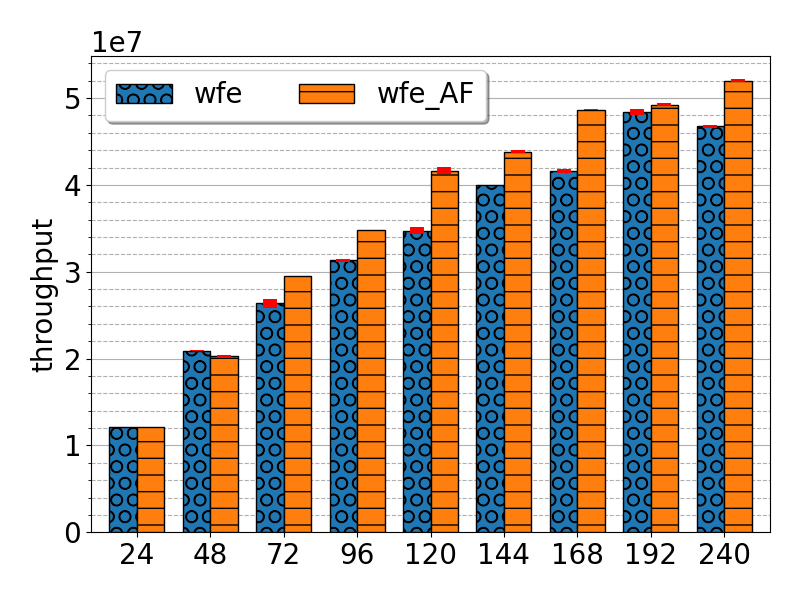}
        \subcaption{wfe}
        \label{fig:indiv-dgt9}
    \end{subfigure}        
    \caption{Throughput (operations/second) across varying threads of amortized free versions of reclaimers with their original implementations. Data structure: DGT Tree. Workload: 50\% inserts and 50\% deletes. Size: 2M. Allocator:JEmalloc.}
    \label{fig:indiv-dgt-exp2}
\end{figure}

\begin{figure}[!ht]
     % \begin{minipage}{\textwidth}
        % \begin{subfigure}{0.49\textwidth}
            \centering
            \includegraphics[width=1\linewidth, height=6cm, keepaspectratio]{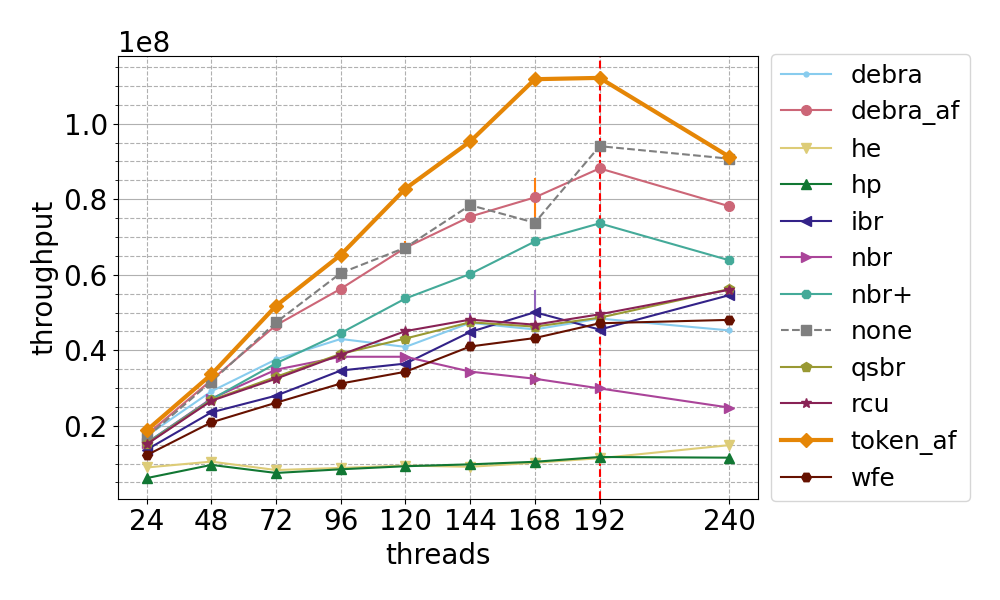}
            % \subcaption{Experiment 1: Comparison of proposed \aftebr{} (token\_af) with other reclamation techniques across threads.}
            % \label{fig:exp1-dgt}
        % \end{subfigure}
    \caption{Comparison of proposed token\_af with other reclamation techniques across threads. Data structure: DGT Tree. Workload: 50\% inserts and 50\% deletes. Size: 2M. Allocator:JEmalloc.
    }
    \label{fig:exp1-dgt}
\end{figure}

\clearpage

\section{Performance on other machines}

\subsection{Intel 4 socket 144 core}

\begin{figure}[!ht]
    \begin{subfigure}{0.48\textwidth}
        \includegraphics[width=\textwidth, height=6cm]{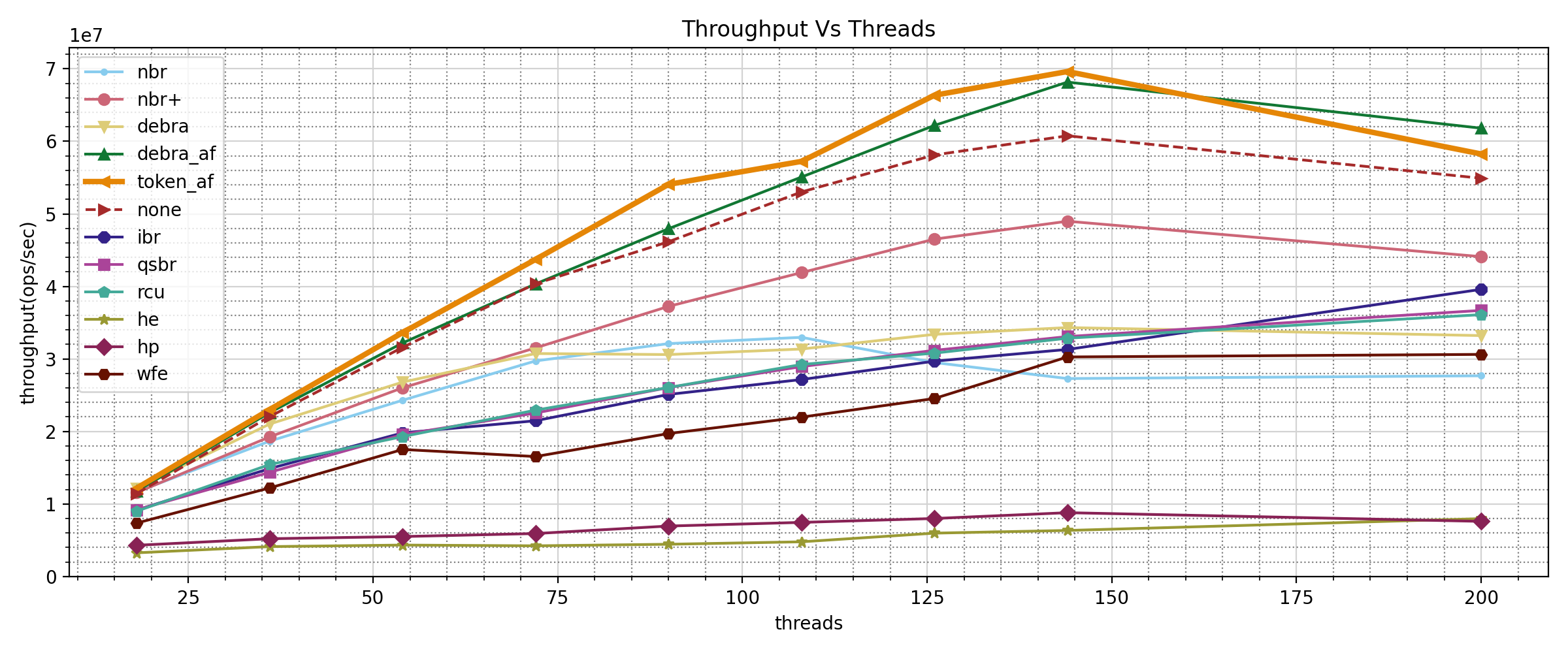}
        \caption{Comparison of proposed token\_af with other reclamation techniques across threads. Data structure: ABtree}
        \label{subfig:Zyra_ABtree_fig11a}
    \end{subfigure}
    \hspace{0.03\linewidth}
    \begin{subfigure}{0.48\textwidth}
        \includegraphics[width=\textwidth, height=6cm]{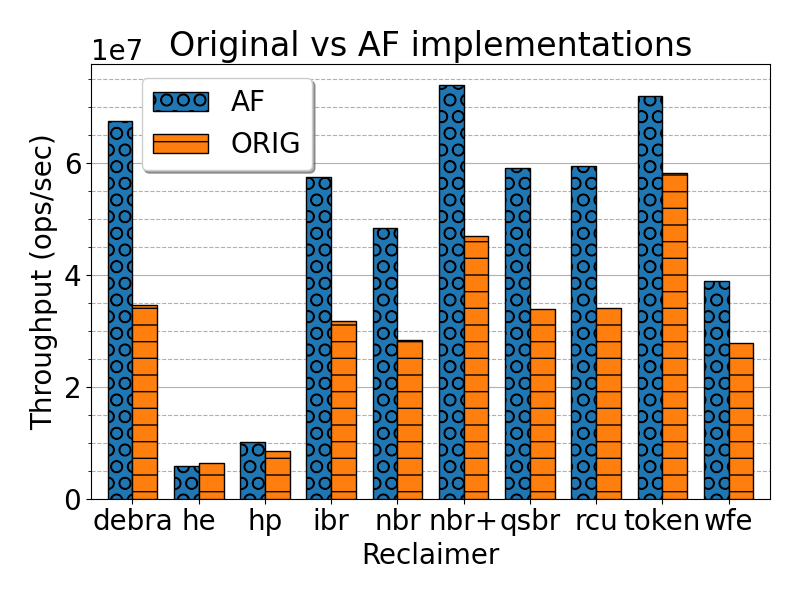}
        \caption{Comparison with amortized free versions of various reclaimers at 192 threads. Data structure: ABtree}
        \label{subfig:Zyra_ABtree_fig11b}
    \end{subfigure}
    \begin{subfigure}{0.48\textwidth}
        \includegraphics[width=\textwidth, height=6cm]{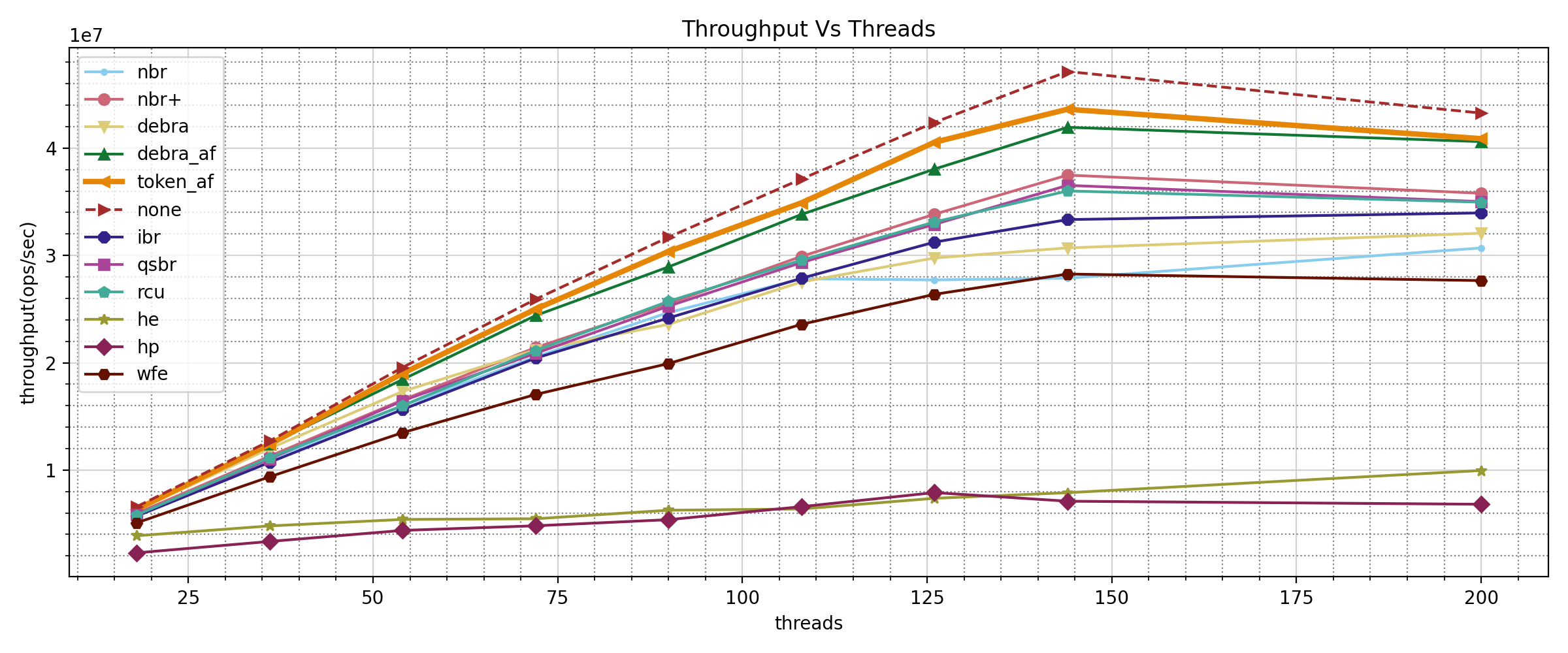}
        \caption{Comparison of proposed token\_af with other reclamation techniques across threads. Data structure: DGT Tree}
        \label{subfig:Zyra_DGT_fig11a}
    \end{subfigure}
    \hspace{0.03\linewidth}
    \begin{subfigure}{0.48\textwidth}
        \includegraphics[width=\textwidth, height=6cm]{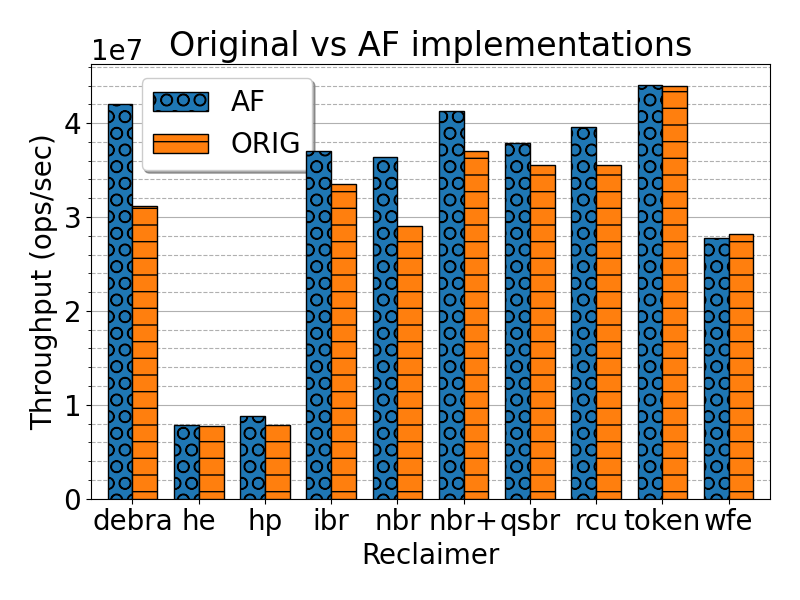}
        \caption{Comparison with amortized free versions of various reclaimers at 192 threads. Data structure: DGT Tree}
        \label{subfig:Zyra_DGT_fig11b}
    \end{subfigure}
    \caption{Intel 4 socket 144 core machine. Workload: 50\% inserts and 50\% deletes. Size: 20M. Allocator:JEmalloc}
\end{figure}

\clearpage

\subsection{AMD 2 socket 256 core}

\begin{figure}[!ht]
    \begin{subfigure}{0.48\textwidth}
        \includegraphics[width=\textwidth, height=6cm]{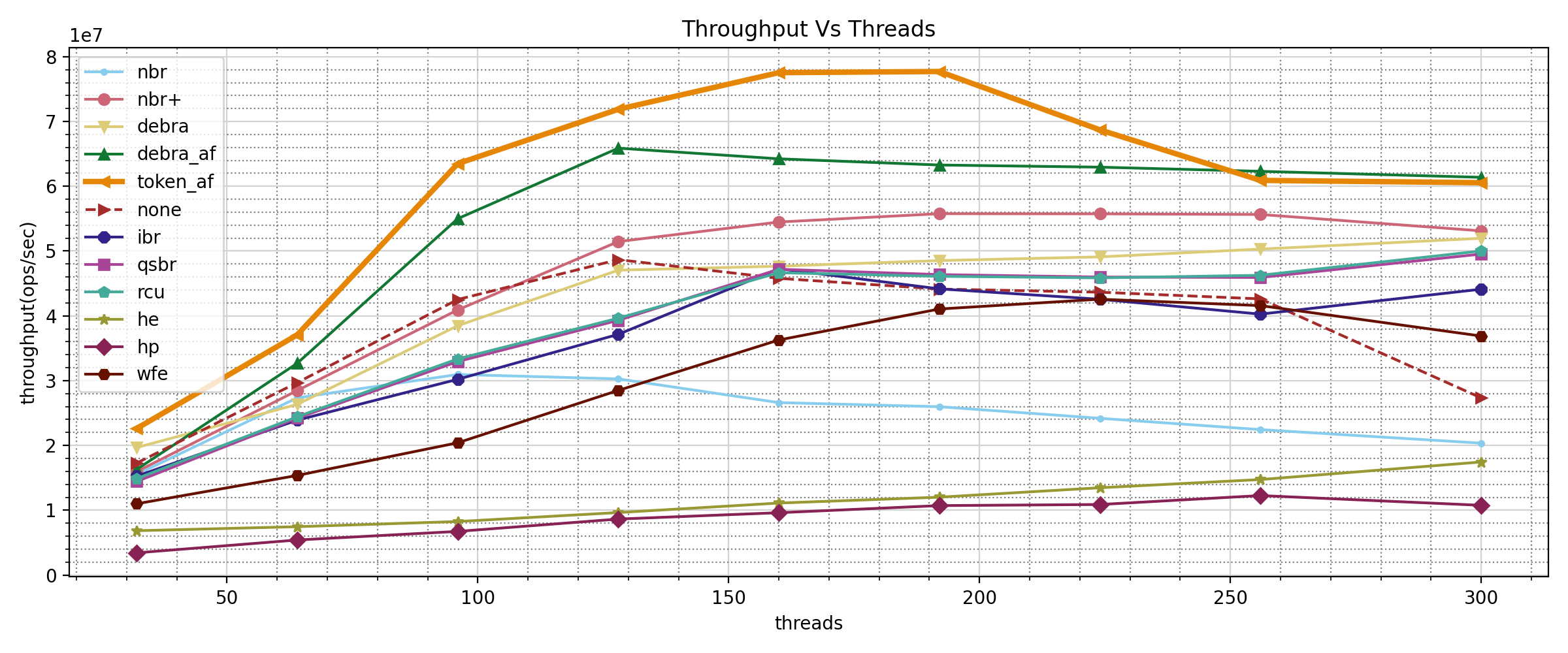}
        \caption{Comparison of proposed token\_af with other reclamation techniques across threads. Data structure: ABtree}
        \label{subfig:Nasus_ABtree_fig11a}
    \end{subfigure}
    \hspace{0.03\linewidth}
    \begin{subfigure}{0.48\textwidth}
        \includegraphics[width=\textwidth, height=6cm]{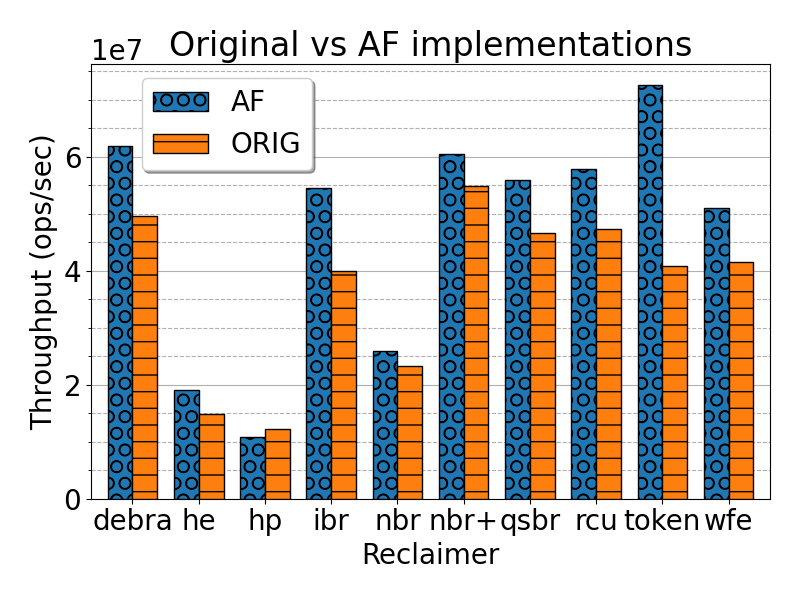}
        \caption{Comparison with amortized free versions of various reclaimers at 192 threads. Data structure: ABtree}
        \label{subfig:Nasus_ABtree_fig11b}
    \end{subfigure}
    \begin{subfigure}{0.48\textwidth}
        \includegraphics[width=\textwidth, height=6cm]{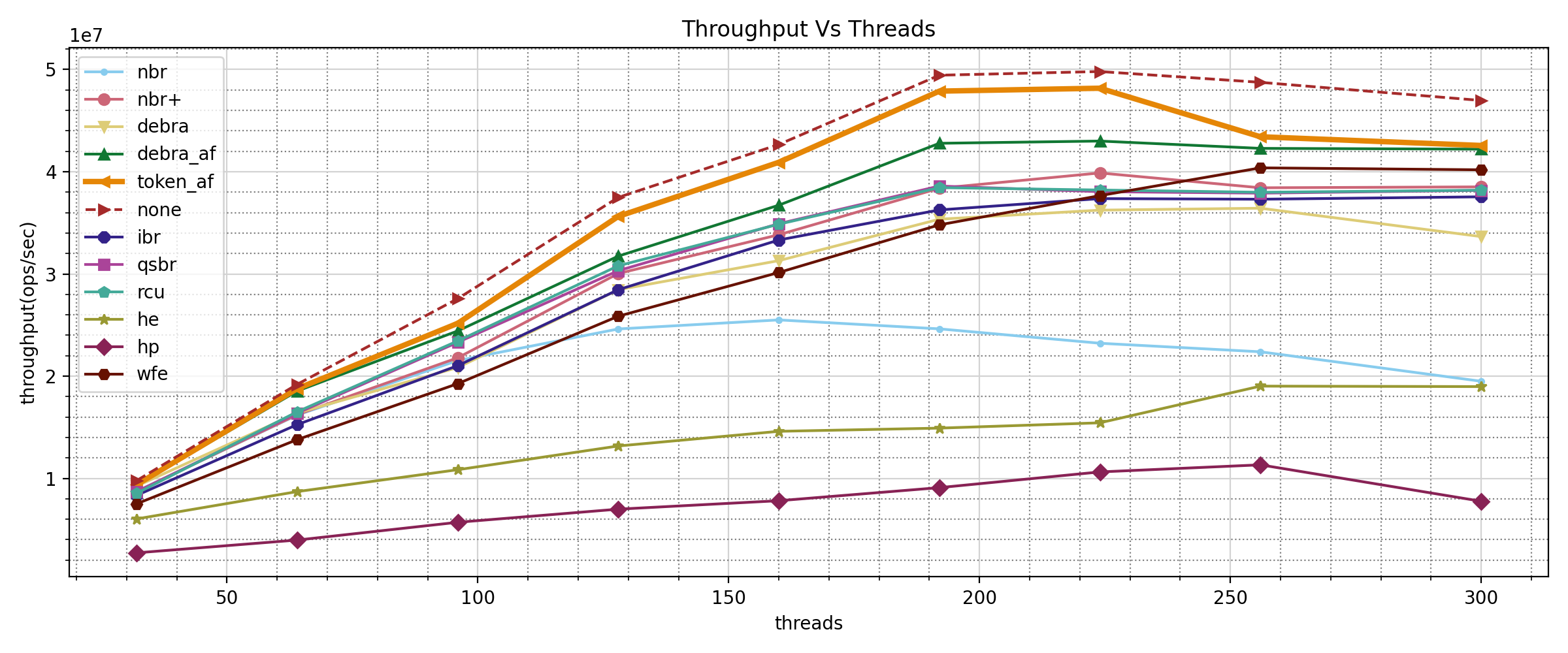}
        \caption{Comparison of proposed token\_af with other reclamation techniques across threads. Data structure: DGT Tree}
        \label{subfig:Zyra_DGT_fig11a}
    \end{subfigure}
    \hspace{0.03\linewidth}
    \begin{subfigure}{0.48\textwidth}
        \includegraphics[width=\textwidth, height=6cm]{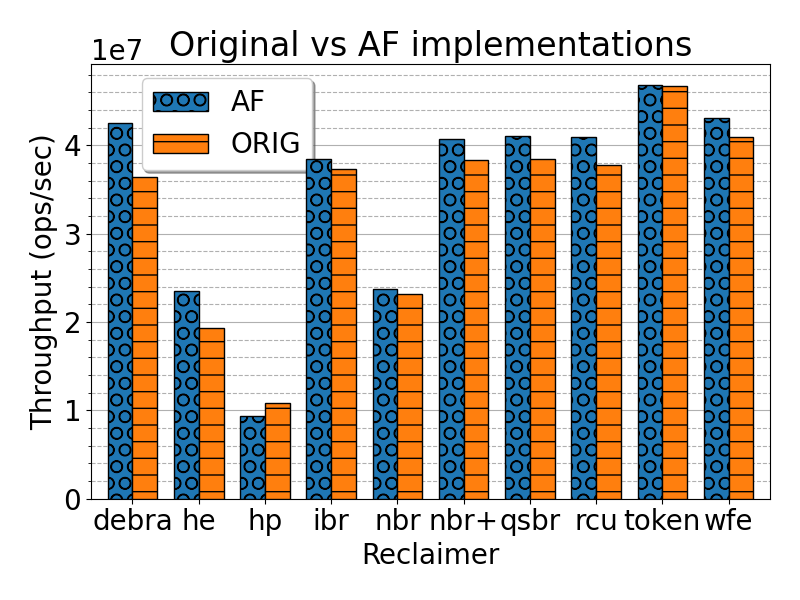}
        \caption{Comparison with amortized free versions of various reclaimers at 192 threads. Data structure: DGT Tree}
        \label{subfig:Nasus_DGT_fig11b}
    \end{subfigure}
    \caption{AMD 2 socket 256 core machine. Workload: 50\% inserts and 50\% deletes. Size: 20M. Allocator:JEmalloc}
\end{figure}

\clearpage
\section{Analysing the Visible Free Calls}
\begin{figure}[h]
    \begin{subfigure}{0.99\linewidth}
        \includegraphics[width=\linewidth]{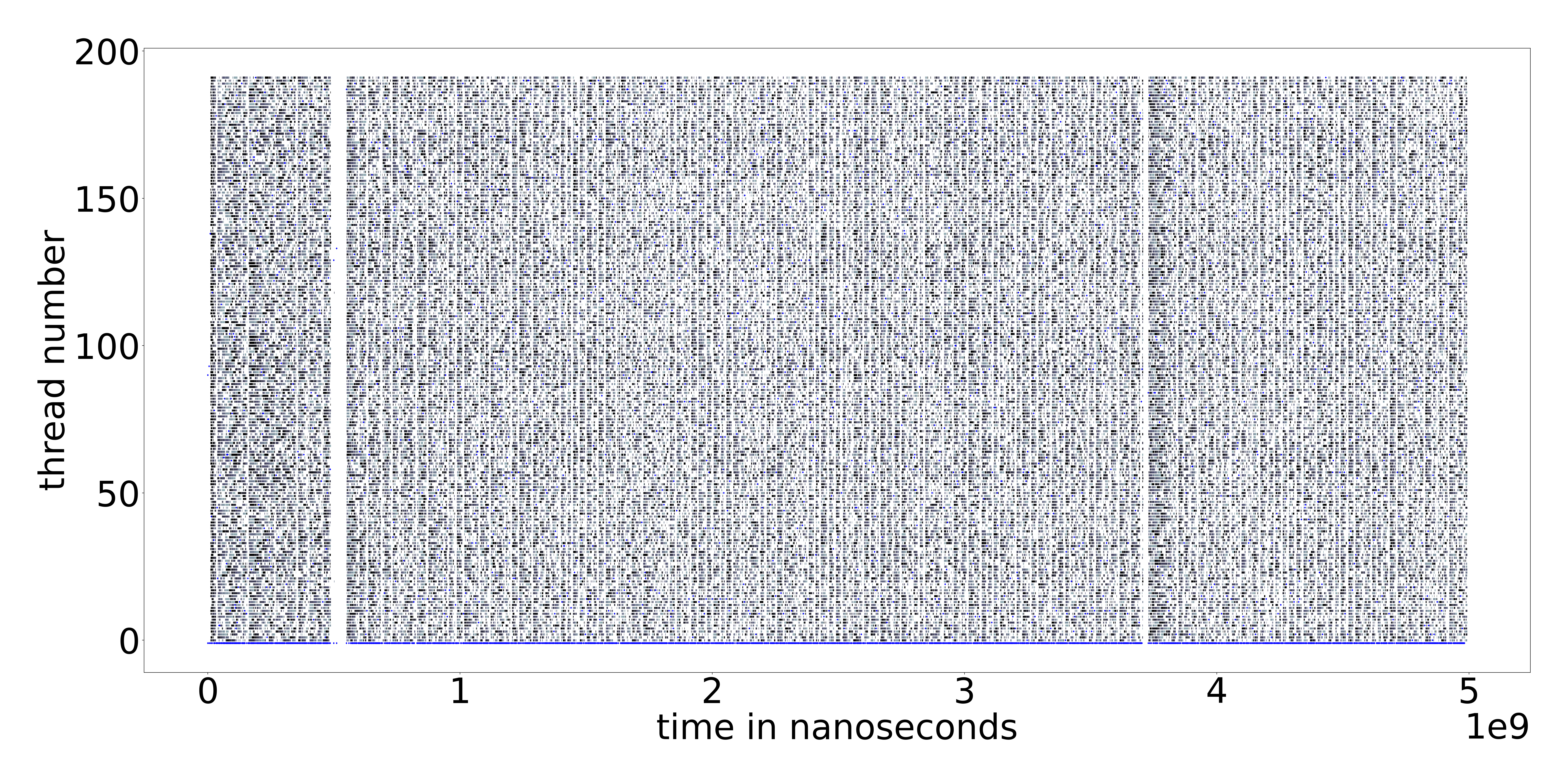}
    \end{subfigure}
    \begin{subfigure}{0.99\linewidth}
        \includegraphics[width=\linewidth]{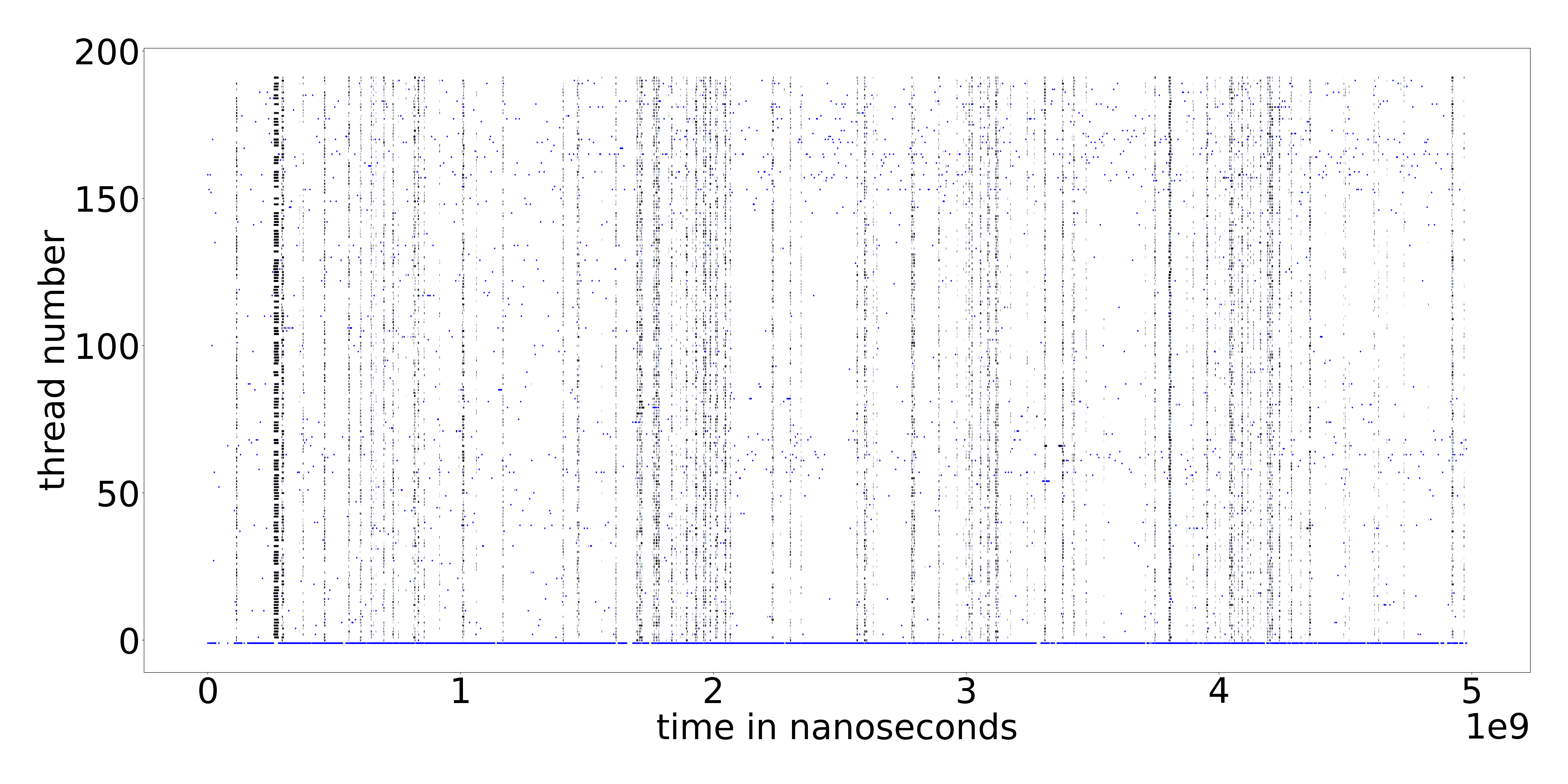}
    \end{subfigure}
    \caption{JEmalloc 192 threads. timeline graph for batch free (upper) and timeline graph for amortized free (lower). each box represents the time spent freeing an object}
    \label{fig:freeone-je192}
\end{figure}

\paragraph{Analysing the Visible Free Calls in Figure~\ref{fig:freeone-je192}}
Interestingly, the tiny fraction of calls that last long enough to be visible are arranged neatly into columns (as opposed to being randomly dispersed), which suggests they may have a common cause.
We gathered profiling data on the specific time intervals when these long \texttt{free} calls occur using Perftools, and found that they are correlated with large drops in CPU utilization.
Since there is no I/O in our benchmark, these drops in CPU utilization indicate that threads are either context switched out or sleeping on mutex locks.

With nothing else running on the system, the probability that all threads are simultaneously context switched out is vanishing.
Moreover, since we see references to the function \texttt{pthread\_mutex\_lock\_slow} in the profiling data, it is likely that all threads in a column are sleeping, waiting for a global mutex lock to be released.
%As we will see below, even though our amortized free approach still incurs some lock overhead, it is approximately 8$\times$ lower than when performing batch frees.
It would be an interesting direction for future work to identify the exact lock and its purpose. %, and to see if it manifests with other allocators or operating systems.

\clearpage
\section{Timeline graphs for DEBRA with JEmalloc, TCmalloc and MImalloc (48, 96, 192 and 240 threads)}
\begin{figure}[h]
    \begin{subfigure}{0.99\linewidth}
        \includegraphics[width=\linewidth]{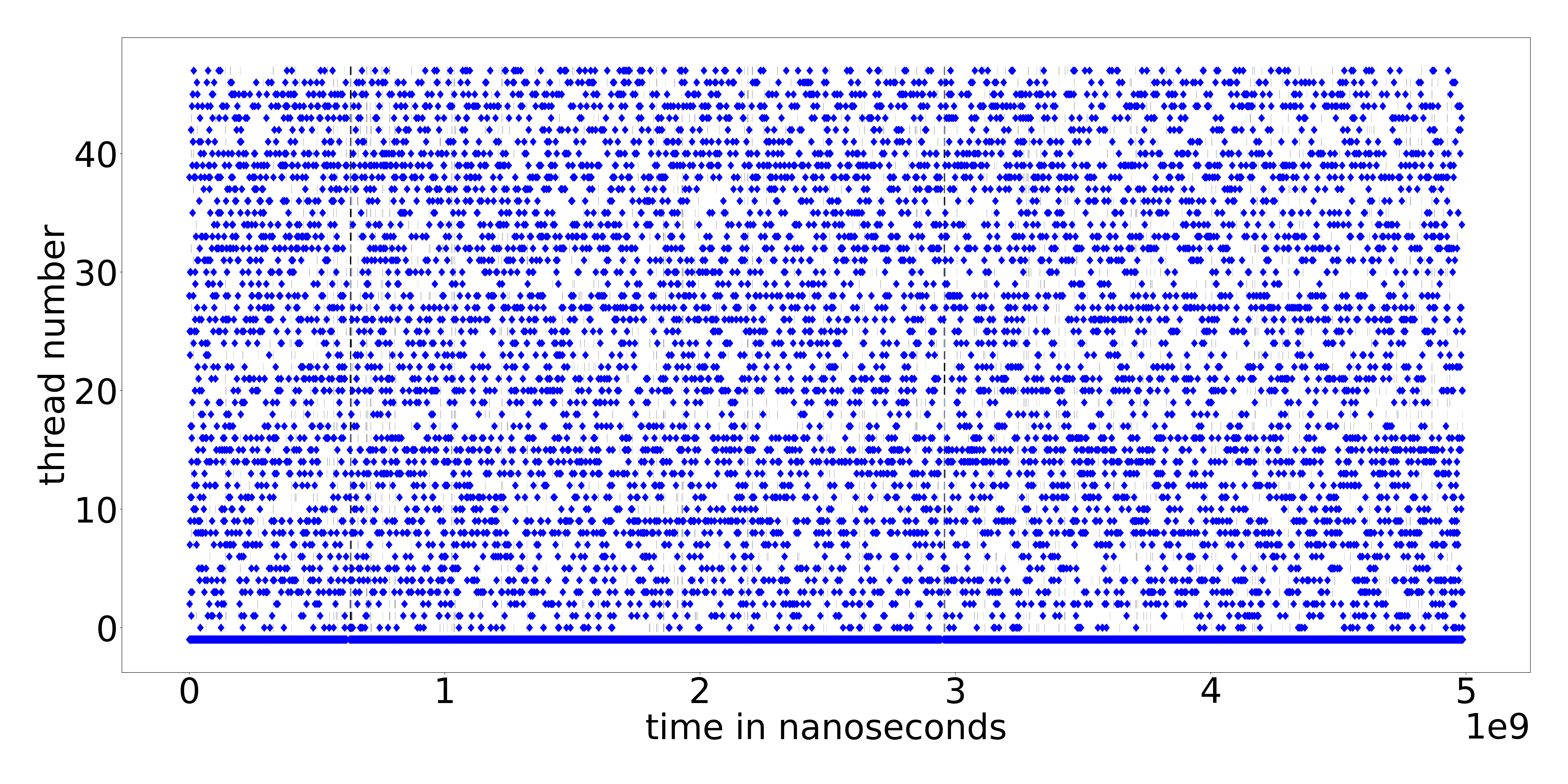}
    \end{subfigure}
    \begin{subfigure}{0.99\linewidth}
        \includegraphics[width=\linewidth]{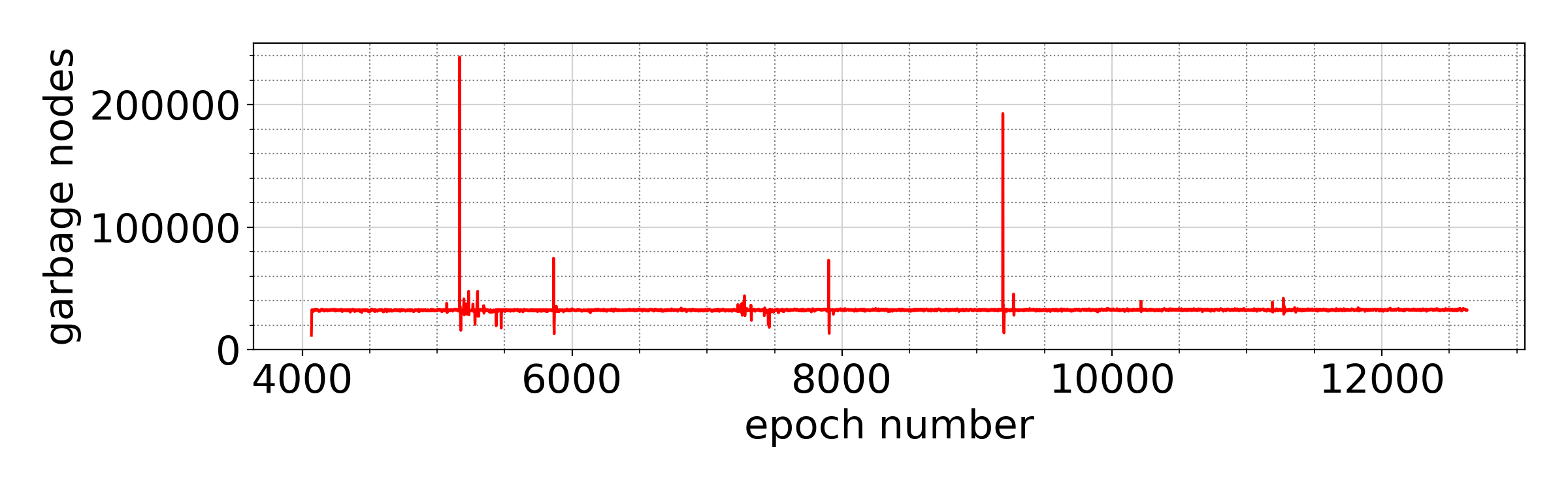}
    \end{subfigure}
    \caption{JEmalloc. DEBRA. 48 threads. timeline graph (upper) and average number of garbage nodes (lower)}
\end{figure}

\begin{figure}[h]
    \begin{subfigure}{0.99\linewidth}
        \includegraphics[width=\linewidth]{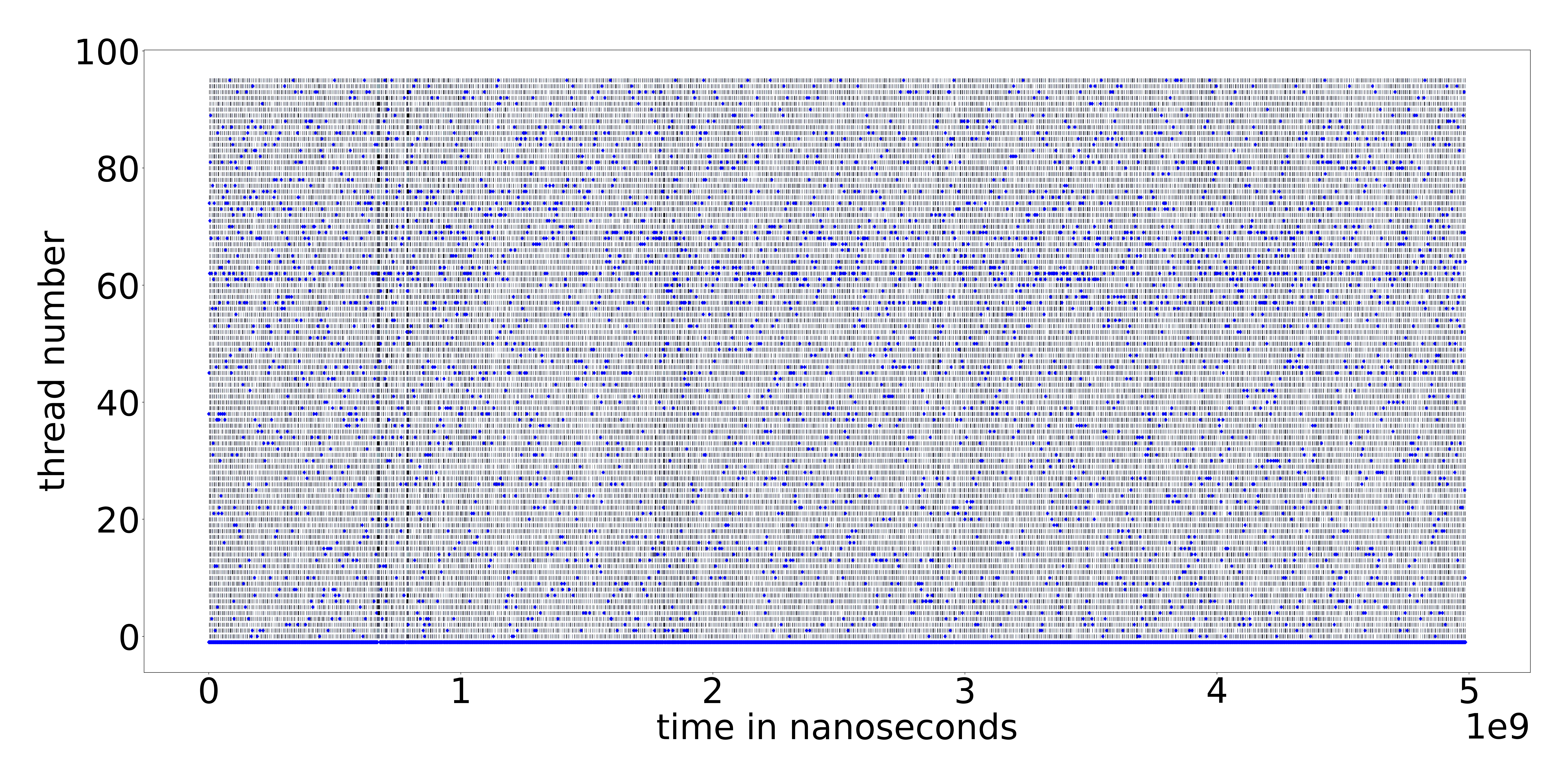}
    \end{subfigure}
    \begin{subfigure}{0.99\linewidth}
        \includegraphics[width=\linewidth]{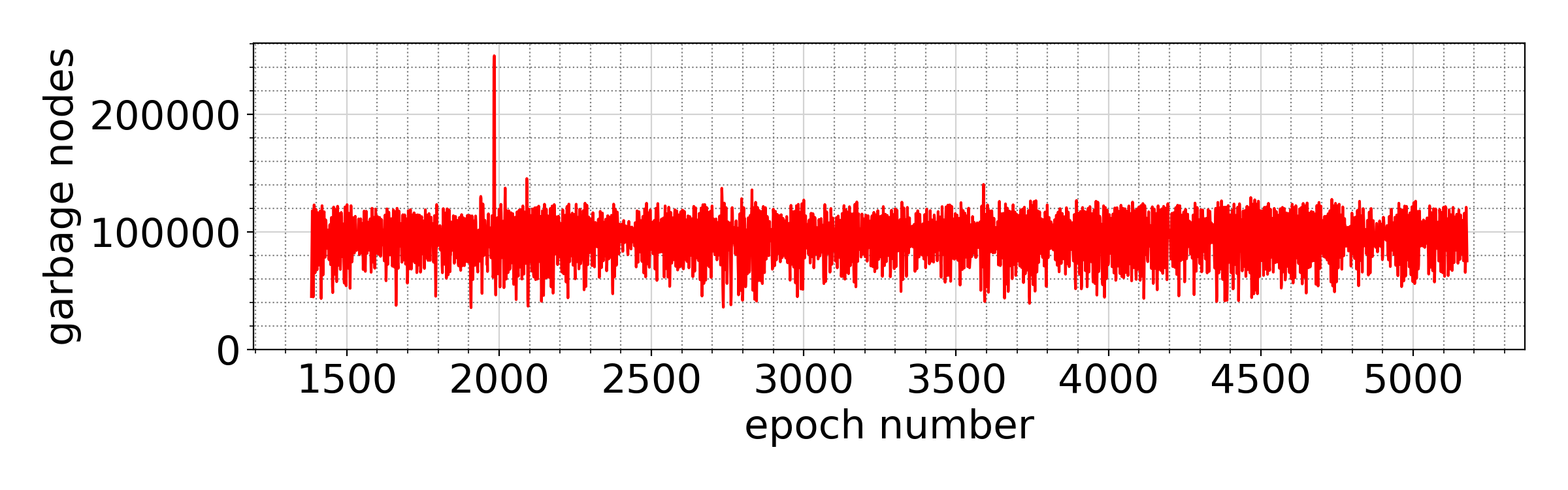}
    \end{subfigure}
    \caption{JEmalloc. DEBRA. 96 threads. timeline graph (upper) and average number of garbage nodes (lower)}
\end{figure}

\begin{figure}[h]
    \begin{subfigure}{0.99\linewidth}
        \includegraphics[width=\linewidth]{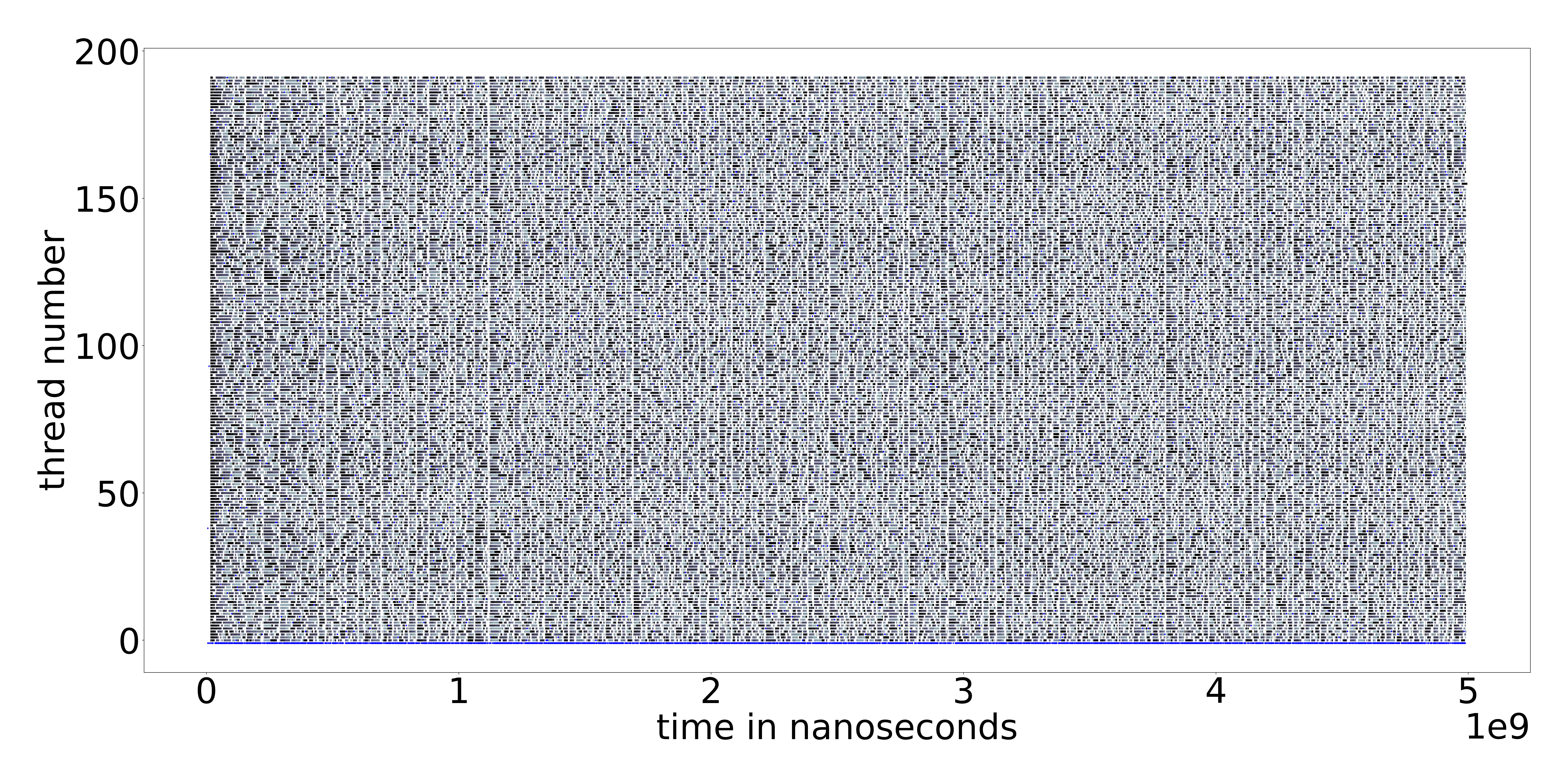}
    \end{subfigure}
    \begin{subfigure}{0.99\linewidth}
        \includegraphics[width=\linewidth]{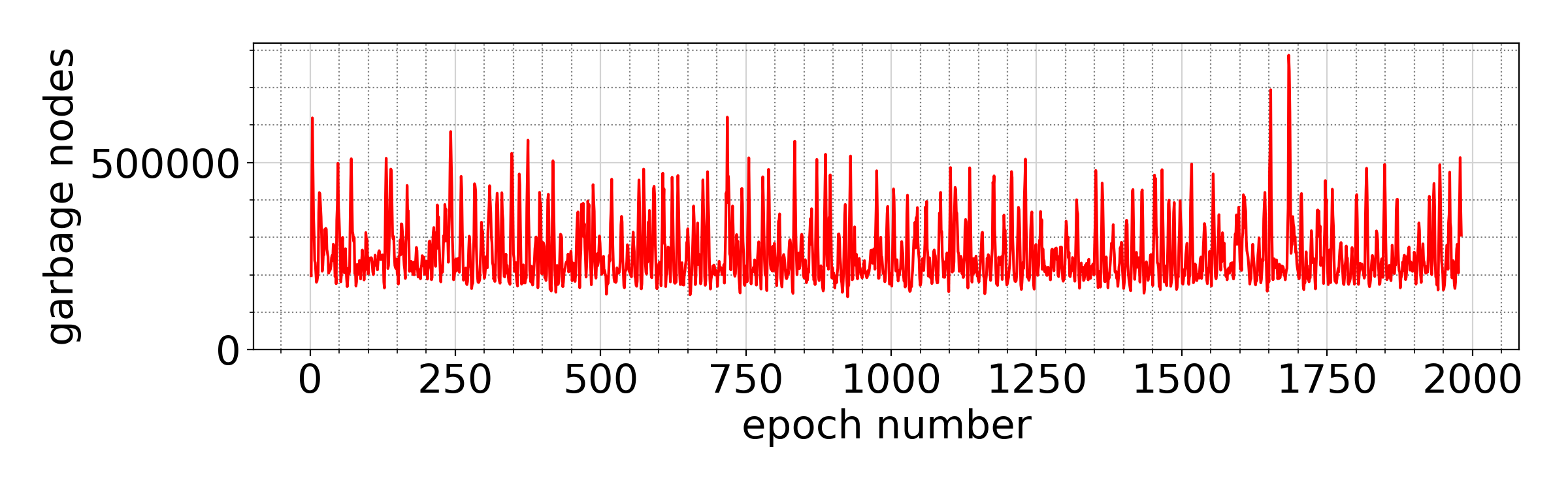}
    \end{subfigure}
    \caption{JEmalloc. DEBRA. 192 threads. timeline graph (upper) and average number of garbage nodes (lower)}
\end{figure}

\begin{figure}[h]
    \begin{subfigure}{0.99\linewidth}
        \includegraphics[width=\linewidth]{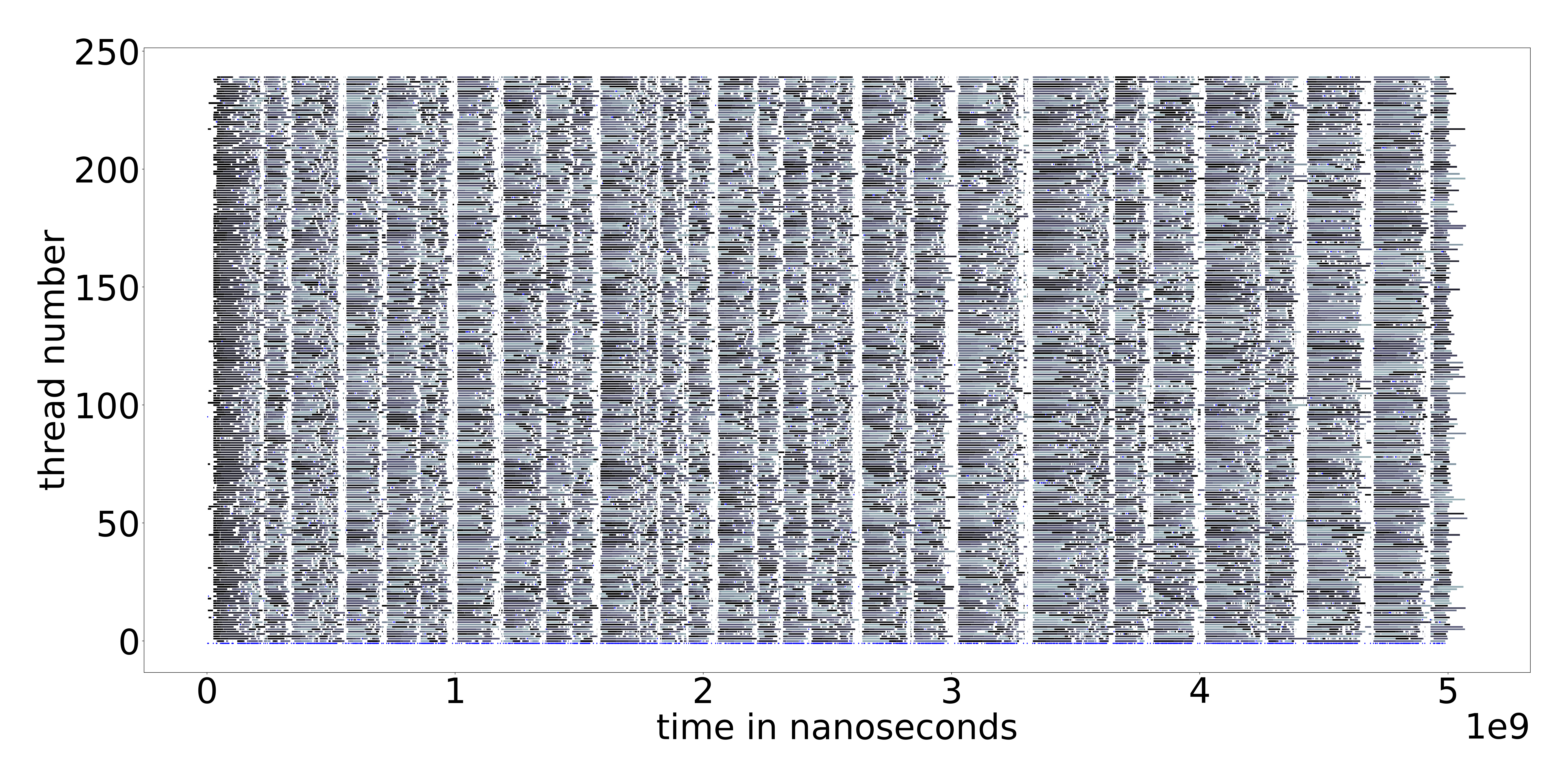}
    \end{subfigure}
    \begin{subfigure}{0.99\linewidth}
        \includegraphics[width=\linewidth]{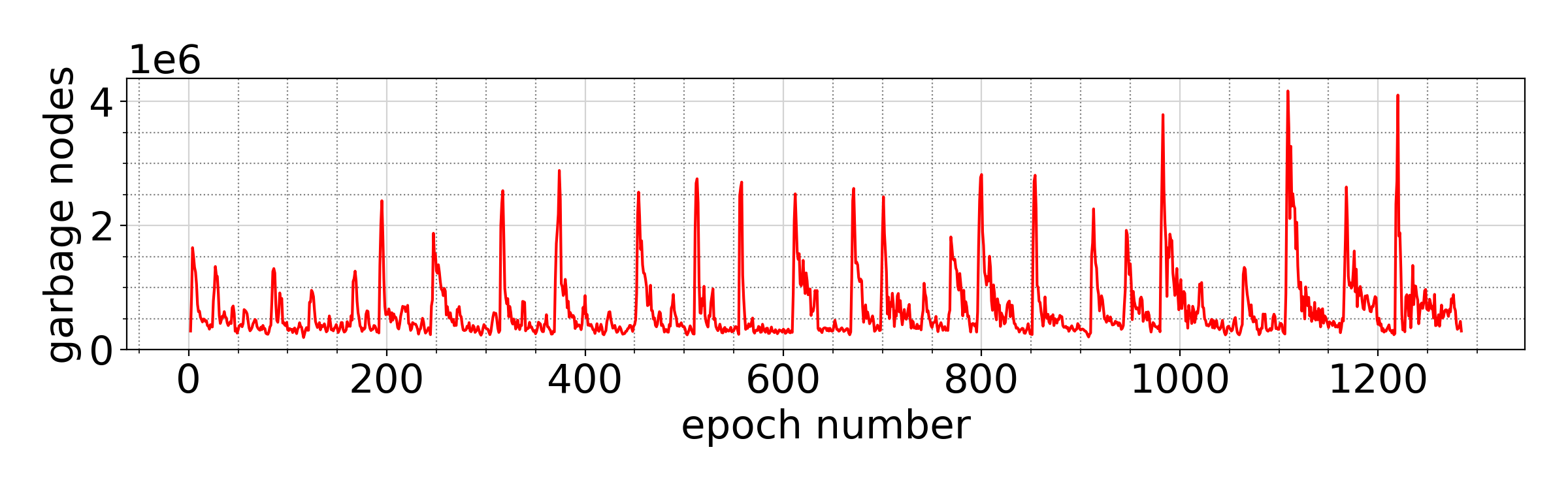}
    \end{subfigure}
    \caption{JEmalloc. DEBRA. 240 threads. timeline graph (upper) and average number of garbage nodes (lower)}
\end{figure}

\begin{figure}[h]
    \begin{subfigure}{0.99\linewidth}
        \includegraphics[width=\linewidth]{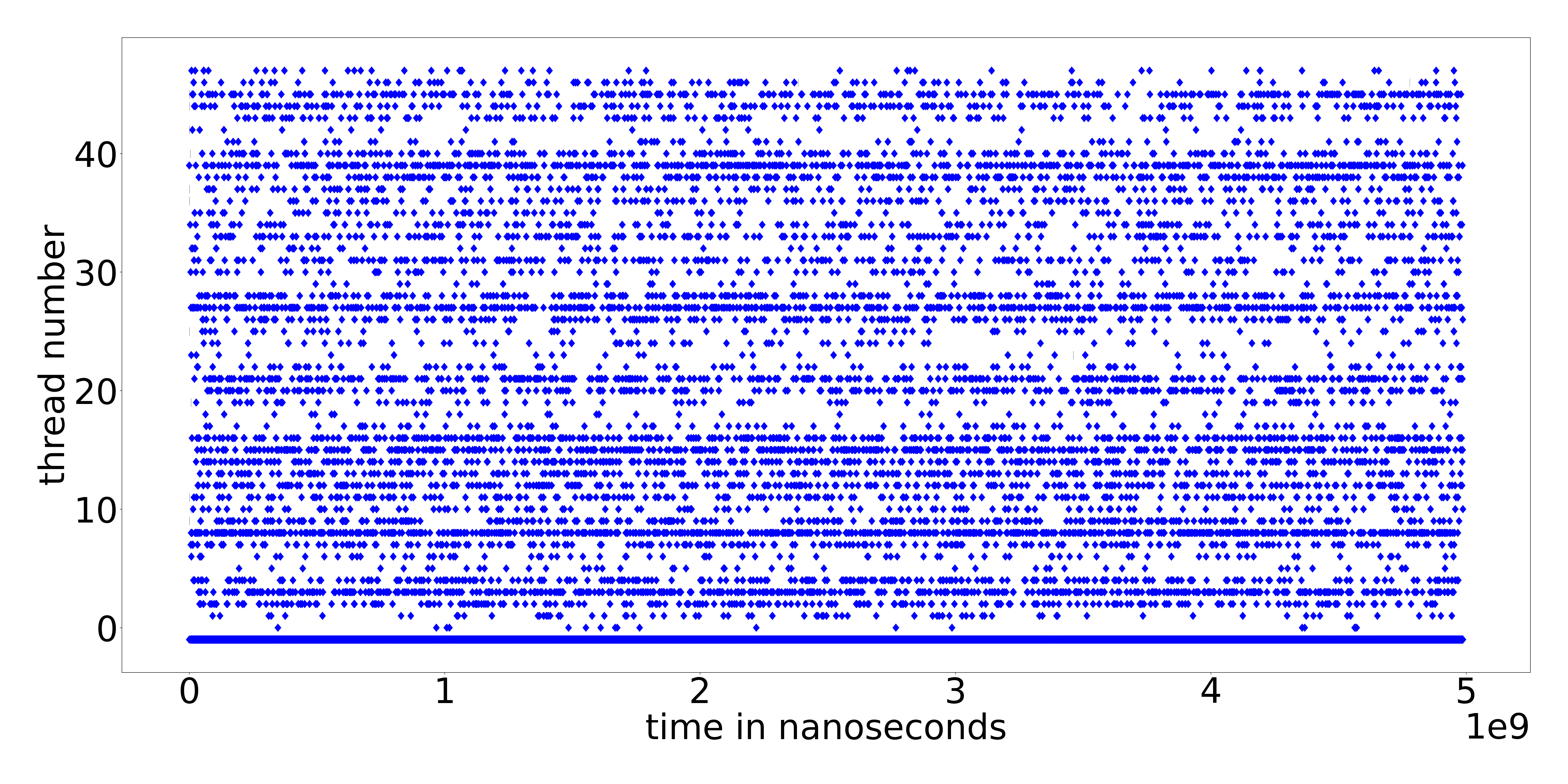}
    \end{subfigure}
    \begin{subfigure}{0.99\linewidth}
        \includegraphics[width=\linewidth]{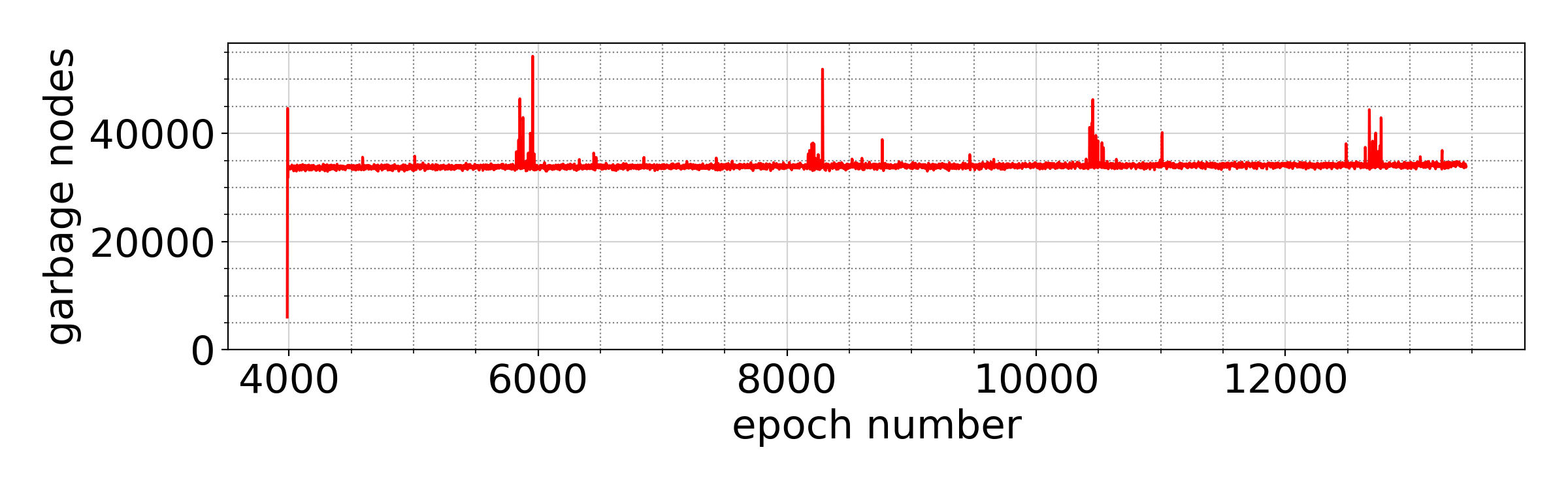}
    \end{subfigure}
    \caption{TCmalloc. DEBRA. 48 threads. timeline graph (upper) and average number of garbage nodes (lower)}
\end{figure}

\begin{figure*}[h]
    \begin{subfigure}{0.99\linewidth}
        \includegraphics[width=\linewidth]{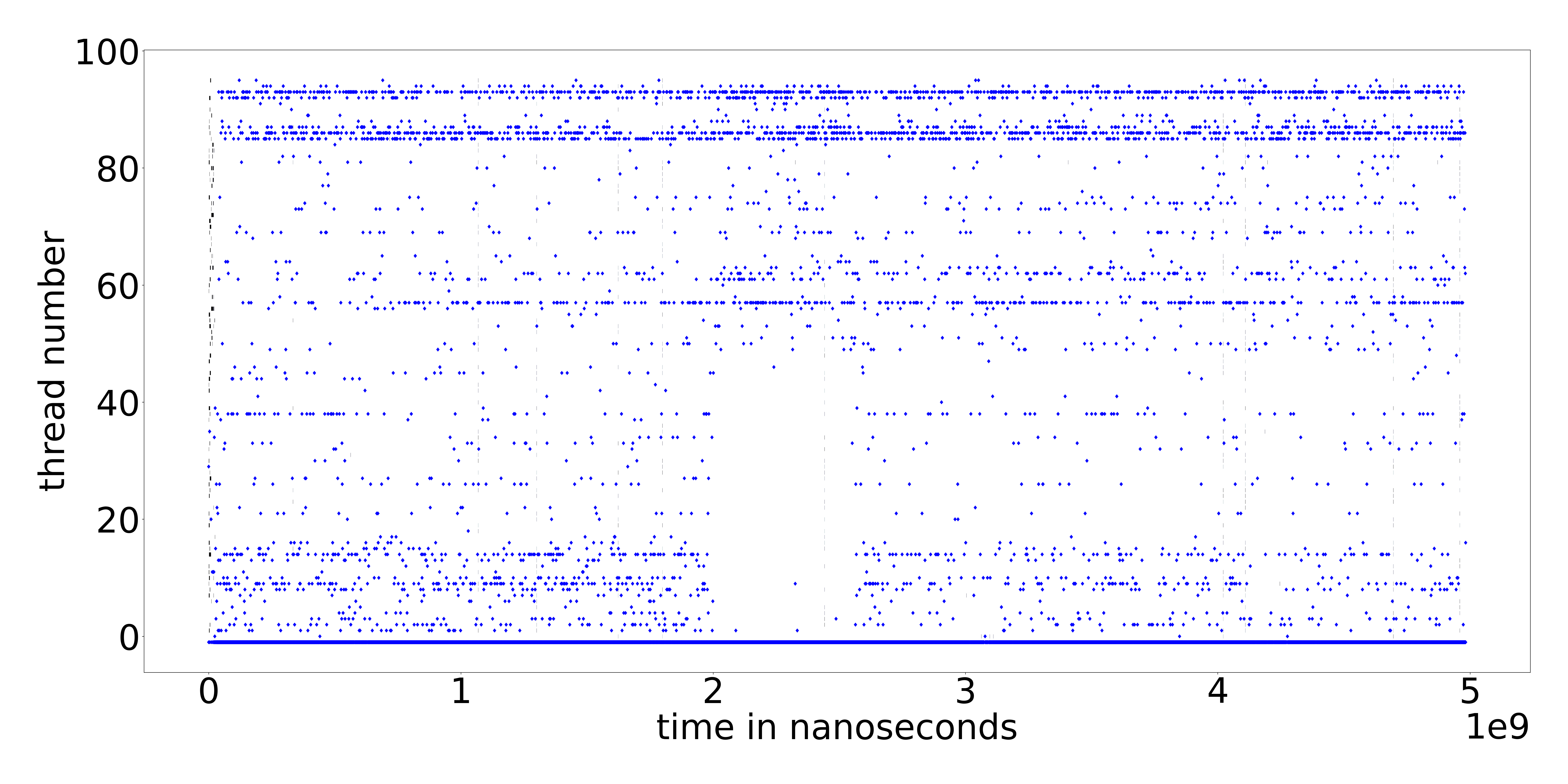}
    \end{subfigure}
    \begin{subfigure}{0.99\linewidth}
        \includegraphics[width=\linewidth]{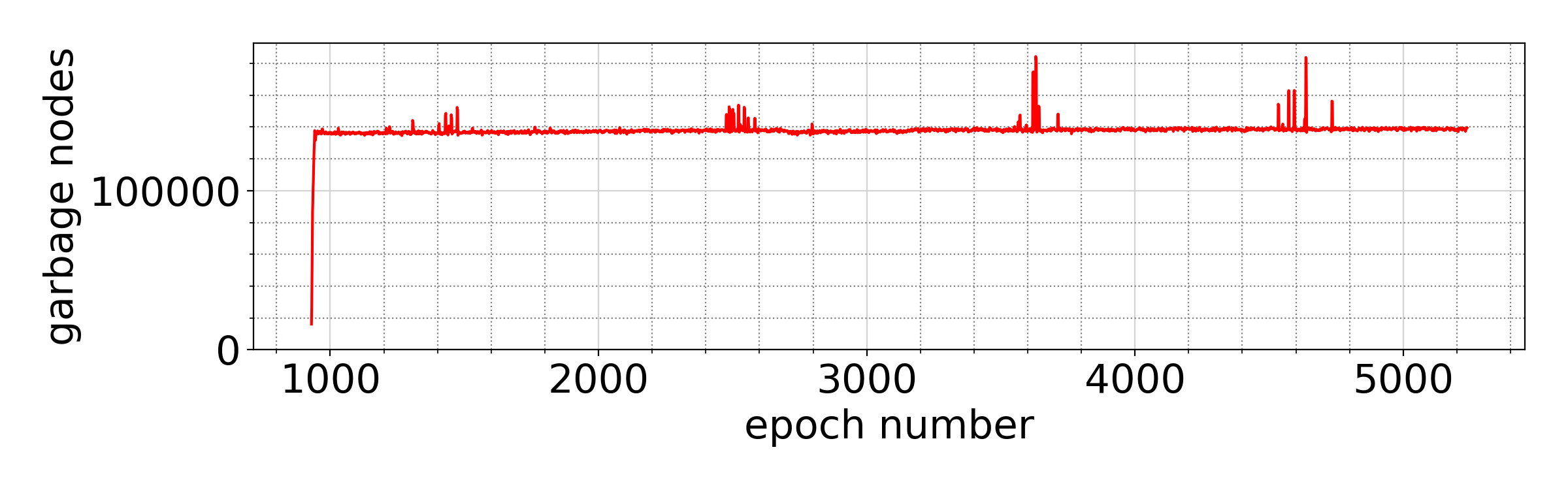}
    \end{subfigure}
    \caption{TCmalloc. DEBRA. 96 threads. timeline graph (upper) and average number of garbage nodes (lower)}
\end{figure*}

\begin{figure}[h]
    \begin{subfigure}{0.99\linewidth}
        \includegraphics[width=\linewidth]{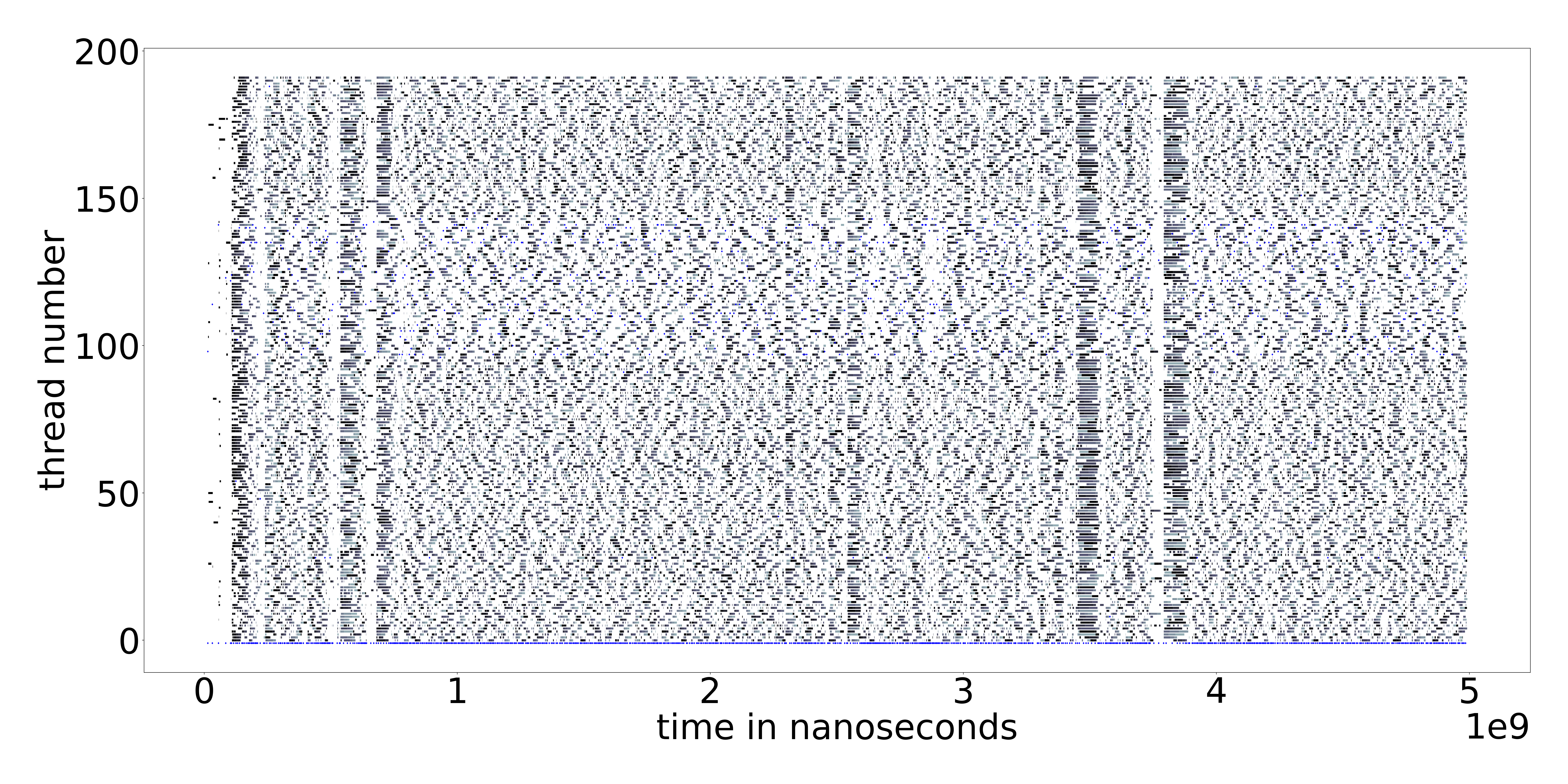}
    \end{subfigure}
    \begin{subfigure}{0.99\linewidth}
        \includegraphics[width=\linewidth]{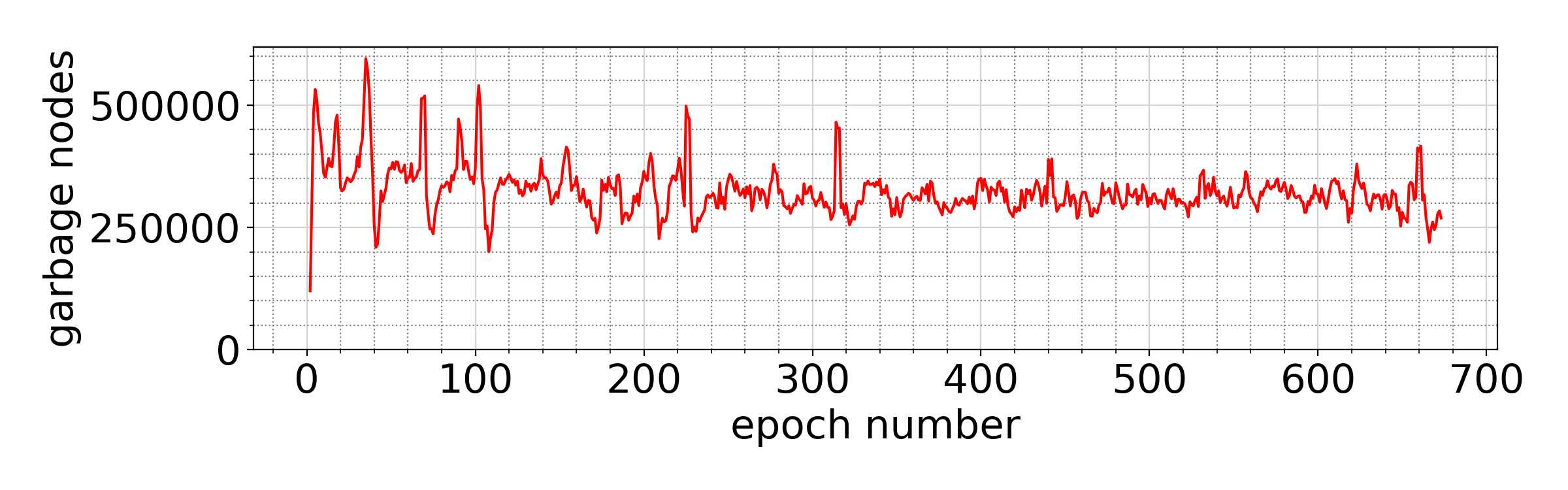}
    \end{subfigure}
    \caption{TCmalloc. DEBRA. 192 threads. timeline graph (upper) and average number of garbage nodes (lower)}
\end{figure}

\begin{figure}[h]
    \begin{subfigure}{0.99\linewidth}
        \includegraphics[width=\linewidth]{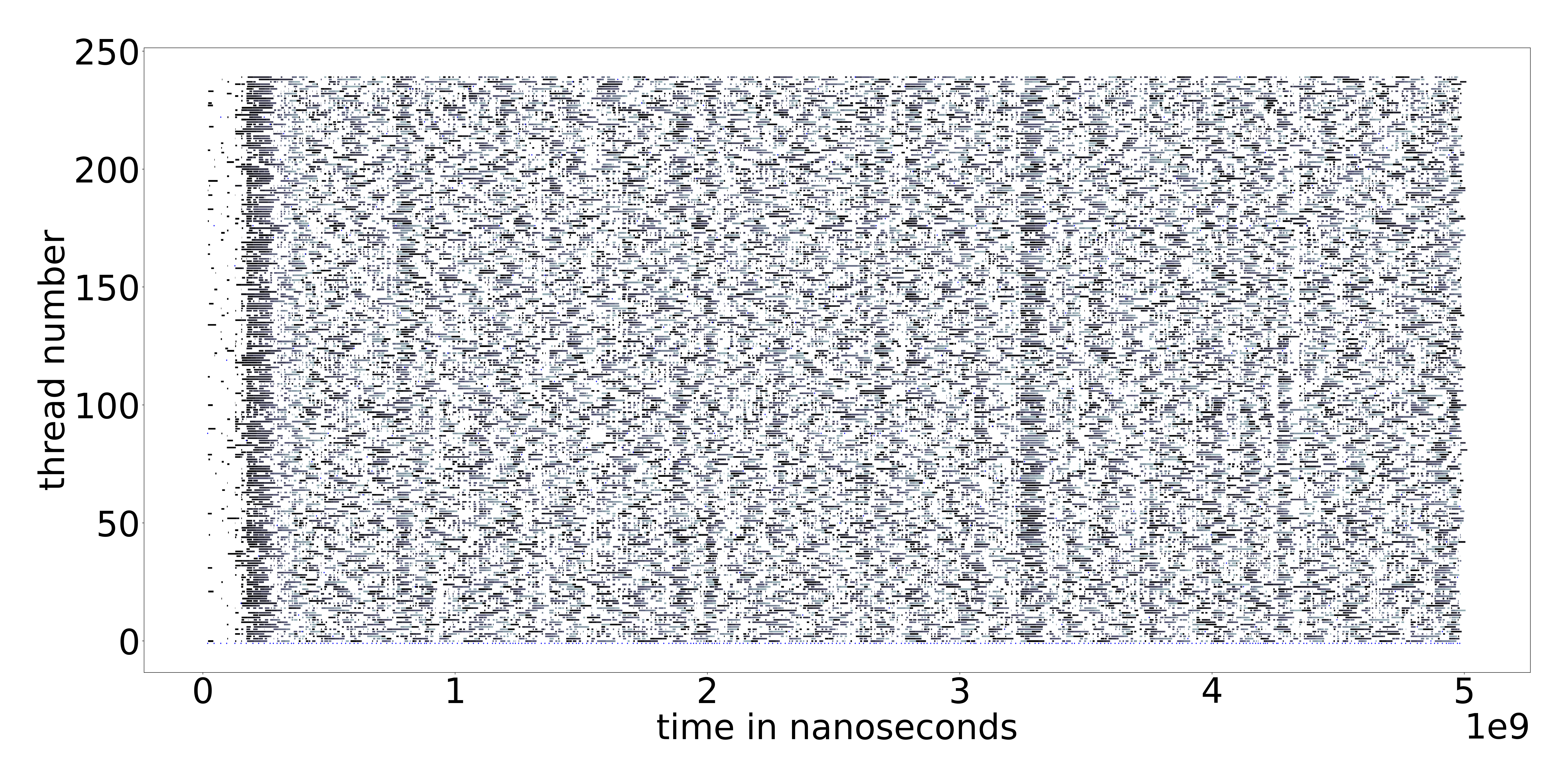}
    \end{subfigure}
    \begin{subfigure}{0.99\linewidth}
        \includegraphics[width=\linewidth]{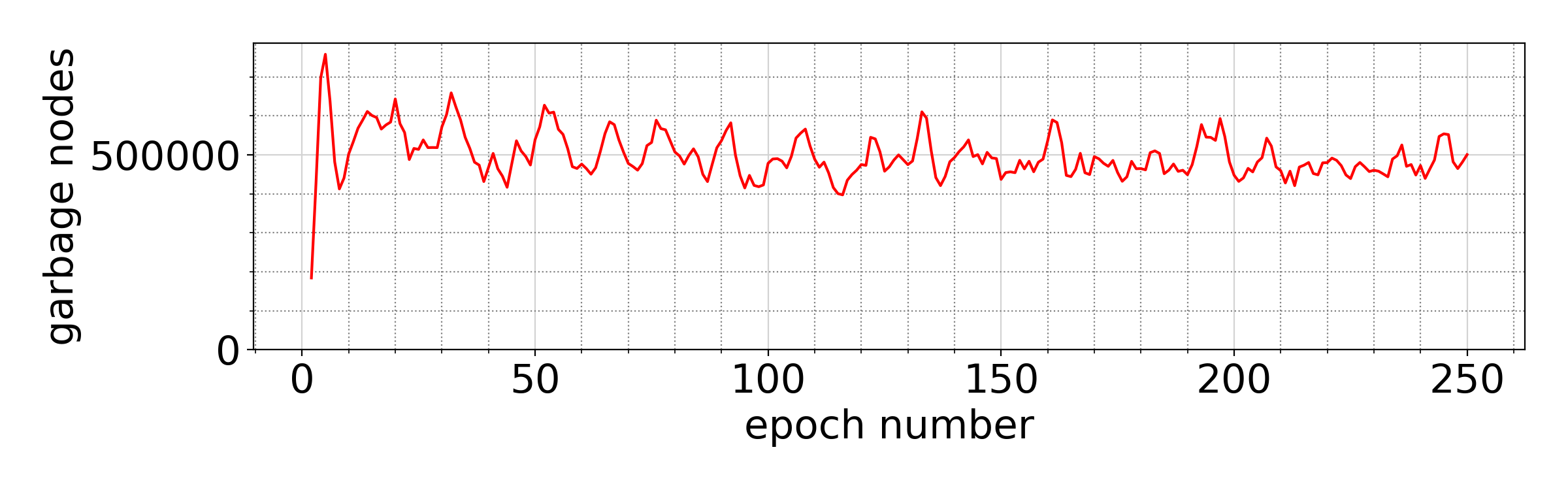}
    \end{subfigure}
    \caption{TCmalloc. DEBRA. 240 threads. timeline graph (upper) and average number of garbage nodes (lower)}
\end{figure}

\begin{figure}[h]
    \begin{subfigure}{0.99\linewidth}
        \includegraphics[width=\linewidth]{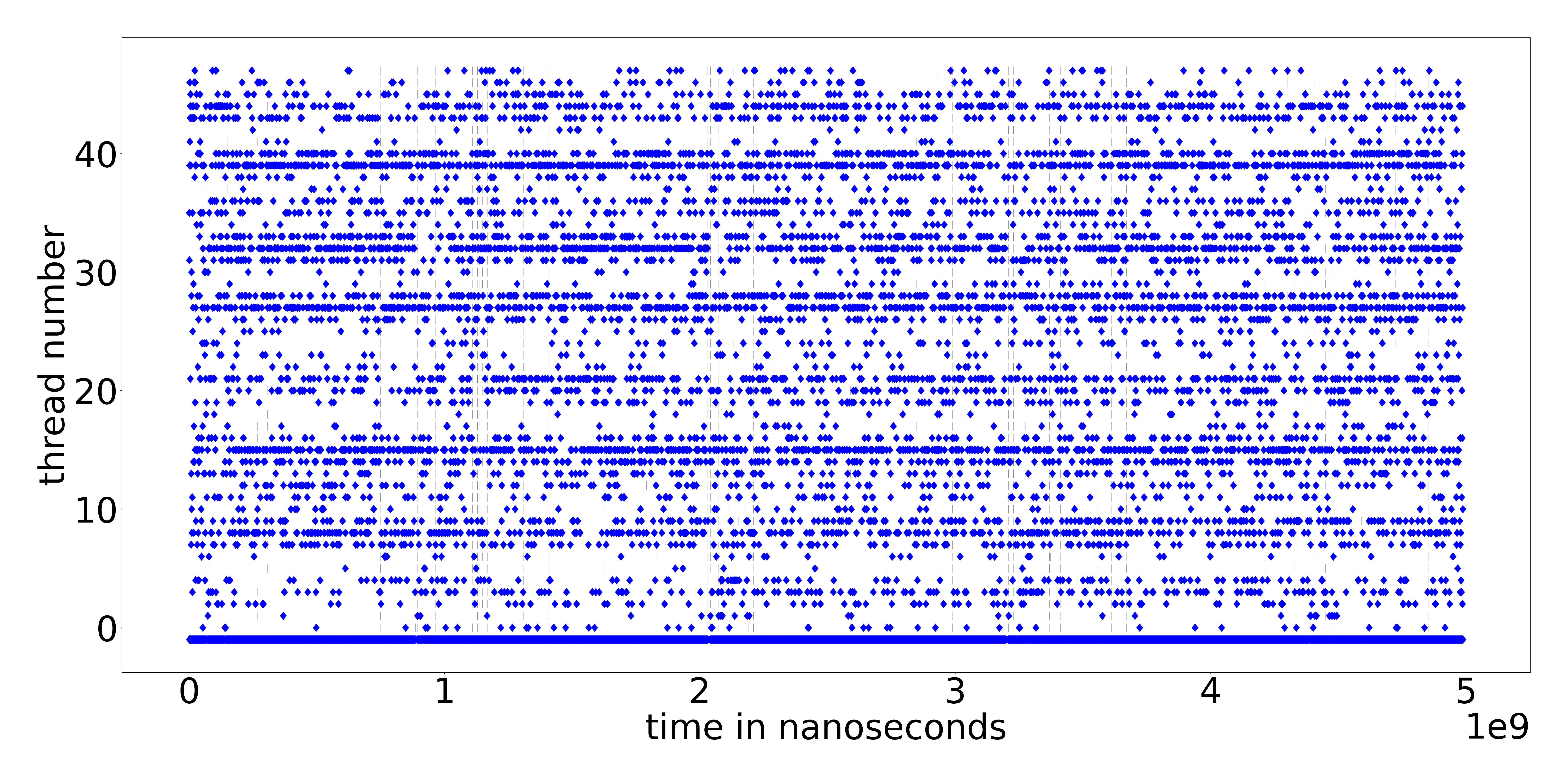}
    \end{subfigure}
    \begin{subfigure}{0.99\linewidth}
        \includegraphics[width=\linewidth]{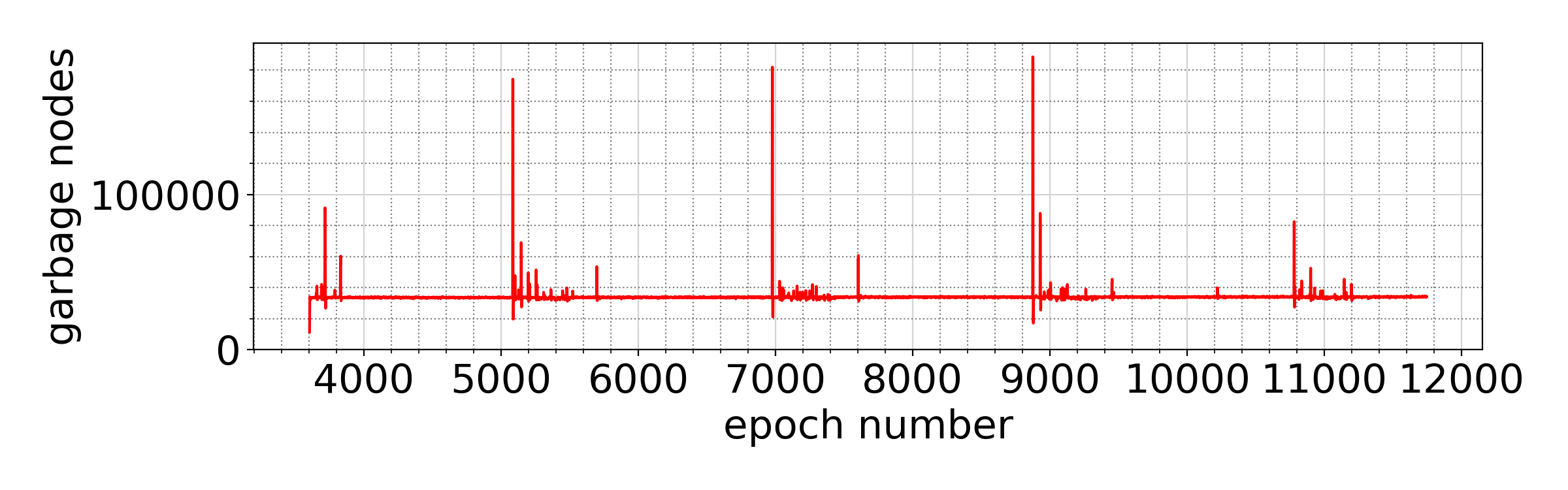}
    \end{subfigure}
    \caption{MImalloc. DEBRA. 48 threads. timeline graph (upper) and average number of garbage nodes (lower)}
\end{figure}

\begin{figure}[h]
    \begin{subfigure}{0.99\linewidth}
        \includegraphics[width=\linewidth]{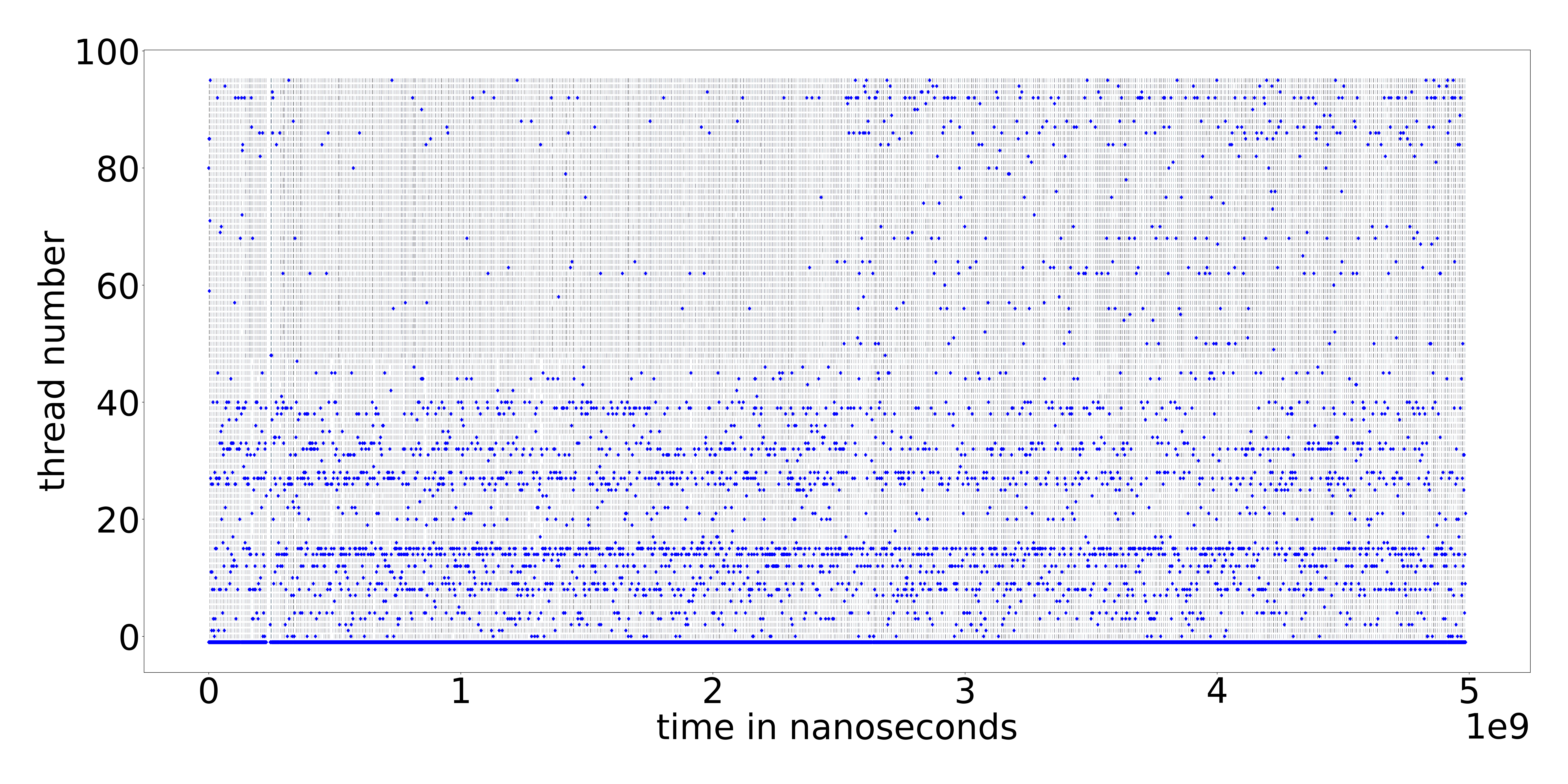}
    \end{subfigure}
    \begin{subfigure}{0.99\linewidth}
        \includegraphics[width=\linewidth]{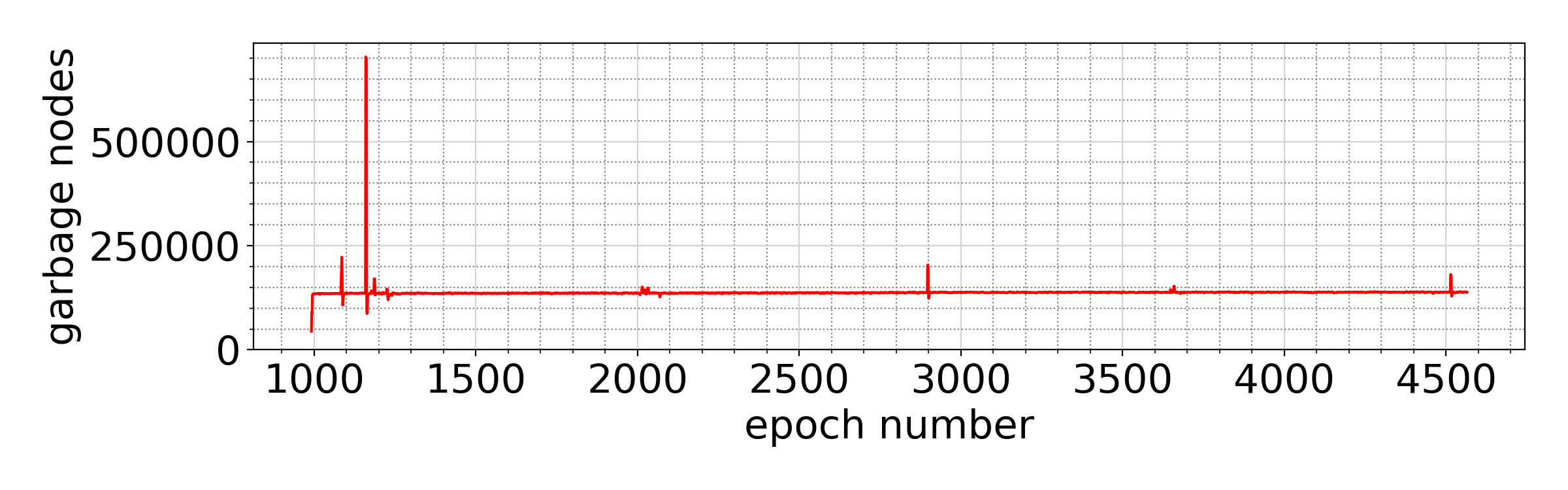}
    \end{subfigure}
    \caption{MImalloc. DEBRA. 96 threads. timeline graph (upper) and average number of garbage nodes (lower)}
\end{figure}

\begin{figure}[h]
    \begin{subfigure}{0.99\linewidth}
        \includegraphics[width=\linewidth]{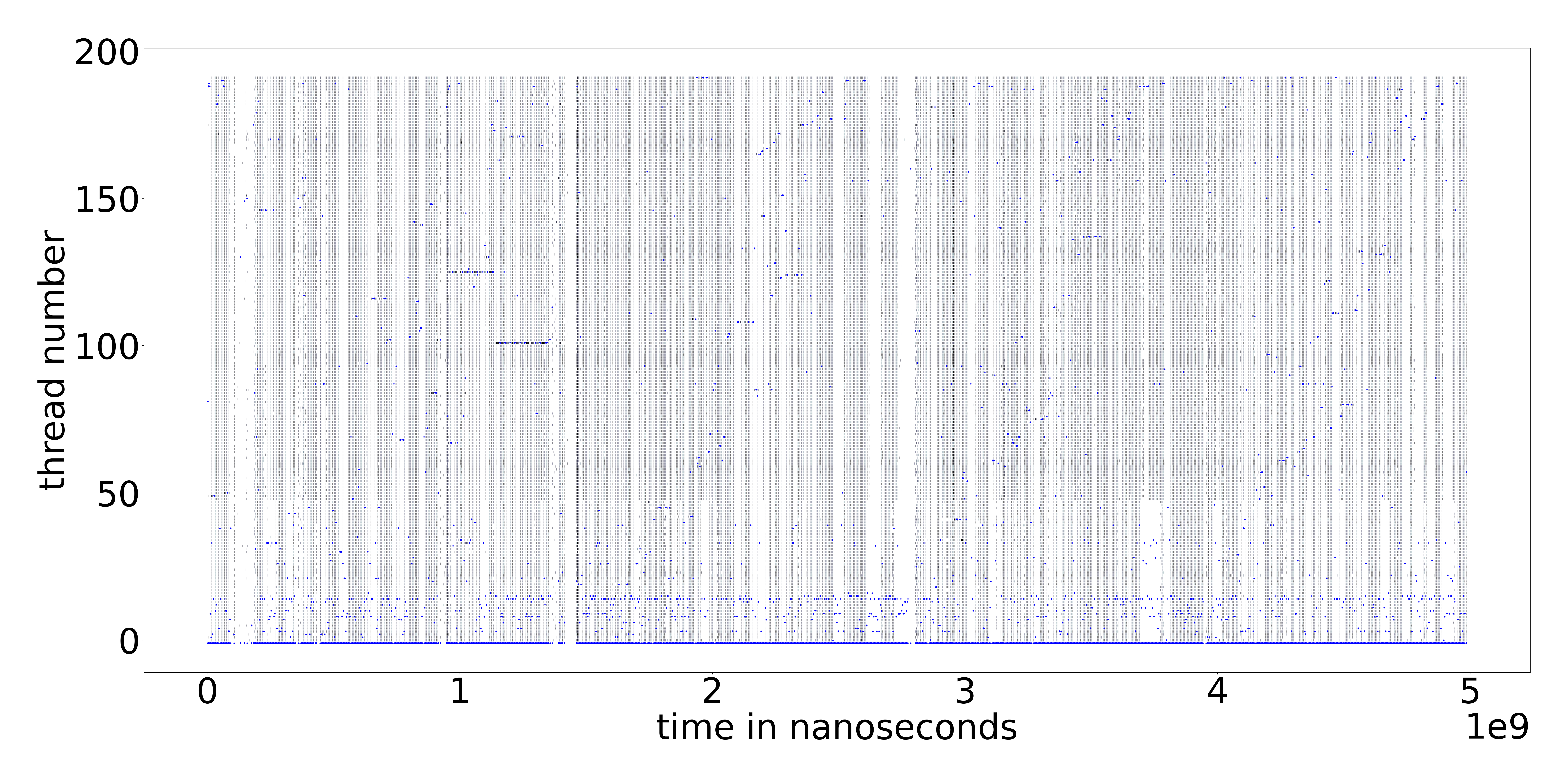}
    \end{subfigure}
    \begin{subfigure}{0.99\linewidth}
        \includegraphics[width=\linewidth]{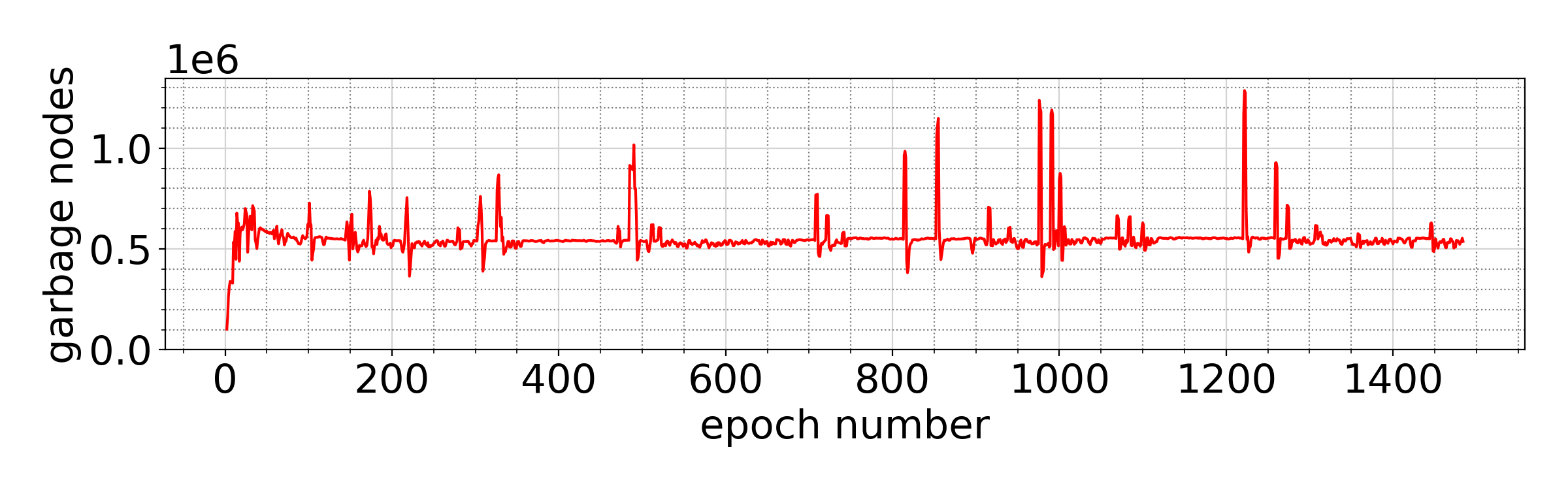}
    \end{subfigure}
    \caption{MImalloc. DEBRA. 192 threads. timeline graph (upper) and average number of garbage nodes (lower)}
\end{figure}

\begin{figure}[h]
    \begin{subfigure}{0.99\linewidth}
        \includegraphics[width=\linewidth]{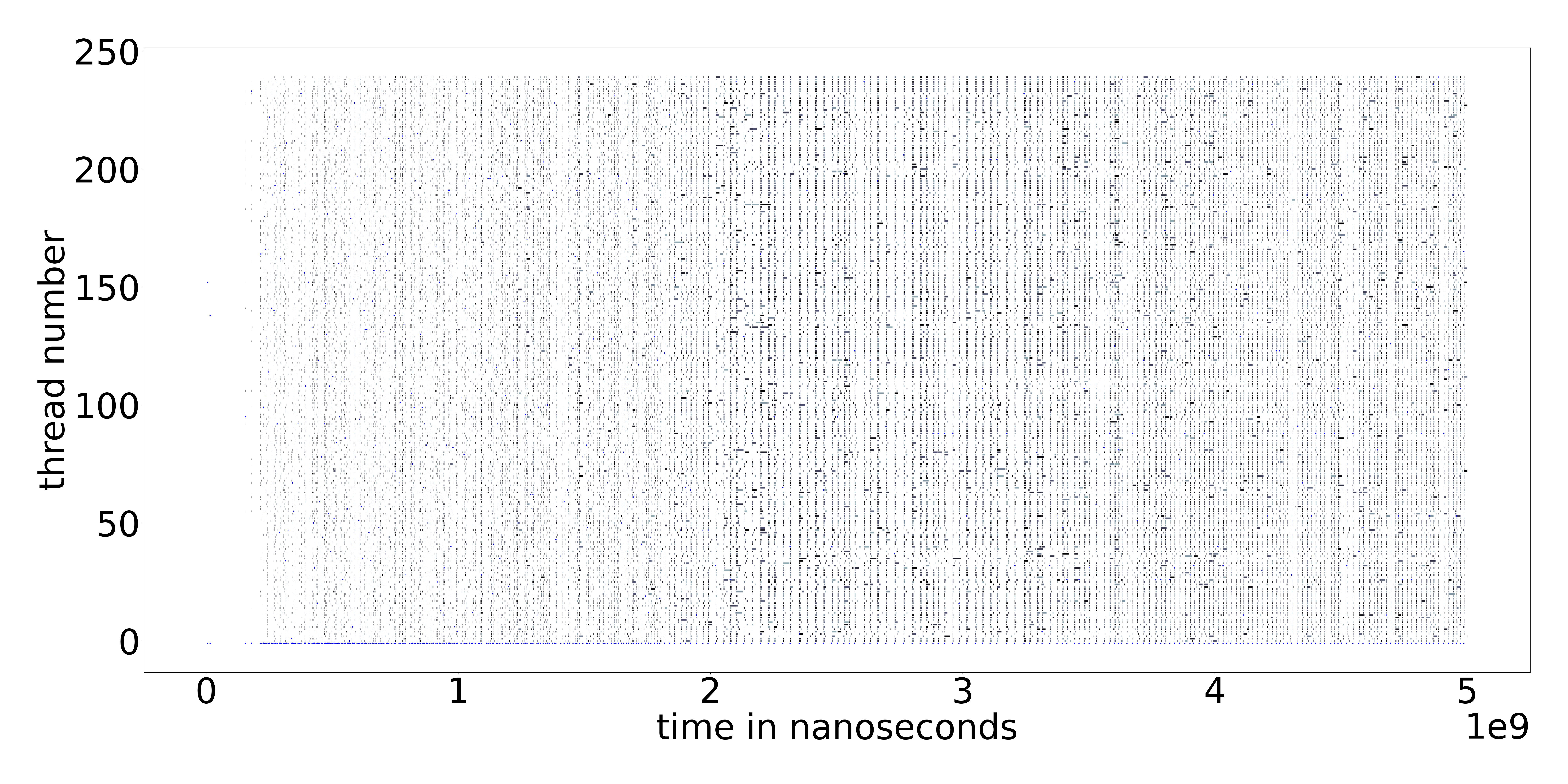}
    \end{subfigure}
    \begin{subfigure}{0.99\linewidth}
        \includegraphics[width=\linewidth]{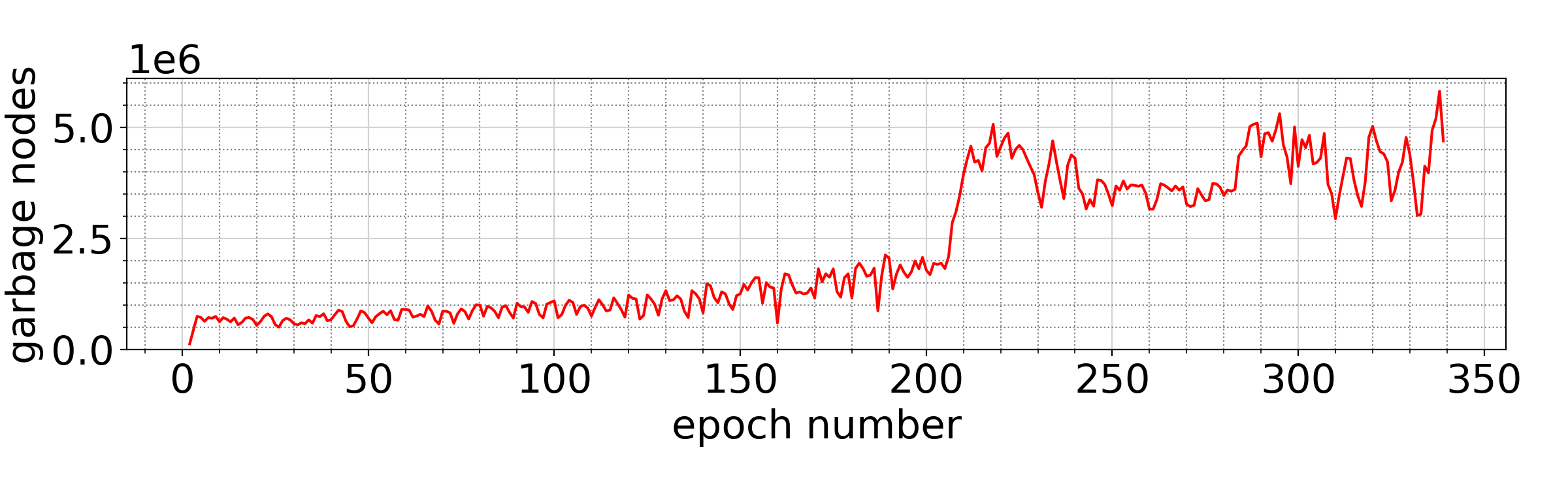}
    \end{subfigure}
    \caption{MImalloc. DEBRA. 240 threads. timeline graph (upper) and average number of garbage nodes (lower)}
\end{figure}

\end{document}